\documentclass[12pt]{article}
\usepackage[letterpaper,margin=1in,headsep=0.1in,headheight=1in]{geometry}

\usepackage[T1]{fontenc}
\usepackage{ae,aecompl}
\usepackage[utf8]{inputenc}
\usepackage{charter}
\usepackage{lipsum}

\usepackage[hidelinks]{hyperref}
\usepackage{graphicx}
\usepackage[sort&compress,numbers]{natbib}

\usepackage{siunitx}
\usepackage{amsmath,amssymb,amstext}
\usepackage{xspace}
\usepackage{aas_macros}

\usepackage{xcolor,colortbl,multirow}

\newcommand\snowmass{
\begin{center}
  \rule[-0.2in]{\hsize}{0.01in}\\
  \rule{\hsize}{0.01in}\\
  \vskip 0.1in
  Submitted to the Proceedings of the US Community Study\\ 
  on the Future of Particle Physics (Snowmass 2021)\\
  \rule{\hsize}{0.01in}\\
  \rule[+0.2in]{\hsize}{0.01in}\\[-2em]
\end{center}
}

\usepackage[firstpage=true]{background}
\backgroundsetup{contents={\parbox{6.5in}{\snowmass}}, scale=1,placement=top,opacity=1,color=black,position={3.25in,1.2in}}

\usepackage{fancyhdr}
\fancypagestyle{plain}{%
  \fancyhf{}%
  \fancyfoot[C]{\thepage}
}

\fancypagestyle{empty}{%
  \fancyhf{}%
  \fancyhead[C]{{\it Snowmass2021 Theory Frontier White Paper: Data-Driven Cosmology}}
  \fancyfoot[C]{\thepage}
}

\title{Snowmass2021 Theory Frontier White Paper: Data-Driven Cosmology}
\date{}

\usepackage{authblk}

\author[1]{Mustafa A. Amin}
\author[2]{Francis-Yan Cyr-Racine}
\author[3]{Tim Eifler}
\author[4]{Raphael Flauger}
\author[5]{Mikhail M. Ivanov}
\author[6]{Marilena LoVerde}
\author[6]{Caio B. de S. Nascimento}
\author[7]{Annika H. G. Peter}
\author[8]{Mark Vogelsberger}
\author[9]{Scott Watson}
\author[10]{Risa Wechsler}

\affil[1]{Department of Physics and Astronomy, Rice University, Houston, TX, USA
}
\affil[2]{Department of Physics and Astronomy, University of New Mexico, NM, USA}
\affil[3]{Department of Astronomy and Steward Observatory, University of Arizona, Tucson, AZ, USA}
\affil[4]{UC San Diego, Department of Physics, La Jolla, CA, USA}
\affil[5]{School of Natural Sciences, Institute for Advanced Study, Princeton, NJ, USA}
\affil[6]{Department of Physics, University of Washington, Seattle, WA, USA}
\affil[7]{CCAPP, Department of Physics, and Department of Astronomy, The Ohio State University, Columbus, OH, USA}
\affil[8]{Department of Physics \& Kavli Institute for Astrophysics and Space Research, MIT, Cambridge, MA, USA}
\affil[9]{Department of Physics, Syracuse University, Syracuse, NY, USA}
\affil[10]{Kavli Institute for Particle Astrophysics and Cosmology, Stanford University, Stanford, CA, USA}

\begin{document}

\maketitle

\begin{abstract}

Over the past few decades, astronomical and cosmological data sets firmly established the existence of physics beyond the Standard Model of particle physics by providing strong evidence for the existence of dark matter, dark energy, and non-zero neutrino mass. In addition, the generation of primordial perturbations most likely also relies on physics beyond the Standard Model of particle physics. Theory work, ranging from models of the early universe in string theory that that led to novel phenomenological predictions to the development of effective field theories to large-scale cosmological simulations that include the physics of galaxy evolution, has played a key role in analyzing and interpreting these data sets and in suggesting novel paths forward to isolate the origins of this new physics. Over the next decade, even more sensitive surveys are beginning to take data and are being planned.  In this white paper, we describe key areas of the theory program that will be needed to optimize the physics return on investment from these new observational opportunities.

\end{abstract}

\section{Introduction}

Cosmology is by now firmly established as a precision science. Different cosmological observations, ranging from observations of distant supernovae, to large scale structure (LSS) surveys, to measurements of the cosmic microwave background (CMB), have established a standard model of cosmology, referred to as $\Lambda$CDM. It describes the evolution of our universe from a time when it was only fractions of a second old until the present, and it does so with only six parameters. Four of the six parameters characterize the homogeneous solution, the remaining two characterize the power spectrum of primordial density perturbations. As a phenomenological model, $\Lambda$CDM has proven extremely successful and its parameters are now known at the percent level. However, the underlying microphysics behind these parameters remains completely unknown. How was the asymmetry between particles and anti-particles created? What is the nature of dark matter? Is dark energy just a cosmological constant or is it dynamical? When and how did the first stars form and how did reionization occur? What generated the primordial density perturbations that grew into the temperature anisotropies in the cosmic microwave background and eventually into the stars and galaxies? 

Implicitly the model makes additional assumptions, like the existence of three species of neutrinos with the sum of their masses assumed to be the smallest mass consistent with neutrino oscillation experiments. But what really is the sum of their masses? Are there three species of relativistic degrees of freedom present at the time of recombination? Or does the number of relativistic degrees of freedom deviate from the standard model prediction, as expected in many extensions of the standard model? 

Over the next decade, CMB experiments and LSS surveys are becoming powerful enough to begin to answer several of these questions. However, making full use of the upcoming data sets will require theoretical progress in several areas to ensure that our measurements are limited by the statistical power, not our theoretical understanding. 

In this white paper we present an overview of the main areas where observational progress is expected as well as the theoretical challenges associated with each of these areas that have to be overcome to fully utilize the next-generation data sets to reveal the physics of the primordial universe (\S\ref{sec:primordial}), dark matter (\S\ref{sec:darkmatter}), neutrinos and other possible light relics (\S\ref{sec:neutrinos}), dark energy (\S\ref{sec:darkenergy}), and the nature of the Hubble tension (\S\ref{sec:expansion}).  Finally, in \S\ref{sec:guide}, we demonstrate the potential of theory to go beyond interpreting observations, but guiding new physics searches.

\section{Primordial Universe}\label{sec:primordial}

One of the biggest open questions in cosmology is what generated the primordial perturbations that seeded the stars and galaxies around us. Observations have established that the primordial perturbations are dominated by density perturbations, and that, within observational uncertainties, these are adiabatic, Gaussian, nearly but not exactly scale-invariant, and well-described by a power law that is conventionally parameterized by the amplitude $A_s$ and spectral index $n_s$~\cite{Planck:2018nkj}.

All these properties of the primordial perturbations are consistent with inflation~\cite{Starobinsky:1980te,Guth:1980zm,Linde:1981mu,Albrecht:1982wi}, the idea that the very early universe underwent a period of nearly exponential expansion driven by one or several scalar fields, and have ruled out various competing ideas, such as perturbations seeded by monopoles, strings, or textures~\cite{Kibble:1976sj,Zeldovich:1980gh,Vilenkin:1981iu,Pen:1997ae,Planck:2013mgr}. Inflation is the most widely studied scenario for the early universe, but there are less explored alternative scenarios~\cite{Brandenberger:1988aj,Tseytlin:1991xk,Ijjas:2016tpn,Ijjas:2019pyf}. Future observations that constrain or detect the amplitude of primordial gravitational waves, measure the primordial power spectrum of density perturbations with higher precision, and further constrain departures from Gaussianity will provide stringent tests for any theory of the early universe.

In addition to a nearly scale invariant spectrum of primordial density perturbations, inflation also predicts a nearly scale invariant spectrum of primordial gravitational waves. According to the simplest models of inflation, the expansion is driven by a single scalar field, the inflaton $\phi$, and the theory is specified by an a priori arbitrary function of the inflaton, the potential $V(\phi)$. In this context, the search for primordial gravitational waves will help answer several key questions, such as what the energy scale of inflation is, how far the inflaton traveled in field space, and over what range the scalar potential varies. A more detailed discussion of inflation and its predictions is available in~\cite{Snowmass2021:inflation}.

A nearly scale-invariant background of gravitational waves is most easily detected through its imprint on the CMB, where the signal is characterized by a plateau on angular scales larger than a degree in the temperature anisotropies and a recombination peak on degree angular scales and a reionization bump on large angular scales in the polarization anisotropies. The primordial density perturbations generate temperature anisotropies and anisotropies in so-called $E$-mode polarization, primordial gravitational waves additionally generate so-called $B$-mode polarization. In the context of inflation, a detection of primordial $B$ modes would provide evidence for quantum fluctuations in the spacetime metric itself, would imply that inflation took place at energy scales comparable to those associated with grand unified theories, and would imply a Planckian field range, just to mention some of the consequences. Because of these far-reaching implications, several CMB experiments are currently searching for this polarization pattern~\cite{BICEP2:2018kqh,SPTpol:2020rqg,BicepKeck:2022dtc,SPIDER:2021ncy,POLARBEAR:2022dxa,Dahal:2021uig} and more will begin taking data soon.

One of these experiments, {\sc Bicep}2, reported a detection of $B$-mode polarization on degree scales at 150 GHz and interpreted this detection as evidence for the  existence  of  primordial  gravitational  waves~\cite{BICEP2:2014owc}. Subsequent analyses demonstrated that the {\sc Bicep}2 measurements are consistent with polarized  emission  from  dust  inside the Milky Way~\cite{Flauger:2014qra,Mortonson:2014bja,BICEP2:2015nss}. However, the measurement has both highlighted that experiments are reaching the sensitivities needed to detect the $B$ modes expected if the simplest models of inflation describe the first fraction of our universe, and that accounting for astrophysical foregrounds is crucial for a convincing detection.

Over the next years the {\sc Bicep}/{\it Keck} collaboration will continue to improve their measurements from the South Pole, and Simons Observatory will begin observations from Chile~\cite{SimonsObservatory:2018koc}, both improving the sensitivity to the amplitude of power in primordial gravitational waves by an order of magnitude over current limits. By the end of the decade a larger NSF and DOE funded effort, CMB-S4, will further increase the reach by a factor of around five and either detect a primordial gravitational wave signal or exclude many of the best-motivated models of inflation~\cite{CMB-S4:2016ple,Abazajian:2019eic,CMB-S4:2020lpa}. Planning for space-based probes is also well-underway, for example, for the JAXA led LiteBIRD satellite~\cite{LiteBIRD:2022cnt} and for PICO~\cite{NASAPICO:2019thw}.

These experiments will begin to cross critical thresholds in the search for primordial gravitational waves from inflation. For example, they will either detect gravitational waves or exclude one of the two classes of potentials that naturally predict a value of the spectral index $n_{\rm s}$ consistent with observations~\cite{CMB-S4:2016ple}. In this class of models the potential during the inflationary period is well approximated by a monomial potential $V(\phi)\approx \mu^{4-2p}\phi^{2p}$. The value of $\mu$ is constrained by the observed amplitude of density perturbations $A_{\rm s}$, and for plausible values of $p$ the models predict an amplitude of the gravitational wave signal, measured in terms of the tensor-to-scalar ratio $r$, of $r> 0.01$. This class of models already appears in tension with observations~\cite{BICEP:2021xfz}. However, the tension is predominantly based on constraints on $n_{\rm s}$, and the theoretical prediction is modified by the presence of additional degrees of freedom~\cite{Wenren:2014cga}. In addition, a detection of a primordial gravitational wave signal with an amplitude in excess of $r\simeq 0.01$ are of interest because they imply that the distance in field space traveled by the inflaton during the inflationary period exceeds the Planck scale~\cite{Lyth:1996im}, which would have profound theoretical implications, like the existence of a symmetry that protects the inflaton potential in quantum gravity. See the Snowmass white paper~\cite{Snowmass2021:stringinflation} for more details.

An additional critical threshold that CMB experiments will begin to cross over the next decade is $r\simeq 0.001$. This threshold provides information about the structure of the inflationary potential rather than the field displacement. The second class of single-field models that naturally predict a value of the spectral index $n_{\rm s}$ consistent with observations are hilltop and plateau models. Prominent examples in this class include Starobinsky's $R^2$ inflation~\cite{Starobinsky:1980te}, models in which the Higgs boson takes on the additional role of the inflaton~\cite{Bezrukov:2007ep,Ballesteros:2016xej}, $\alpha$-attractors~\cite{Kallosh:2013yoa,Kallosh:2014rga,Carrasco:2015pla}, fibre inflation~\cite{Cicoli:2008gp}, and Poincar\'e disk models~\cite{Ferrara:2016fwe,Kallosh:2017ced}. These potentials contain an additional parameter compared to the first class, the scale in field space over which the potential appreciably changes from its hilltop or plateau value. The absence of a detection by CMB-S4 and LiteBIRD would constrain the tensor-to-scalar ratio to be below $r\simeq 0.001$, and would exclude all models in this class in which this scale exceeds the Planck scale. 

Primordial $B$ modes generated by primordial gravitational waves are not the only source of $B$-mode polarization. As we already mentioned, astrophysical foregrounds also create $B$ modes. In addition, the presence of matter along the line of sight between us and the so-called last-scattering surface at which the CMB is emitted deflects the CMB photons through weak gravitational lensing. This converts primordial $E$ modes into a mixture of $E$ modes and so-called lensing $B$ modes~\cite{Zaldarriaga:1998ar,Hu:2000ee}, which are brightest on angular scales around ten arcminutes. 

Ground-based experiments will probe tensor-to-scalar ratios as small as $r\simeq 0.001$ by searching for the recombination peak on a carefully selected part  of  the  sky  with  minimal  foreground  contamination.  To reduce the sample variance from lensing $B$ modes, these experiments will rely on high-precision measurements of polarization on arcminute angular scales that will allow the removal of the lensing contribution on degree scales~\cite{Knox:2002pe,Kesden:2002ku,Seljak:2003pn,Smith:2010gu}. This process is referred to as delensing. Satellite missions measure primordial $B$ modes over a large  fraction of the sky. As a consequence, they are less dependent on delensing both because the sample variance of the lensing $B$ modes is reduced, simply because more modes are observed, and because they also target the reionization bump on large angular scales where the ratio of primordial B-mode power to lensing B-mode power is largest. At the same time, because they observe the full sky, satellite missions must, on average, deal with significantly higher levels of foregrounds. As a consequence, the different approaches are highly complementary. 

Contamination of the primordial $B$-mode signal by $B$ modes of astrophysical origin remains one of the main challenges for both ground- and space-based observations. The two main sources of astrophysical $B$ modes are polarized emission by dust grains that are aligned with the Galactic magnetic field, and synchroton radiation emitted by relativistic electrons in the Galactic magnetic field. The amplitude of the astrophysical signal exceeds the amplitude of the primordial signal at all frequencies and on all angular scales, even in the cleanest regions of the sky. As a consequence all experiments searching for primordial $B$ modes necessarily rely on multifrequency observations that use together the different frequency dependence of the CMB and foreground signal. Ground-based observations can target the cleanest regions of the sky, but the atmosphere severely restricts the viable observing frequencies. All-sky observations from space are not subject to limitations imposed by the earth's atmosphere but must deal with higher levels of astrophysical $B$ modes.\footnote{Of course, this also provides an opportunity to those interested in a better understanding of the processes that produce the foreground emission.} Independent of the observing platform, a better understanding of these astrophysical foregrounds is critical, both for the analysis and interpretation of current and upcoming data sets as well as for the design and planning of experiments. There has been recent progress both in the form of ab initio magnetohydrodynamic (MHD) simulations~\cite{Kritsuk:2017aab,Kim:2019xov}, phenomenological models~\cite{Hervias-Caimapo:2021zue}, and in the form of generative models~\cite{Thorne:2021nux,Krachmalnicoff:2020rln}, but additional work in this direction, in particular a coordinated effort to make full use of the different approaches is needed.

While the frequency dependence of astrophysical foregrounds differs from that of the CMB signal, lensing $B$ modes have the same frequency dependence as the primordial signal and cannot be reduced by multi-frequency observations. However, since the statistical properties of the CMB are well understood, high-precision measurements of the polarization on scales near the lensing peak can be used to reconstruct the lensing potential, and ultimately to remove the lensing $B$ modes~\cite{Knox:2002pe,Kesden:2002ku,Seljak:2003pn,Smith:2010gu}.  
This has only recently been demonstrated on data~\cite{Larsen:2016wpa, SPT:2017ddy, Carron:2017vfg, Planck:2018lbu, POLARBEAR:2019snn, SPTpol:2020rqg, ACT:2020goa}, and work is ongoing to develop techniques appropriate for the stringent requirements of future ground-based surveys~\cite{Carron:2017mqf,Lizancos:2021wpy, Millea:2020cpw, Namikawa:2021gyh, Millea:2021had}. A key aspect that remains to be better understood is the effect of foregrounds on delensing. Because delensing relies on the correlations between modes on arcminute and degree scales, this motivates higher resolution ab initio MHD simulations with a larger dynamic range than currently available. 

Precision measurements of the polarization of the cosmic microwave background on small scales~\cite{SimonsObservatory:2018koc,CMB-S4:2016ple,Abazajian:2019eic,CMB-S4:2020lpa} will also lead to improved measurements of the power spectrum of primordial density perturbations that will provide interesting constraints on inflation and fundamental physics more generally. Over the next decade, constraints on the scalar spectral index $n_{\rm s}$ will improve by a factor of two. Similarly, constraints on its scale dependence, referred to as the running of the scalar spectral index, will improve by a factor of two to three~\cite{CMB-S4:2016ple,Abazajian:2019eic,CMB-S4:2020lpa}. In combination with the increased precision on $r$, this will significantly reduce the space of viable inflation models. 

In addition, the improved measurements of $n_{\rm s}$ have implications for the aftermath of inflation. As inflation ends, the potential energy density stored in the inflaton is transferred into kinetic energy and eventually into the energy density of a thermal plasma of standard-model particles. This process is referred to as `reheating'~\cite{Kofman:1994rk,Kofman:1997yn,Lozanov:2019jxc}. How reheating occurs in detail remains unknown, but at least for the simplest models of inflation the observational predictions related to inflationary physics only depend weakly on these details of reheating~\cite{Bardeen:1980kt,Weinberg:2003sw,Weinberg:2004kr}. The dependence only arises because observables depend on the relation between physical scales today and physical scales during inflation, which depends on the amount by which the Universe expands during reheating. More quantitatively, the details of the reheating process lead to small changes of the spectral index $n_{\rm s}$. With the increase in precision on $n_{\rm s}$, observations will begin to distinguish between different reheating histories for a given model of inflation~\cite{Adshead:2010mc,Mortonson:2010er, Allahverdi:2020bys}. The physics that is probed in this way is very rich. For example, the expansion history and the duration to radiation domination after inflation can depend on the self-interactions of the inflaton~\cite{Amin:2011hj,Lozanov:2016hid}, inflaton couplings to other fields and the efficiency of the energy transfer between the inflaton and daughter fields~\cite{Antusch:2020iyq,Fan:2021otj}. Constraints on the expansion history can also impact predictions for dark sector abundances~\cite{Mambrini:2013iaa,Baumann:2015rya,Daido:2017tbr,Hooper:2018buz,Bernal:2018hjm,Garcia:2018wtq,Green:2019glg}, provide insights into the possibility of producing primordial black holes~\cite{Green:2000he,Carr:2020gox} and  additional small scale structure in the early universe~\cite{Jedamzik:2010dq,Easther:2010mr,Erickcek:2011us}. For a recent review, see~\cite{Allahverdi:2020bys}. 

There are several well-motivated classes of inflationary models that predict departures from a power law in the form of oscillatory or sharp features in the primordial power spectrum. For example, as we mentioned earlier, a detection of primordial gravitational waves with an amplitude above $r\simeq 0.01$ would imply the existence of a symmetry that protects the inflaton potential. In this case axions are natural inflaton candidates because they enjoy a shift symmetry to all orders in perturbation theory. Non-perturbatively, the inflaton acquires a potential that may contain both a non-periodic contribution suitable to drive inflation and a subdominant periodic contribution. The small periodic contribution leads to features in the primordial power spectrum~\cite{Flauger:2009ab}. These contributions have been searched for and have been constrained using CMB data~\cite{Flauger:2009ab,Aich:2011qv,Peiris:2013opa,Easther:2013kla,Flauger:2014ana} and more recently using the BOSS data, which currently provides the strongest constraints on these models for a significant part of parameter space~\cite{Beutler:2019ojk}. Constraints from LSS data will improve further as new data from DESI~\cite{DESI:2016fyo} and Euclid~\cite{EUCLID:2011zbd} becomes available~\cite{Beutler:2019ojk}.

If the inflaton is the only light degree of freedom during inflation, the primordial density perturbations are adiabatic~\cite{Weinberg:2004kr}. The improved measurements of the power spectrum of primordial density perturbations will tightly constrain departures from adiabaticity, referred to as isocurvature modes, that would be expected in theories with additional light degrees of freedom like axions~\cite{Peccei:1977hh,Peccei:1977ur,Weinberg:1977ma}, or in the curvaton scenario~\cite{Mollerach:1989hu,Mukhanov:1990me,Moroi:2001ct,Lyth:2001nq,Lyth:2002my}. Just like the inflaton, light degrees of freedom present during inflation experience quantum fluctuations and contribute to the density perturbations. Since the two fields fluctuate independently, their contributions are uncorrelated. If the density perturbations are predominantly sourced by the inflaton and departures from adiabaticity are associated with additional light fields, the adiabatic and isocurvature modes are uncorrelated. Over the next decade limits on this type of departures from adiabaticity will improve by a factor five~\cite{CMB-S4:2016ple}. The curvaton scenario is an alternative to single-field models of inflation in which the observed density perturbations are dominated by the vacuum fluctuations in a second field, the curvaton, that subsequently decays. Depending on the details of the decay process, this scenario allows for a wide variety of departures from adiabaticity. Since the density perturbations are dominated by the curvaton and the departures from adabaticity are set by the curvaton as well, in this scenario the adiabatic and isocurvature components are fully correlated (or anti-correlated). Constraints on these departures from adiabaticity will improve by as much as an order of magnitude~\cite{CMB-S4:2016ple}. 

The upcoming precision measurements of CMB polarization will also tighten constraints on departures from Gaussianity. The constraints are most commonly presented as constraints on amplitudes of different functional forms, typically referred to as shapes, of low-order correlation functions. Constraints on the amplitudes of the most widely studied shapes of the 3-point function will improve by a factor of two to three compared to existing constraints~\cite{CMB-S4:2016ple,Abazajian:2019eic}. These constraints can, for example, help answer the questions how strongly the inflaton interacts with itself, and more generally whether there was a single light degree of freedom or several. The constraints achievable with CMB observations alone just fall short of key theoretical targets~\cite{Meerburg:2019qqi}, and further improvements will require measurements of higher order correlation functions from galaxy surveys or intensity mapping~\cite{CosmicVisions21cm:2018rfq,PUMA:2019jwd}.

Nominally, LSS data provides access to orders of magnitude more Fourier modes than the CMB. However, as already briefly mentioned, the analysis of LSS data is more challenging because of nonlinear effects of matter clustering, galaxy formation physics, and redshift space distortions. On sufficiently large scales (larger than 1 Mpc), these effects can be systematically described within the Effective Field Theory (EFT) of Large Scale Structure~\cite{Ivanov:2019pdj,DAmico:2019fhj,Snowmass2021:EFT}, and there has recently been significant progress in deriving constraints on departures from non-Gaussianity from large scale structure~\cite{Castorina:2019wmr,Mueller:2021tqa,Cabass:2022wjy,DAmico:2022gki}.

The EFT framework provides robust first-principle theoretical models for the late-time non-Gaussian patterns in the galaxy distribution, which act like a background noise that complicates the extraction of the primordial non-Gaussian signal. The other important ingredients necessary for the analysis of primordial non-Gaussianity (PNG) in galaxy surveys are optimal estimators of summary statistics~\cite{Philcox:2021ukg,Philcox:2021kcw}, efficient data compression and covariance matrix estimation techniques~\cite{Wadekar:2020hax,Philcox:2020zyp}, and codes for EFT calculations~\cite{Chudaykin:2020aoj,Chen:2020zjt,DAmico:2020kxu}. The recent application of these tools to the galaxy power spectrum and bispectrum data from the BOSS~\cite{SDSS-III:2015hof} is a proof of principle that measurements of PNG from galaxy surveys are feasible, and there is a systematic program that aims to reach the level of precision necessary to answer key questions about inflation.

Over the next few years the upcoming surveys like DESI~\cite{DESI:2016fyo} and Euclid~\cite{EUCLID:2011zbd} will create a detailed map of our Universe up to redshift of $z\approx 2$, which will permit the improvement of the current limits on PNG from galaxy surveys at least by a factor of four~\cite{Sailer:2021yzm}. An even more impressive improvement will become possible with future surveys like MegaMapper~\cite{Ferraro:2019uce,Sailer:2021yzm,Ferraro:2022ee}, which will map our Universe up to $z\approx 5$ and will reach unprecedented precision in measuring PNG. In addition, the local type of non-Gaussianity will soon be probed with the SPHEREx mission~\cite{Dore:2014cca}.

To make full use of the data, it will be important to improve the accuracy of the EFT calculations (higher order $n$-point functions and high loop orders), and to obtain inputs from high fidelity hydrodynamical simulations. These simulations will yield tight priors on the Wilson coefficients of the EFT (nuisance parameters) that capture the details of galaxy formation on large scales. This will break the degeneracy between PNG and galaxy formation physics, and hence reduce the error bars on the potential PNG signal~\cite{Wadekar:2020hax,Cabass:2022wjy}. On the experimental side, the biggest challenge will be imaging systematics. This issue can be addressed, e.g. with recently developed network-based techniques~\cite{Rezaie:2021voi}. 

Given the leaps in sensitivity and data quality for both CMB experiments and LSS surveys, cross-correlations between the data sets are an important additional avenue to constrain PNG and cosmological parameters more generally. For example, upcoming CMB experiments like Simons Observatory and CMB-S4 will provide exquisite measurements of the lensing convergence that contains information about the projection of matter along the line of sight. Correlations between the CMB lensing maps and deep LSS surveys can provide complementary and highly competitive constraints on PNG~\cite{Schmittfull:2017ffw}. Secondary CMB anisotropies, caused by interactions of CMB photons with electrons in non-linear structures along the line of sight similarly correlate with LSS surveys and provide yet another route to constrain PNG~\cite{Munchmeyer:2018eey,Smith:2018bpn}.

Most theoretical work and analyses have focused on scale-invariant shapes of the 3-point correlation function. However, the inflationary models mentioned earlier that predict oscillatory or sharp features in the primordial power spectrum also predict corresponding features in higher order correlation functions~\cite{Chen:2006xjb,Flauger:2009ab}. Since searches for these shapes are computationally more challenging than the searches for the scale-invariant shapes, at present only constraints from the CMB exist~\cite{Planck:2013wtn,Fergusson:2014hya,Fergusson:2014tza,Meerburg:2015owa,Planck:2015zfm,Planck:2019kim}, and the information available in LSS data remains to be extracted.

Finally, there are several examples of physical processes for which departures from Gaussianity are not well described by the first few moments of the probability distribution function~\cite{Bond:2009xx,Flauger:2016idt,Chen:2018uul,Panagopoulos:2019ail,Panagopoulos:2020sxp,Celoria:2021vjw}. Signatures associated with these processes might be missed in traditional searches. For the example of reference~\cite{Flauger:2016idt} the optimal estimator has been found and is naturally formulated in real space~\cite{Munchmeyer:2019wlh}. This raises the more general question of how to systematically and optimally extract the information stored in the data beyond the power spectrum and low-order correlation functions~\cite{Baumann:2021ykm}. See~\cite{Snowmass2021:stringinflation} for additional discussion.

\section{Dark matter}\label{sec:darkmatter}

The discovery of dark matter in galaxy clusters \cite{Zwicky:1933gu} and individual galaxies \cite{Rubin:1970zza,Rubin:1980zd} was one of the first signs of physics beyond what we now know as the Standard Model \cite{Glashow:1961tr,Weinberg:1967tq,Salam:1968rm}.  By the 1980's it was clear from astronomical observations what dark matter could NOT be, namely neutrinos \cite{Huchra:1983wy,Peebles:1982ib,Peebles:1982ib,White:1983fcs,Geller:1989da}.  \emph{Simulations} showed that the clustering of halos in a neutrino-dominated universe could not be reconciled with observations for masses consistent with matching the relic abundance of neutrinos.  Instead, simulations suggested that another, colder form of dark matter could be consistent with both the cosmological abundance of dark matter and its small-scale clustering \cite{Blumenthal:1984bp,Davis:1985rj}.  Since then, it has been recognized that the physics of dark matter shapes the homogeneous evolution of the Universe and the evolution of perturbations.  Particle theorists have drawn inspiration for dark matter model building from astronomical observations (e.g., \cite{Pagels:1981ke}), and the community is using observations paired with high-resolution simulations to illuminate dark-matter particle properties.

In fact, dark matter astrophysics is becoming a precision science \cite{Buckley:2017ijx,bullock2017,LSSTDarkMatterGroup:2019mwo,Green:2021jrr}.  On the observational side, there are many different probes of dark matter on a variety of scale, from the expansion history to LSS to dark-matter halos so small that they may not contain luminous matter.  Importantly, new wide-field surveys, from radio to optical to gamma-ray, are enabling the discovery of new targets for dark-matter searches, with well-quantified statistical and systematic uncertainties (e.g., \cite{2004MNRAS.350.1210Z,2009ApJ...696.2179K,2009AJ....137..450W,Fermi-LAT:2015att,2020ApJ...893...47D}).  When these observations are paired with a commensurate cosmological simulation and theory program (e.g., \cite{rocha2013,Wang:2013rha,Robertson:2016qef,Despali:2018zpw,2019MNRAS.490..962F,2020JCAP...02..024B,2020MNRAS.498..702L,2020MNRAS.497.2393L,2021ApJ...906...96A,2021MNRAS.507.4826L,2021ApJ...923...35M,2021arXiv211101158S,2021MNRAS.500.1531C,2022MNRAS.tmp..348M}), we as a community are obtaining stringent constraints on the WIMP annihilation cross section, the momentum distribution of dark matter at its production era, dark matter self-interactions and interactions with Standard Model particles, and the dark matter particle mass (e.g., \cite{Kennedy:2013uta,Kim:2017iwr,Gilman:2019nap,Banik:2019smi,Nadler:2021dft,Dekker:2021scf,StenDelos:2019xdk,Newton:2020cog,Nguyen:2021cnb,Kim:2021zzw,Mau:2022sbf}).  The constraints inform dark matter model builders and complement dark matter searches in the lab.  

As detailed in other Snowmass contributions, the observational facilities of the next decade or two can provide tremendous insight into the nature of dark matter (e.g., \cite{DMfacilitiesWP,DMRubinWP,DMDESIWP,DMCMB64WP,DMsimsWP,DMhalosWP,DMPBHWP,DMextremeWP,TFastroDMWP}).  However, this opportunity can only be realized with a strong theory and simulation program.  The opportunities and challenges are detailed in Ref. \cite{DMsimsWP} (see also Ref. \cite{TFastroDMWP}), which we summarize here.  In brief, collaboration between particle theorists and simulators is desirable to translate from the Lagrangian model level to phenomenological cosmologically relevant parameter space (e.g., as in the ETHOS framework \cite{Cyr-Racine:2015ihg,Vogelsberger:2015gpr}). Part of this process is figuring out how best to map specific physics into simulation algorithms.  For example, for self-interactions, what matters is simulating the transfer of momentum and energy in dark-matter halos, and so careful thought must go into determining which cross section is relevant for the coarse-graining of the transfer \cite{Tulin:2017ara}.  Simulations must include the physics of galaxy formation, and the simulation outputs need to be rendered in the space of real astronomical observations (see Ref. \cite{Brooks2017} for an application to dwarf H\textsc{i} single-dish observations).  Because simulations are slow, we will use simulations to train emulators and semi-analytic models (e.g., \cite{Benson:2010kx,Pullen:2014gna,2016MNRAS.461...60L,Sameie:2018juk}).  Thus, likelihood function approaches to constraining dark-matter parameter space will become possible in finite compute time, and we will have a unified theoretical framework to consider all astronomical probes of dark matter together (e.g., \cite{Kaplinghat:2015aga,Enzi:2020ieg,Nadler:2021dft}).  Simulations can also point to completely new types of observables \cite{StenDelos:2019xdk,Cruz:2020rit,2021Natur.592..534C}, including ones that affect lab dark matter searches \cite{Read:2008fh,Necib:2018iwb,Besla:2019xbx}.  This mapping between dark matter particle models and astronomical observables, including the effects of galaxy evolution physics, enables sharp tests of dark matter microphysics with telescopes, and a connection to terrestrial experiments.

There is an enormous discovery potential for dark matter physics with the next generation of experiments on telescopes and in the lab.  Revealing the particle nature of dark matter from these experiments requires a theoretical and simulations program to unite all probes of dark matter into a consistent interpretation framework.  

\section{Neutrinos and other light relics}\label{sec:neutrinos}
Standard cosmology predicts that the Universe is filled with a sea of relic neutrinos produced during the Hot Big Bang. As the Universe expands and cools, the neutrino momenta redshift along with photons and other particles leaving a relic background characterized by a temperature $T_\nu \propto 1/a \approx 10^{-4}$ eV today. In the early Universe, when $T_\nu(a) \gg m_{\nu i}$, these particles were relativistic and contributed to the radiation energy budget. Today, we expect that at least two of the three neutrino mass eigenstates have masses $m_{\nu} \gg T_\nu$. Cosmology therefore probes neutrinos across a range of epochs from the era of decoupling ($T \sim 10$ MeV) through the non-relativistic transition and to today. 
Measurements of the radiation density in the early universe provide constraints on the number of neutrino states and the energy density carried by each. Measurements of the Universe at late times characterizing the matter budget and amplitude of large-scale structures provide constraints on the neutrino energy density at late times, and therefore the sum of the neutrino masses. Both provide powerful constraints on the thermal history of the Universe and new physics beyond the Standard Model. For a thorough discussion of the science of light relics, see these Snowmass papers \cite{LRWP, NSWP}.

The radiation energy budget is conventionally parameterized by $N_{\rm eff} \equiv (\rho_{radiation}/\rho_\gamma - 1)/(\frac{7}{8}(\frac{4}{11})^{4/3})$, where $\rho_{\gamma}$ is the CMB photon energy density and $\frac{7}{8}(\frac{4}{11})^{4/3}\rho_\gamma$ is the expected energy density of a single species of neutrino and anti-neutrino that decouples instantaneously. The standard model prediction of three light neutrino and anti-neutrino states translates into a prediction of $N_{\rm eff} = 3.044$, where the additional digits after the decimal are due to residual heating of neutrinos due to electron-positron annihilation \cite{Gnedin:1997vn,Mangano:2001iu,Mangano:2005cc,deSalas:2016ztq, Froustey:2020mcq}. Current constraints on $N_{\rm eff}$ from CMB and BAO data are $N_{\rm eff} = 2.99 \pm 0.17$ \cite{Planck:2018vyg}, in remarkable agreement with the Standard Model expectation. CMB data is expected to continue to provide evermore stringent constraints on $N_{\rm eff}$, large-scale structure is an emerging probe of $N_{\rm eff}$ \cite{Baumann:2017gkg, Baumann:2017lmt} that can also produce interesting limits on neutrinos. 

A variety of well-motivated beyond-the-Standard-Model scenarios predict additional light degrees of freedom such as axions, gravitinos, gravitational waves, or other dark radiation (\cite{LRWP}) that, at some point in the early Universe, would have been in thermal equilibrium with the rest of the Standard Model particles. These very same measurements of neutrinos in the early and late Universe can be used to infer the presence of these new particles. There are firm theoretical predictions for the additional contribution to $N_{\rm eff}$ from any light ($\lesssim$eV) thermal relic particle that was ever in equilibrium with the primordial plasma, specifically $\Delta N_{\rm eff} = 0.027, 0.047, 0.054$ for a single scalar, Weyl fermion, or vector Boson that decoupled at epochs when all Standard Model degrees of freedom were in equilibrium.  Remarkably, experiments in the next decade are approaching these thresholds of detection \cite{CMB-S4:2016ple,Abazajian:2019eic,SimonsObservatory:2018koc,LRWP, CFCMBWP, CFCCMBS4WP}. For particles that decoupled at later epochs, the contribution to $N_{\rm eff}$ is larger because those particles would have experienced the same heating as the photon bath when heaver particles fell out of equilibrium. 

Simple counting of particles and spins gives a prediction for $\Delta N_{\rm eff}(T_{{\rm freeze-out}})$, the function specifying the contribution to $N_{\rm eff}$ from a species that freezes out at $T_{{\rm freeze-out}}$, that is accurate to the $\%$-level at epochs when the relativistic degrees of freedom are not changing. For particles that decouple during the QCD or electroweak phase transitions, for instance, computing $\Delta N_{\rm eff}$ is considerably more complicated. During this epochs perturbative techniques and lattice gauge theory are required (for a summary, see, e.g. \cite{CMB-S4:2016ple, Saikawa:2018rcs}) and theoretical uncertainties are currently present at the $10\%$-level. Reaching sub-percent-level accuracy for the neutrino contribution to the energy density requires detailed computations including non-instantaneous decoupling and, to a lesser extent, neutrino oscillations \cite{Gnedin:1997vn,Mangano:2001iu,Mangano:2005cc,deSalas:2016ztq, Froustey:2020mcq}.

CMB and LSS datasets sensitive to $N_{\rm eff}$, a measure of the total energy density in relativistic particles, are also able to infer properties of the perturbations in relativistic particles. This allows these experiments to set limits on the existence of non-standard neutrino self interactions \cite{Cyr-Racine:2013jua, Lancaster:2017ksf,Song:2018zyl, Kreisch:2019yzn,Das:2020xke, RoyChoudhury:2020dmd, Brinckmann:2020bcn, Berryman:2022hds} and interactions among other new contributions to the relativistic energy budget \cite{Baumann:2015rya, Baumann:2016wac} such as self-interacting dark radiation \cite{Blinov:2020hmc} or dark radiation that is tightly coupled to other dark sector particles \cite{Choi:2018gho}. 

At present the strongest constraints on $N_{\rm eff}$ come from CMB temperature and polarization anisotropies. The physical effects can be understood as follows: the radiation density in the early Universe dictates the Hubble rate, which characterizes lengths and timescales that impact features in the CMB power spectra (for a review see, e.g. \cite{Hou:2011ec,CMB-S4:2016ple}). For free-streaming contributions to $N_{\rm eff}$, there is an additional change to the power spectra, a phase shift in the peaks, due to the different propagation speeds of perturbations in free-streaming particles ($c$) and perturbations in the photon-baryon fluid ($c_{s} = c/\sqrt{3}$) \cite{Bashinsky:2003tk}. These signatures on CMB primary power spectra are well-understood and straightforward to model, for instance using publicly available Boltzmann codes CLASS and CAMB \cite{2011JCAP...07..034B,Lewis:1999camb}. On the other hand, reaching target constraints on $N_{\rm eff}$ will require removing the changes to the CMB power spectra induced by gravitational lensing from matter along the line of sight \cite{Green:2016cjr, ACT:2020goa} as well as cleanly separating out any foreground emission contaminating the measured power spectra.  While galactic and extragalactic foregrounds are not expected to be a limiting factor for CMB polarization data at $\ell \lesssim 5000$, future experiments will be measuring CMB polarization anisotropies on those scales for the first time. Quantifying the impact of foregrounds, and delensing in the presence of foregrounds, on measurements of $N_{\rm eff}$ is an active area of research that requires accurate simulations of high-resolution maps of galactic and extragalactic foreground emission, as well as nonlinear CDM and baryon structure.

Neutrino oscillation data specifies the splitting of the square of the neutrino masses to be $\Delta m_{12}^2 = 7.42 ^{+ 0.21}_{-0.20}\times 10^{-5}$ eV$^2$ and $\Delta m_{13}^2 = |2.51 \pm 0.027\times 10^{-3}|$ eV$^2$ \cite{Gonzalez-Garcia:2021dve}. As the relic neutrino temperature is $T_\nu \sim 10^{-4}$eV today, at least two of the three mass eigenstates are non-relativistic. These particles then contribute to the matter budget of the Universe today, with $\Omega_{\nu}h^2 \approx \sum_{i} m_{\nu i} /94 eV$ (e.g. \cite{Lesgourgues:2012uu}). This contribution has yet to be detected, but remains the only unknown parameter in the simplest $\Lambda$CDM cosmology. As the mass splittings are known, a cosmological measurement of $\Omega_{\nu}$ translates into a constraint on the lightest of the neutrino mass states\footnote{Unless $m_{\nu {\rm lightest}}$ is sufficiently large (e.g. $\gtrsim$ few meV), next generation cosmological detections of the neutrino mass sum will only provide an upper bound on $m_{\nu {\rm lightest}}$. Detecting the value of $m_{\nu {\rm lightest}}$ directly would require $\sigma_{\Sigma m_{\nu}} \lesssim m_{\nu {\rm lightest}}$.}. Detecting the neutrino mass scale, and finding consistency with laboratory experiments, would be a triumph of cosmology and particle physics. Determining the neutrino mass scale would also set a benchmark for neutrinoless double-$\beta$ decay experiments: if the neutrino mass sum is detected at $\gtrsim 0.1$ eV via cosmology and that process is not observed, the simplest interpretation is that neutrinos are Dirac particles (see, e.g. \cite{CMB-S4:2016ple, NSWP}). In the event that neutrinoless double beta decay is detected, a cosmological measurement of the neutrino mass sum can help to constrain the Majorana phases. Pinning down the neutrino mass scale is also important for studies of new physics.  Dark energy constraints, for example, can be affected by degeneracies with the neutrino mass. 

Neutrinos were relativistic for much of the history of the Universe and therefore kinematically forbidden from participating in gravitational clustering until late times. This manifests as a strong suppression in the amplitude of neutrino perturbations on scales smaller than the neutrino free-streaming scale, a length scale characterizing the typical distance neutrinos travel in a Hubble time. The absence of neutrinos on these scales weakens the gravitational potentials and slows the overall growth of cold dark matter and baryon structures. The net result is a suppression in the amplitude of structures, which can be parameterized by $\sigma_8$. The suppression in structure is detectable via a variety of methods, from gravitational lensing of galaxies or CMB \cite{2009arXiv0912.0201L, Sherwin:2016tyf, Planck:2018vyg, SPT:2019fqo, Euclid:2019clj}, to redshift space distortions to galaxy clustering \cite{Font-Ribera:2013rwa, DESI:2016fyo,Euclid:2019clj}, and galaxy cluster counts (for a summary, see \cite{NSWP}). The transition of relic neutrinos from relativistic to non-relativistic also alters the evolution of the neutrino energy density, and therefore the Hubble rate, but this signature is expected to be too small to detect \cite{Hou:2011ec}. The physical processes and observables described here will also occur for other light relic species. Consequently, constraints on neutrino mass and $N_{\rm eff}$ can be generalized to constrain the mass of other light relic species \cite{Xu:2021rwg}.

If neutrino masses are described by the minimal mass normal or inverted orderings, $\sum m_\nu \approx 0.06$ eV or $\approx 0.1$ eV. In these scenarios the primary observable for neutrino mass -- a suppression in the matter power spectrum, relative to what would be seen from CMB predictions for the amplitude of structure in a universe with massless neutrinos -- is a small effect ($\sim 3-6\%$). To robustly detect the neutrino signal and confidently limit the masses of any other new  light relic particles, will require exquisite control over theoretical and observational systematics. Achieving that control will require strong efforts in the theory and simulations of structure formation, astrophysical processes, and survey data. There are also opportunities to identify observables or techniques that may help isolate a signature of neutrino or other light relic particle masses. 

Current constraints on the neutrino masses from primary CMB, CMB lensing, and LSS power spectrum measurements combined are $\sum m_\nu < 0.16$ eV at $95\%$ confidence \cite{Ivanov:2019hqk}. This bound on the mass puts the neutrino free-streaming scales well into the linear regime of structure formation. Yet, datasets probing the suppression in structure due to neutrinos receive contributions from quasilinear and nonlinear scales where simulations, or advanced techniques such as EFT \cite{Ivanov:2019pdj,DAmico:2019fhj,Snowmass2021:EFT}, are typically used to model nonlinear gravitational evolution. Neutrinos are fast-moving particles that travel over cosmological distances and have a significant velocity dispersion, accurately incorporating them into studies of gravitational evolution can therefore pose challenges. A number of different approaches exist in the literature.  

On the simulations side, a popular technique consists in including neutrinos as particles, while adding a thermal component to their initial velocities \cite{bird2012massive, villaescusa2014cosmology, villaescusa2018imprint, rossi2020sejong,adamek2017relativistic,brandbyge2008effect, villaescusa2013non, castorina2015demnuni,Emberson:2016ecv}. This method naturally takes into account neutrino nonlinearities, which can be important in some scenarios \cite{baldi2014cosmic, viel2010effect, LoVerde:2013lta, chen2021cosmic}. However, it also suffers from a few challenges that are associated  to the large thermal velocities of neutrinos, such as the need for a special relativistic description \cite{de2021neutrinos} and shot noise. Shot noise arises as a problem when treating neutrinos as $N$-body particles because the neutrino density field lacks intrinsic power on small scales so shot noise due to finite sampling of the density field quickly dominates. The shot noise can be reduced by increasing the number of neutrino particles, at the expense of significantly increasing the use of computational 
resources. Alternative approaches have been developed to mitigate this problem \cite{banerjee2018reducing,elbers2021optimal, Inman:2020oda, Yoshikawa:2020ehd}. To achieve robust constraints with future survey, simulations will also need to accurately account for baryonic feedback processes \cite{Natarajan:2014xba}. 

Another simulations-based approach consists in treating the neutrinos in linear theory, while coupling to the non-linear gravitational potential of the cold dark matter \cite{brando2021relativistic,tram2019fully,chiang2019first,brandbyge2009grid, archidiacono2015efficient, ali2013efficient}. This can make simulations with massive neutrinos only as computationally expensive as in the case of cold dark matter alone, while also accounting for all relativistic corrections \cite{heuschling2022minimal, fidler2019new, partmann2020fast}. This approach of treating neutrinos in linear theory, while accounting for nonlinear evolution of CDM, is also adopted in separate universe simulations, which allow for precise calculations of a subset of nonlinear statistics such as halo bias and the squeezed-limit bispectrum \cite{Chiang:2017vuk}. However, in all of these approaches the effects of nonlinear clustering of neutrinos are neglected. While neglecting nonlinear clustering of massive neutrinos should be adequate for $\sum m_\nu \lesssim 0.3$ eV and studies of structure on large scales \cite{LoVerde:2013lta}, for heavier neutrinos or observables on halo scales one should account for them. There are also some hybrid schemes that aim to combine the advantages in both methods \cite{bird2018efficient, brandbyge2010resolving, Inman:2016qmg}.

Finally, there continues to be rapid progress in the development of analytic or hybrid methods for large-scale structure.  For example, multi-component perturbation theories to compute the power spectra and bispectra of matter fields \cite{Shoji:2010hm,Lesgourgues:2009am,Dupuy:2013jaa, Fuhrer:2014zka, Blas:2014hya, Peloso:2015jua, Levi:2016tlf, Senatore:2017hyk}, spherical collapse models to compute halo formation \cite{Ichiki:2011ue, LoVerde:2014rxa}, and peak-background split / separate universe approaches to halo clustering statistics \cite{LoVerde:2014pxa, Chiang:2017vuk}. At this stage there continues to be a strong interplay between simulations and analytic approaches to modeling large-scale structure in the presence of massive neutrinos, and therefore observables sensitive to neutrino mass. 

\section{Dark energy}\label{sec:darkenergy}

Discovering the mechanism that drives cosmic acceleration, whether it is a cosmological constant $\Lambda$, a time-dependent scalar field, or modifications of the laws of gravity, is a core science goal of ongoing and future DOE experiments and NASA missions. Dark energy as a term describes our lack of understanding of the physical concepts that underlie cosmic acceleration. As such it encompasses a wide variety of fundamental physics topics including modified gravity, neutrino physics, dark matter-dark energy coupling, early dark energy, and more. A joint analysis of multiple cosmological probes across multiple experiments is required to control the systematics budget and to increase the constraining power such that the community can discriminate between the different physical concepts that explain cosmic acceleration.

Two complementary avenues emerge in order to constrain the underlying physics model driving cosmic acceleration: 1) Measuring tensions between different experiments within the same underlying model and 2) combining the constraining power of different experiments that are not in tension in order to compare different models. 

Major progress on this topic is made by the current (Stage 3) generation of photometric surveys, such as Kilo-Degree Survey (KiDS)~\cite{Kuijken:2019gsa}, the Hyper Suprime Cam (HSC)~\cite{Aihara:2021jwb}, the Dark Energy Survey (DES)~\cite{DES:2021wwk} and spectroscopic surveys, such as the Baryon Oscillation Spectroscopic Survey (BOSS)~\cite{eBOSS:2020yzd}. These low-redshift constraints of the $\Lambda$CDM model can be contrasted with CMB measurements from the early Universe made e.g., by the Planck satellite~\cite{Planck:2018nkj}, the Atacama Cosmology Telescope (ACT)~\cite{ACT:2020frw,ACT:2020gnv}, and the South Pole Telescope (SPT)~\cite{SPTpol:2020rqg}. 

These initial results will become more exciting in the near future with the decreasing statistical uncertainty and better systematics control. With the advent of so-called Stage 4 surveys, e.g., the Dark Energy Spectroscopic Instrument (DESI)~\cite{DESI:2016fyo}, the Prime Focus Spectrograph (PFS)~\cite{tec14}, the Vera C. Rubin Observatory~\cite{LSST19}, Euclid~\citep{EUCLID:2011zbd}, the Spectro-Photometer for the History of the Universe, Epoch of Reionization, and Ices Explorer (SPHEREx)~\cite{Dore:2014cca}, and the Nancy Grace Roman Space Telescope \citep[][]{sgb15} the science community can expect an abundance of data to study the late-time Universe at increased precision. Similarly, the next generation of CMB surveys, such as the Simons Observatory (SO)~\cite{SimonsObservatory:2018koc} and CMB-S4~\cite{CMB-S4:2016ple} will enable us to contrast high and low redshift at increased precision and to combine information from both eras to increase the constraining power on cosmological models. 

Below we list the main focus areas in theory and numerical modeling that need to be addressed in order to fully extract the cosmological information from multi-probe, multi-survey analyses (see e.g.~\cite{Alonso:2021nzr,Battaglia:2020cch} and the Snowmass Computational Frontier white paper \cite{Banerjee:2022ljx} for more details):

\noindent
{\bf Observational modeling uncertainties:} For example, photo-z errors, shear calibration, depth variations need to be parameterized consistently across the different probes and surveys if the datasets are combined.  

\noindent
{\bf Astrophysical modeling uncertainties:} For example, nonlinear modeling of the density field, baryonic physics, intrinisic alignment, galaxy bias and Halo Occupation Distribution models are key astrophysical uncertainties that need to be modeled through a combination of numerical simulations, analytical models and combinations thereof. Consistent parameterizations and coordination of priors is important if datasets or probes are to be combined.

\noindent
{\bf Statistical uncertainties:} For example, the functional form of the likelihood and, if a multivariate Gaussian is assumed, the computation of data covariances are key uncertainties in a joint CMB-LSS analysis.

\noindent
{\bf Simulated likelihood analyses:} Simulated likelihood analyses are important early on to design survey strategy, and at later stages to inform costly numerical simulation campaigns, and to optimize the final analyses on the measured survey data. These simulated mock analyses need to be run to quantify the error budget as a function of the analysis choices (scales, redshifts, galaxy samples, summary statistics) for the different probe and experiment combinations. At later stages, mock analyses are required to quantify tensions between different probes and/or experiments and to do model comparison.   

\noindent
{\bf Numerical simulations:} Nonlinear modeling of the density field and exploring the statistical uncertainties mentioned above requires numerical simulations. The initial conditions of these simulations should be coordinated across all survey collaborations to enable a better comparison. 

\noindent
{\bf Hydrodynamic simulations:} Baryonic physics, intrinsic alignment, galaxy bias and Halo Occupation Distribution models require a hydrodynamic simulation campaign that is computationally extremely expensive. In order to utilize the available computing resources most effectively this simulation campaign must be informed by the composition of the error budget of a joint analysis. In other words, the requirements for a simulation campaign will be different when analyzing data from a single survey as opposed to data from multiple surveys. Simulated cosmological likelihood analyses of multi-survey data can identify the main contributors to the overall error budget and can inform a corresponding simulation campaign. A close connection between the simulated analyses and the simulation effort is required.

\section{Expansion rate of the universe}\label{sec:expansion}

A current cosmological mystery that could have  significant implications for our understanding of the Universe in the near future is the observed large discrepancy between different inferences of the Hubble constant $H_0$. Indeed, measurements of the luminosity distance of Cepheid-calibrated Type Ia supernovae \cite{Riess:2021jrx} differ at $ \sim 5\sigma$ from the predictions of our current standard $\Lambda$CDM cosmological model once its parameters are fitted to observations of the temperature and polarization anisotropies of the CMB \cite{SPT-3G:2021wgf,Planck:2018vyg,aiola:2020}. This inconsistency is known as the ``$H_0$ tension" since distances in cosmology are inherently linked to the Hubble constant. At the moment, there is a strong case that this discrepancy is not caused by systematics in CMB data \cite{Aylor:2018drw,Bernal:2016gxb,Lemos:2019}. The situation is more ambiguous for distance-ladder-based inferences, for which different calibration techniques yield somewhat conflicting distance-redshift relations. Of particular note, a distance-ladder calibration based on the Tip of the Red Giant Branch (TRGB) \cite{Beaton:2016nsw,Hatt:2017rxl,Hatt:2018opj,Hatt:2018zfv,CSP:2018rag,Hoyt_2019,Beaton_2019,freedman2019,Jang_2021,Hoyt_2021} find important inconsistencies with the Cepheid-based SH$_0$ES team \cite{Riess:2011,Riess:2016jrr,Riess2019,riess2021} in estimates of luminosity distances to neighboring supernovae. Nonetheless, several independent attempts to determine $H_0$ from distance and redshift measurements from a suite of different observations \cite{freedman2012,Suyu:2016qxx,Birrer:2018vtm, Wong:2019kwg,Huang:2019yhh, Kourkchi:2020iyz, Reid:2019tiq, Freedman:2020dne, Freedman:2021ahq, Pesce:2020xfe, Khetan:2020hmh, Blakeslee:2021rqi} also find higher values than inferred from the CMB, albeit with larger uncertainties \cite{DiValentino:2020vnx}. To shed new light on this observational puzzle, several new ideas for probing the recent expansion history of our Universe have been proposed (see e.g.~Ref.~\cite{Moresco:2022phi}). Turning these ideas into actual observational realities is a key priority in the coming decade to help provide more clues into the fundamental nature of the tension. 

While this discrepancy is commonly referred to as the $H_0$ tension, it is important to realize that what is actually in ``tension'' is cosmological distance measurements \cite{Efstathiou:2020wxn,Camarena:2021jlr,Efstathiou:2021ocp}. For instance, one could rephrase the current tension by saying that a $\Lambda$CDM model fit to Planck CMB data \cite{Planck:2018vyg} places the Hubble-flow Type Ia supernovae further away from us than the Cepheid-calibrated distance ladder does. Turning the problem around, we could also phrase the issue by saying that a $\Lambda$CDM model fitt to Cepheid-calibrated Type Ia supernovae places the CMB last-scattering surface closer to us than what is required by CMB observations. This emphasis on distances is important to identity physics-based solutions that can actually address the root cause of the problem. In other words, simply finding a cosmological model that has a value of $H_0$ compatible with that quoted in Ref.~\cite{Riess:2021jrx} is not sufficient \cite{Benevento:2020fev,Greene:2021shv}; the model must instead provide a good fit to all distance measurements available (including supernovae, BAO, time-delay strong lenses, etc.), in addition to CMB data. Therefore, a better characterization of the current situation would be that we have a cosmological ``distance crisis'' on our hands.

For theoretical physics, this apparent discrepancy presents an opportunity to carefully reexamine all the different assumptions that go into our current cosmological model. As a starting point, one could ask how well we understand the late-time expansion history of our Universe. Observations of the \emph{relative} luminosity distances to Type Ia supernovae at $0.02 \lesssim z\lesssim 2$ \cite{Scolnic:2017caz,Brout:2022vxf} strongly constrain deviations from the standard $\Lambda$CDM expansion history at those redshifts, giving us confidence that our understanding of the Universe is on solid ground at these epochs. Given these constraints, one might be tempted to instead change the expansion history at very late times ($z < 0.02$). Such models, while technically able to accommodate large value of $H_0$ (typically at the price of a phantom dark energy equation of state), do not provide good fits the actual measured \emph{distances} to low-redshift supernovae and therefore do not address the root cause of the tension \cite{Benevento:2020fev,Efstathiou:2021ocp,Greene:2021shv}. Given the variety of low-redshift distance measurements available, such ``late solutions'' do seem to face an uphill battle in resolving the current discrepancy. Future measurements of low-redshift cosmological distances, such as those from multi-messenger astronomy, will play an important in determining whether late solutions are at all viable.

Another possibility is that we are missing some important physics in the early Universe. Since the CMB is fundamentally observed in angular space, making it compatible with a larger value of $H_0$ (which brings the last-scattering surface closer to us) requires shrinking all physical length scales present near photon decoupling to leave the observed angles invariant. In particular, the angular size of the baryon-photon sound horizon $\theta_*$ is one of the most precisely measured quantities in all of cosmology. Since $\theta_*\propto r_{\rm s} H_0$, where $r_s$ is the physical size of the sound horizon, keeping this angle constant with a larger $H_0$ value requires a smaller $r_{\rm s}$. Not too surprisingly, most proposed ``early times'' solutions (see e.g.~Refs.~\cite{Poulin:2018cxd,Agrawal:2019lmo,Smith:2010gu,Niedermann:2020dwg,Hill:2021yec,Smith:2022hwi}) to the current tension effectively work by reducing the size of the baryon-photon sound horizon. As $r_{\rm s}$ is mathematically given by an integral over the sound speed $c_{\rm s}$,
\begin{equation}
    r_{\rm{s}} = \int_{z_{\star}}^{\infty}\frac{c_{\rm{s}} dz}{H(z)},
\end{equation}
several models shrink the sound horizon by \emph{increasing} $H(z)$ in the pre-recombination universe, which suppresses the integrand. Others do so by changing $z_*$ (the photon decoupling redshift) to an earlier epoch (see e.g.~Refs.~\cite{Jedamzik:2020abc,Pogosian:2020ded,Rashkovetskyi:2021rwg}). Whichever mechanism is used, the difficulty lies in doing so without ruining the detailed fit to the temperature and polarization power spectra of the CMB. Indeed, another important length scale to the CMB is the photon diffusion length (also called the photon mean free path). Shrinking the baryon-photon sound horizon without also reducing the photon diffusion length by \emph{the same factor} nearly guarantees either a poor fit to CMB data, or the introduction of new tensions with other data sets, especially those from large-scale structure, or both. Thus, any successful ``early solution'' needs to include a mechanism to properly adjust this diffusion length. 

The centrality of the photon diffusion length (or its inverse, the photon scattering rate) to the Hubble tension as a whole was recognized in Ref.~\cite{Cyr-Racine:2021alc}. Modifying this quantity is highly non-trivial as it involves low-energy Standard Model physics, which is well understood. While this represents a significant model-building challenge, it also provides a clear target for future studies on which kind of new physics is required. One possibility that has been explored is a variation of the fine-structure constant and of the electron mass between the epoch of last scattering and today \cite{Hart:2020,Hart:2021kad}. Such an approach has had significant phenomenological success in a fair model-to-model comparison \cite{Sekiguichi:2021,Schoneberg:2021qvd}. However, significant model-building is required to explain the required percent-level changes in these quantities (see e.g.~Refs.~\cite{Burgess:2021obw,Burgess:2021qti}). Another possibility is modify the helium abundance near the epoch of recombination, which would affect the free-electron fraction in the cosmic plasma and thus change the photon diffusion length. Such an approach would require modifying Big Bang Nucleosynthesis predictions of the helium and deuterium abundances, which is challenging given their current consistency with direct light-element abundance measurements \cite{Fields:2019pfx}. Whichever physical mechanism is proposed to adjust the photon diffusion length to make the CMB compatible with a larger Hubble constant, it will leave subtle signatures in the data that could be detected in future observations, such as those from CMB-S4 \cite{CMB-S4:2020lpa}.

Whether it is the result of unknown systematics or new physics, the Hubble tension presents a golden opportunity to scrutinize both our theoretical beliefs and our data analysis techniques with the hope that they can be reconciled. As more high-precision data become available, our leading cosmological model might have to be amended, ushering in a new era of fundamental physics understanding.

\section{Theory as guide for the development of future experiments}\label{sec:guide}

As we consider future opportunities, it is useful to consider past successes at the intersection of particle theory, particle experiment, astronomical observations, and simulations as a guide for what might be possible. 

The simplest models of inflation predicted a universe with primordial perturbations that are dominated by adiabatic, Gaussian, and nearly but not exactly scale-invariant density perturbations with a spectral index $n_{\rm s}\lesssim 1$ at a time when measurements with a precision that could test these predictions were a distant dream. These predictions have now all been confirmed to high precision~\cite{Planck:2018nkj}. In addition to the density perturbations, many of the simplest models of inflation predict primordial gravitational waves within reach of the next generation of experiments~\cite{SimonsObservatory:2018koc,CMB-S4:2016ple}. The detection of this characteristic signature of inflation is one of the main science goals for upcoming CMB experiments, and both the planning and design relies on close collaboration between theorists and experimentalists. 

While this example was one about inflation, other examples exist for the other fundamental physics topics in this work.  Theorists can guide the development of new experiments (e.g., SPHEREx).

As another example of fruitful interplay that leads to the development of a novel class of experiments, consider the case of dark matter with a hidden-sector Yukawa coupling (see Ref. \cite{Tulin:2017ara} for a comprehensive review).  In the early 2000's, the ``missing satellites problem'' \cite{1999ApJ...522...82K,1999ApJ...524L..19M}---the apparent mismatch between the number of luminous satellite galaxies in the Milky Way relative to simulated dark matter subhalos, now recognized to not, in fact, be a problem \cite{2018PhRvL.121u1302K,Kim:2021zzw,Nadler191203303}---motivated physicists to consider that dark matter may have a strong self-interaction cross section \cite{spergel2000}.  As direct-detection and collider experiments continued to search for WIMP and axion dark matter without success\footnote{All the while, they continue to place strong constraints on particle parameters and open the window on solar neutrino searches (e.g.,\cite{Cooley:2021rws,Aalbers:2022dzr}).}, attention turned to the anomalous ratio of cosmic-ray positrons to electrons as observed with the ATIC \cite{Chang:2008aa} and PAMELA \cite{PAMELA:2008gwm} experiments.  Many particle theorists suggested that such an excess could arise via enhanced dark matter annihilation from light dark-sector mediators that could be kinetically mixed with elecroweak gauge bosons (see, e.g., \cite{Pospelov:2008jd,Cholis:2008qq,2009PhRvD..79a5014A}).  

Several authors pointed out that this could lead to enhanced elastic dark matter self-interactions as well \cite{Buckley:2009in,Feng:2009mn,2011PhRvL.106q1302L}, leading to significant theory work to characterize the cross section as a function of velocity \cite{Tulin:2012wi,Tulin:2013teo}, and many cosmic numerical simulation studies for signatures of this kind of self-interaction on a wide variety of scales \cite{Vogelsberger:2012ku,rocha2013,Dooley:2016ajo,Robertson:2016qef,Correa:2020qam,Cruz:2020rit}.  

All the while, new annihilation searches in gamma rays and direct detection searches on Earth constrained the coupling of these ``hidden sector" models to the Standard Model \cite{essig2009,DelNobile:2015uua}, and new searches for the ``dark matter photon" in these models commenced at a variety of colliders \cite{Bjorken:2009mm,Essig:2009nc,Battaglieri:2017aum}.  

To this day, simulators and observers are working to sharpen predictions and tests for Yukawa coupling of dark matter in the smallest halos to nearly horizon scales \cite{Vogelsberger:2012ku,Cyr-Racine:2013fsa,Buckley:2014hja,Robertson:2016qef,Despali:2018zpw,Ren:2018jpt,2019PhRvD.100l3006P,2020JCAP...02..024B,Correa:2020qam,Turner:2020vlf,Gilman:2021sdr,2020ApJ...896..112N,2021MNRAS.507..720S,Zeng:2021ldo}.  More broadly, if dark matter exists in a rich hidden sector, its physics may be primarily accessible through cosmic probes if its interaction with the Standard Model is small.  But, only when measurements of dark matter in the sky are coupled with terrestrial experiments can we fully characterize dark matter's particle properties.

For most applications (perhaps most notably for neutrinos and dark matter), there is a strong foundation of interdisciplinary work among observational cosmology and laboratory experiments, united by theory, to reveal new physics.  We expect this interaction among fields to be even more critical to suss new physics out of the next generation of cosmological and laboratory data sets.

\bibliographystyle{unsrt}
\bibliography{main.bib,extrarefs.bib}

\begin{thebibliography}{100}

\bibitem{Planck:2018nkj}
N.~Aghanim et~al.
\newblock {Planck 2018 results. I. Overview and the cosmological legacy of
  Planck}.
\newblock {\em Astron. Astrophys.}, 641:A1, 2020.

\bibitem{Starobinsky:1980te}
Alexei~A. Starobinsky.
\newblock {A New Type of Isotropic Cosmological Models Without Singularity}.
\newblock {\em Phys. Lett. B}, 91:99--102, 1980.

\bibitem{Guth:1980zm}
Alan~H. Guth.
\newblock {The Inflationary Universe: A Possible Solution to the Horizon and
  Flatness Problems}.
\newblock {\em Phys. Rev. D}, 23:347--356, 1981.

\bibitem{Linde:1981mu}
Andrei~D. Linde.
\newblock {A New Inflationary Universe Scenario: A Possible Solution of the
  Horizon, Flatness, Homogeneity, Isotropy and Primordial Monopole Problems}.
\newblock {\em Phys. Lett. B}, 108:389--393, 1982.

\bibitem{Albrecht:1982wi}
Andreas Albrecht and Paul~J. Steinhardt.
\newblock {Cosmology for Grand Unified Theories with Radiatively Induced
  Symmetry Breaking}.
\newblock {\em Phys. Rev. Lett.}, 48:1220--1223, 1982.

\bibitem{Kibble:1976sj}
T.~W.~B. Kibble.
\newblock {Topology of Cosmic Domains and Strings}.
\newblock {\em J. Phys. A}, 9:1387--1398, 1976.

\bibitem{Zeldovich:1980gh}
Ya.~B. Zeldovich.
\newblock {Cosmological fluctuations produced near a singularity}.
\newblock {\em Mon. Not. Roy. Astron. Soc.}, 192:663--667, 1980.

\bibitem{Vilenkin:1981iu}
A.~Vilenkin.
\newblock {Cosmological Density Fluctuations Produced by Vacuum Strings}.
\newblock {\em Phys. Rev. Lett.}, 46:1169--1172, 1981.
\newblock [Erratum: Phys.Rev.Lett. 46, 1496 (1981)].

\bibitem{Pen:1997ae}
Ue-Li Pen, Uros Seljak, and Neil Turok.
\newblock {Power spectra in global defect theories of cosmic structure
  formation}.
\newblock {\em Phys. Rev. Lett.}, 79:1611--1614, 1997.

\bibitem{Planck:2013mgr}
P.~A.~R. Ade et~al.
\newblock {Planck 2013 results. XXV. Searches for cosmic strings and other
  topological defects}.
\newblock {\em Astron. Astrophys.}, 571:A25, 2014.

\bibitem{Brandenberger:1988aj}
Robert~H. Brandenberger and C.~Vafa.
\newblock {Superstrings in the Early Universe}.
\newblock {\em Nucl. Phys. B}, 316:391--410, 1989.

\bibitem{Tseytlin:1991xk}
Arkady~A. Tseytlin and C.~Vafa.
\newblock {Elements of string cosmology}.
\newblock {\em Nucl. Phys. B}, 372:443--466, 1992.

\bibitem{Ijjas:2016tpn}
Anna Ijjas and Paul~J. Steinhardt.
\newblock {Classically stable nonsingular cosmological bounces}.
\newblock {\em Phys. Rev. Lett.}, 117(12):121304, 2016.

\bibitem{Ijjas:2019pyf}
Anna Ijjas and Paul~J. Steinhardt.
\newblock {A new kind of cyclic universe}.
\newblock {\em Phys. Lett. B}, 795:666--672, 2019.

\bibitem{Snowmass2021:inflation}
G.~L. Pimentel, B.~Wallisch, W.~L.~K. Wu, et~al.
\newblock {Inflation: Theory and Observations}.
\newblock \emph{Snowmass 2021 White Paper}, 2022.

\bibitem{BICEP2:2018kqh}
P.~A.~R. Ade et~al.
\newblock {BICEP2 / Keck Array x: Constraints on Primordial Gravitational Waves
  using Planck, WMAP, and New BICEP2/Keck Observations through the 2015
  Season}.
\newblock {\em Phys. Rev. Lett.}, 121:221301, 2018.

\bibitem{SPTpol:2020rqg}
P.~A.~R. Ade et~al.
\newblock {A demonstration of improved constraints on primordial gravitational
  waves with delensing}.
\newblock {\em Phys. Rev. D}, 103(2):022004, 2021.

\bibitem{BicepKeck:2022dtc}
P.~A.~R. Ade et~al.
\newblock {BICEP/Keck XV: The Bicep3 Cosmic Microwave Background Polarimeter
  and the First Three-year Data Set}.
\newblock {\em Astrophys. J.}, 927(1):77, 2022.

\bibitem{SPIDER:2021ncy}
P.~A.~R. Ade et~al.
\newblock {A Constraint on Primordial $B$-Modes from the First Flight of the
  SPIDER Balloon-Borne Telescope}.
\newblock 3 2021.

\bibitem{POLARBEAR:2022dxa}
S.~Adachi et~al.
\newblock {Improved upper limit on degree-scale CMB B-mode polarization power
  from the 670 square-degree POLARBEAR survey}.
\newblock 3 2022.

\bibitem{Dahal:2021uig}
Sumit Dahal et~al.
\newblock {Four-year Cosmology Large Angular Scale Surveyor (CLASS)
  Observations: On-sky Receiver Performance at 40, 90, 150, and 220 GHz
  Frequency Bands}.
\newblock {\em Astrophys. J.}, 926(1):33, 2022.

\bibitem{BICEP2:2014owc}
P.~A.~R. Ade et~al.
\newblock {Detection of $B$-Mode Polarization at Degree Angular Scales by
  BICEP2}.
\newblock {\em Phys. Rev. Lett.}, 112(24):241101, 2014.

\bibitem{Flauger:2014qra}
Raphael Flauger, J.~Colin Hill, and David~N. Spergel.
\newblock {Toward an Understanding of Foreground Emission in the BICEP2
  Region}.
\newblock {\em JCAP}, 08:039, 2014.

\bibitem{Mortonson:2014bja}
Michael~J. Mortonson and Uro\v{s} Seljak.
\newblock {A joint analysis of Planck and BICEP2 B modes including dust
  polarization uncertainty}.
\newblock {\em JCAP}, 10:035, 2014.

\bibitem{BICEP2:2015nss}
P.~A.~R. Ade et~al.
\newblock {Joint Analysis of BICEP2/$Keck Array$ and $Planck$ Data}.
\newblock {\em Phys. Rev. Lett.}, 114:101301, 2015.

\bibitem{SimonsObservatory:2018koc}
Peter Ade et~al.
\newblock {The Simons Observatory: Science goals and forecasts}.
\newblock {\em JCAP}, 02:056, 2019.

\bibitem{CMB-S4:2016ple}
Kevork~N. Abazajian et~al.
\newblock {CMB-S4 Science Book, First Edition}.
\newblock 10 2016.

\bibitem{Abazajian:2019eic}
Kevork Abazajian et~al.
\newblock {CMB-S4 Science Case, Reference Design, and Project Plan}.
\newblock 7 2019.

\bibitem{CMB-S4:2020lpa}
Kevork Abazajian et~al.
\newblock {CMB-S4: Forecasting Constraints on Primordial Gravitational Waves}.
\newblock {\em Astrophys. J.}, 926(1):54, 2022.

\bibitem{LiteBIRD:2022cnt}
E.~Allys et~al.
\newblock {Probing Cosmic Inflation with the LiteBIRD Cosmic Microwave
  Background Polarization Survey}.
\newblock 2 2022.

\bibitem{NASAPICO:2019thw}
Shaul Hanany et~al.
\newblock {PICO: Probe of Inflation and Cosmic Origins}.
\newblock 3 2019.

\bibitem{BICEP:2021xfz}
P.~A.~R. Ade et~al.
\newblock {Improved Constraints on Primordial Gravitational Waves using Planck,
  WMAP, and BICEP/Keck Observations through the 2018 Observing Season}.
\newblock {\em Phys. Rev. Lett.}, 127(15):151301, 2021.

\bibitem{Wenren:2014cga}
Danjie Wenren.
\newblock {Tilt and Tensor-to-Scalar Ratio in Multifield Monodromy Inflation}.
\newblock 5 2014.

\bibitem{Lyth:1996im}
David~H. Lyth.
\newblock {What would we learn by detecting a gravitational wave signal in the
  cosmic microwave background anisotropy?}
\newblock {\em Phys. Rev. Lett.}, 78:1861--1863, 1997.

\bibitem{Snowmass2021:stringinflation}
Raphael Flauger, Victor Gorbenko, Austin Joyce, Liam McAllister, Gary Shiu, and
  Eva Silverstein.
\newblock {Cosmology at the Theory Frontier}.
\newblock \emph{Snowmass 2021 White Paper}, 2022.

\bibitem{Bezrukov:2007ep}
Fedor~L. Bezrukov and Mikhail Shaposhnikov.
\newblock {The Standard Model Higgs boson as the inflaton}.
\newblock {\em Phys. Lett. B}, 659:703--706, 2008.

\bibitem{Ballesteros:2016xej}
Guillermo Ballesteros, Javier Redondo, Andreas Ringwald, and Carlos Tamarit.
\newblock {Standard
  Model\textemdash{}axion\textemdash{}seesaw\textemdash{}Higgs portal
  inflation. Five problems of particle physics and cosmology solved in one
  stroke}.
\newblock {\em JCAP}, 08:001, 2017.

\bibitem{Kallosh:2013yoa}
Renata Kallosh, Andrei Linde, and Diederik Roest.
\newblock {Superconformal Inflationary $\alpha$-Attractors}.
\newblock {\em JHEP}, 11:198, 2013.

\bibitem{Kallosh:2014rga}
Renata Kallosh, Andrei Linde, and Diederik Roest.
\newblock {Large field inflation and double $\alpha$-attractors}.
\newblock {\em JHEP}, 08:052, 2014.

\bibitem{Carrasco:2015pla}
John Joseph~M. Carrasco, Renata Kallosh, and Andrei Linde.
\newblock {$\alpha $-Attractors: Planck, LHC and Dark Energy}.
\newblock {\em JHEP}, 10:147, 2015.

\bibitem{Cicoli:2008gp}
M.~Cicoli, C.~P. Burgess, and F.~Quevedo.
\newblock {Fibre Inflation: Observable Gravity Waves from IIB String
  Compactifications}.
\newblock {\em JCAP}, 03:013, 2009.

\bibitem{Ferrara:2016fwe}
Sergio Ferrara and Renata Kallosh.
\newblock {Seven-disk manifold, $\alpha$-attractors, and $B$ modes}.
\newblock {\em Phys. Rev. D}, 94(12):126015, 2016.

\bibitem{Kallosh:2017ced}
Renata Kallosh, Andrei Linde, Timm Wrase, and Yusuke Yamada.
\newblock {Maximal Supersymmetry and B-Mode Targets}.
\newblock {\em JHEP}, 04:144, 2017.

\bibitem{Zaldarriaga:1998ar}
Matias Zaldarriaga and Uros Seljak.
\newblock {Gravitational lensing effect on cosmic microwave background
  polarization}.
\newblock {\em Phys. Rev. D}, 58:023003, 1998.

\bibitem{Hu:2000ee}
Wayne Hu.
\newblock {Weak lensing of the CMB: A harmonic approach}.
\newblock {\em Phys. Rev. D}, 62:043007, 2000.

\bibitem{Knox:2002pe}
Lloyd Knox and Yong-Seon Song.
\newblock {A Limit on the detectability of the energy scale of inflation}.
\newblock {\em Phys. Rev. Lett.}, 89:011303, 2002.

\bibitem{Kesden:2002ku}
Michael Kesden, Asantha Cooray, and Marc Kamionkowski.
\newblock {Separation of gravitational wave and cosmic shear contributions to
  cosmic microwave background polarization}.
\newblock {\em Phys. Rev. Lett.}, 89:011304, 2002.

\bibitem{Seljak:2003pn}
Uros Seljak and Christopher~M. Hirata.
\newblock {Gravitational lensing as a contaminant of the gravity wave signal in
  CMB}.
\newblock {\em Phys. Rev. D}, 69:043005, 2004.

\bibitem{Smith:2010gu}
Kendrick~M. Smith, Duncan Hanson, Marilena LoVerde, Christopher~M. Hirata, and
  Oliver Zahn.
\newblock {Delensing CMB Polarization with External Datasets}.
\newblock {\em JCAP}, 06:014, 2012.

\bibitem{Kritsuk:2017aab}
Alexei~G. Kritsuk, Raphael Flauger, and Sergey~D. Ustyugov.
\newblock {Dust-polarization maps for local interstellar turbulence}.
\newblock {\em Phys. Rev. Lett.}, 121(2):021104, 2018.

\bibitem{Kim:2019xov}
Chang-Goo Kim, Steve~K. Choi, and Raphael Flauger.
\newblock {Dust Polarization Maps from TIGRESS: E/B power asymmetry and TE
  correlation}.
\newblock 1 2019.

\bibitem{Hervias-Caimapo:2021zue}
Carlos Herv\'\i{}as-Caimapo and Kevin Huffenberger.
\newblock {Full-sky, arcminute-scale, 3D models of Galactic microwave
  foreground dust emission based on filaments}.
\newblock 7 2021.

\bibitem{Thorne:2021nux}
Ben Thorne, Lloyd Knox, and Karthik Prabhu.
\newblock {A generative model of galactic dust emission using variational
  autoencoders}.
\newblock {\em Mon. Not. Roy. Astron. Soc.}, 504(2):2603--2613, 2021.

\bibitem{Krachmalnicoff:2020rln}
Nicoletta Krachmalnicoff and Giuseppe Puglisi.
\newblock {ForSE: A GAN-based Algorithm for Extending CMB Foreground Models to
  Subdegree Angular Scales}.
\newblock {\em Astrophys. J.}, 911(1):42, 2021.

\bibitem{Larsen:2016wpa}
Patricia Larsen, Anthony Challinor, Blake~D. Sherwin, and Daisy Mak.
\newblock {Demonstration of cosmic microwave background delensing using the
  cosmic infrared background}.
\newblock {\em Phys. Rev. Lett.}, 117(15):151102, 2016.

\bibitem{SPT:2017ddy}
A.~Manzotti et~al.
\newblock {CMB Polarization B-mode Delensing with SPTpol and Herschel}.
\newblock {\em Astrophys. J.}, 846(1):45, 2017.

\bibitem{Carron:2017vfg}
Julien Carron, Antony Lewis, and Anthony Challinor.
\newblock {Internal delensing of Planck CMB temperature and polarization}.
\newblock {\em JCAP}, 05:035, 2017.

\bibitem{Planck:2018lbu}
N.~Aghanim et~al.
\newblock {Planck 2018 results. VIII. Gravitational lensing}.
\newblock {\em Astron. Astrophys.}, 641:A8, 2020.

\bibitem{POLARBEAR:2019snn}
S.~Adachi et~al.
\newblock {Internal delensing of Cosmic Microwave Background polarization
  $B$-modes with the POLARBEAR experiment}.
\newblock {\em Phys. Rev. Lett.}, 124(13):131301, 2020.

\bibitem{ACT:2020goa}
Dongwon Han et~al.
\newblock {The Atacama Cosmology Telescope: delensed power spectra and
  parameters}.
\newblock {\em JCAP}, 01:031, 2021.

\bibitem{Carron:2017mqf}
Julien Carron and Antony Lewis.
\newblock {Maximum a posteriori CMB lensing reconstruction}.
\newblock {\em Phys. Rev. D}, 96(6):063510, 2017.

\bibitem{Lizancos:2021wpy}
Ant\'on~Baleato Lizancos, Anthony Challinor, Blake~D. Sherwin, and Toshiya
  Namikawa.
\newblock {Delensing the CMB with the cosmic infrared background: the impact of
  foregrounds}.
\newblock 2 2021.

\bibitem{Millea:2020cpw}
Marius Millea, Ethan Anderes, and Benjamin~D. Wandelt.
\newblock {Sampling-based inference of the primordial CMB and gravitational
  lensing}.
\newblock {\em Phys. Rev. D}, 102(12):123542, 2020.

\bibitem{Namikawa:2021gyh}
Toshiya Namikawa et~al.
\newblock {Simons Observatory: Constraining inflationary gravitational waves
  with multitracer B-mode delensing}.
\newblock {\em Phys. Rev. D}, 105(2):023511, 2022.

\bibitem{Millea:2021had}
Marius Millea and Uros Seljak.
\newblock {MUSE: Marginal Unbiased Score Expansion and Application to CMB
  Lensing}.
\newblock 12 2021.

\bibitem{Kofman:1994rk}
Lev Kofman, Andrei~D. Linde, and Alexei~A. Starobinsky.
\newblock {Reheating after inflation}.
\newblock {\em Phys. Rev. Lett.}, 73:3195--3198, 1994.

\bibitem{Kofman:1997yn}
Lev Kofman, Andrei~D. Linde, and Alexei~A. Starobinsky.
\newblock {Towards the theory of reheating after inflation}.
\newblock {\em Phys. Rev. D}, 56:3258--3295, 1997.

\bibitem{Lozanov:2019jxc}
Kaloian~D. Lozanov.
\newblock {Lectures on Reheating after Inflation}.
\newblock 7 2019.

\bibitem{Bardeen:1980kt}
James~M. Bardeen.
\newblock {Gauge Invariant Cosmological Perturbations}.
\newblock {\em Phys. Rev. D}, 22:1882--1905, 1980.

\bibitem{Weinberg:2003sw}
Steven Weinberg.
\newblock {Adiabatic modes in cosmology}.
\newblock {\em Phys. Rev. D}, 67:123504, 2003.

\bibitem{Weinberg:2004kr}
Steven Weinberg.
\newblock {Can non-adiabatic perturbations arise after single-field inflation?}
\newblock {\em Phys. Rev. D}, 70:043541, 2004.

\bibitem{Adshead:2010mc}
Peter Adshead, Richard Easther, Jonathan Pritchard, and Abraham Loeb.
\newblock {Inflation and the Scale Dependent Spectral Index: Prospects and
  Strategies}.
\newblock {\em JCAP}, 02:021, 2011.

\bibitem{Mortonson:2010er}
Michael~J. Mortonson, Hiranya~V. Peiris, and Richard Easther.
\newblock {Bayesian Analysis of Inflation: Parameter Estimation for Single
  Field Models}.
\newblock {\em Phys. Rev. D}, 83:043505, 2011.

\bibitem{Allahverdi:2020bys}
Rouzbeh Allahverdi et~al.
\newblock {The First Three Seconds: a Review of Possible Expansion Histories of
  the Early Universe}.
\newblock 6 2020.

\bibitem{Amin:2011hj}
Mustafa~A. Amin, Richard Easther, Hal Finkel, Raphael Flauger, and Mark~P.
  Hertzberg.
\newblock {Oscillons After Inflation}.
\newblock {\em Phys. Rev. Lett.}, 108:241302, 2012.

\bibitem{Lozanov:2016hid}
Kaloian~D. Lozanov and Mustafa~A. Amin.
\newblock {Equation of State and Duration to Radiation Domination after
  Inflation}.
\newblock {\em Phys. Rev. Lett.}, 119(6):061301, 2017.

\bibitem{Antusch:2020iyq}
Stefan Antusch, Daniel~G. Figueroa, Kenneth Marschall, and Francisco Torrenti.
\newblock {Energy distribution and equation of state of the early Universe:
  matching the end of inflation and the onset of radiation domination}.
\newblock {\em Phys. Lett. B}, 811:135888, 2020.

\bibitem{Fan:2021otj}
JiJi Fan, Kaloian~D. Lozanov, and Qianshu Lu.
\newblock {Spillway Preheating}.
\newblock {\em JHEP}, 05:069, 2021.

\bibitem{Mambrini:2013iaa}
Yann Mambrini, Keith~A. Olive, Jeremie Quevillon, and Bryan Zaldivar.
\newblock {Gauge Coupling Unification and Nonequilibrium Thermal Dark Matter}.
\newblock {\em Phys. Rev. Lett.}, 110(24):241306, 2013.

\bibitem{Baumann:2015rya}
Daniel Baumann, Daniel Green, Joel Meyers, and Benjamin Wallisch.
\newblock {Phases of New Physics in the CMB}.
\newblock {\em JCAP}, 01:007, 2016.

\bibitem{Daido:2017tbr}
Ryuji Daido, Fuminobu Takahashi, and Wen Yin.
\newblock {The ALP miracle revisited}.
\newblock {\em JHEP}, 02:104, 2018.

\bibitem{Hooper:2018buz}
Dan Hooper, Gordan Krnjaic, Andrew~J. Long, and Samuel~D. Mcdermott.
\newblock {Can the Inflaton Also Be a Weakly Interacting Massive Particle?}
\newblock {\em Phys. Rev. Lett.}, 122(9):091802, 2019.

\bibitem{Bernal:2018hjm}
Nicol\'as Bernal, Arindam Chatterjee, and Arnab Paul.
\newblock {Non-thermal production of Dark Matter after Inflation}.
\newblock {\em JCAP}, 12:020, 2018.

\bibitem{Garcia:2018wtq}
Marcos A.~G. Garcia and Mustafa~A. Amin.
\newblock {Prethermalization production of dark matter}.
\newblock {\em Phys. Rev. D}, 98(10):103504, 2018.

\bibitem{Green:2019glg}
Daniel Green et~al.
\newblock {Messengers from the Early Universe: Cosmic Neutrinos and Other Light
  Relics}.
\newblock {\em Bull. Am. Astron. Soc.}, 51(7):159, 2019.

\bibitem{Green:2000he}
Anne~M. Green and Karim~A. Malik.
\newblock {Primordial black hole production due to preheating}.
\newblock {\em Phys. Rev. D}, 64:021301, 2001.

\bibitem{Carr:2020gox}
Bernard Carr, Kazunori Kohri, Yuuiti Sendouda, and Jun'ichi Yokoyama.
\newblock {Constraints on primordial black holes}.
\newblock {\em Rept. Prog. Phys.}, 84(11):116902, 2021.

\bibitem{Jedamzik:2010dq}
Karsten Jedamzik, Martin Lemoine, and Jerome Martin.
\newblock {Collapse of Small-Scale Density Perturbations during Preheating in
  Single Field Inflation}.
\newblock {\em JCAP}, 09:034, 2010.

\bibitem{Easther:2010mr}
Richard Easther, Raphael Flauger, and James~B. Gilmore.
\newblock {Delayed Reheating and the Breakdown of Coherent Oscillations}.
\newblock {\em JCAP}, 04:027, 2011.

\bibitem{Erickcek:2011us}
Adrienne~L. Erickcek and Kris Sigurdson.
\newblock {Reheating Effects in the Matter Power Spectrum and Implications for
  Substructure}.
\newblock {\em Phys. Rev. D}, 84:083503, 2011.

\bibitem{Flauger:2009ab}
Raphael Flauger, Liam McAllister, Enrico Pajer, Alexander Westphal, and Gang
  Xu.
\newblock {Oscillations in the CMB from Axion Monodromy Inflation}.
\newblock {\em JCAP}, 06:009, 2010.

\bibitem{Aich:2011qv}
Moumita Aich, Dhiraj~Kumar Hazra, L.~Sriramkumar, and Tarun Souradeep.
\newblock {Oscillations in the inflaton potential: Complete numerical treatment
  and comparison with the recent and forthcoming CMB datasets}.
\newblock {\em Phys. Rev. D}, 87:083526, 2013.

\bibitem{Peiris:2013opa}
Hiranya Peiris, Richard Easther, and Raphael Flauger.
\newblock {Constraining Monodromy Inflation}.
\newblock {\em JCAP}, 09:018, 2013.

\bibitem{Easther:2013kla}
Richard Easther and Raphael Flauger.
\newblock {Planck Constraints on Monodromy Inflation}.
\newblock {\em JCAP}, 02:037, 2014.

\bibitem{Flauger:2014ana}
Raphael Flauger, Liam McAllister, Eva Silverstein, and Alexander Westphal.
\newblock {Drifting Oscillations in Axion Monodromy}.
\newblock {\em JCAP}, 10:055, 2017.

\bibitem{Beutler:2019ojk}
Florian Beutler, Matteo Biagetti, Daniel Green, An\v{z}e Slosar, and Benjamin
  Wallisch.
\newblock {Primordial Features from Linear to Nonlinear Scales}.
\newblock {\em Phys. Rev. Res.}, 1(3):033209, 2019.

\bibitem{DESI:2016fyo}
Amir Aghamousa et~al.
\newblock {The DESI Experiment Part I: Science,Targeting, and Survey Design}.
\newblock 10 2016.

\bibitem{EUCLID:2011zbd}
R.~Laureijs et~al.
\newblock {Euclid Definition Study Report}.
\newblock 10 2011.

\bibitem{Peccei:1977hh}
R.~D. Peccei and Helen~R. Quinn.
\newblock {CP Conservation in the Presence of Instantons}.
\newblock {\em Phys. Rev. Lett.}, 38:1440--1443, 1977.

\bibitem{Peccei:1977ur}
R.~D. Peccei and Helen~R. Quinn.
\newblock {Constraints Imposed by CP Conservation in the Presence of
  Instantons}.
\newblock {\em Phys. Rev. D}, 16:1791--1797, 1977.

\bibitem{Weinberg:1977ma}
Steven Weinberg.
\newblock {A New Light Boson?}
\newblock {\em Phys. Rev. Lett.}, 40:223--226, 1978.

\bibitem{Mollerach:1989hu}
Silvia Mollerach.
\newblock {Isocurvature Baryon Perturbations and Inflation}.
\newblock {\em Phys. Rev. D}, 42:313--325, 1990.

\bibitem{Mukhanov:1990me}
Viatcheslav~F. Mukhanov, H.~A. Feldman, and Robert~H. Brandenberger.
\newblock {Theory of cosmological perturbations. Part 1. Classical
  perturbations. Part 2. Quantum theory of perturbations. Part 3. Extensions}.
\newblock {\em Phys. Rept.}, 215:203--333, 1992.

\bibitem{Moroi:2001ct}
Takeo Moroi and Tomo Takahashi.
\newblock {Effects of cosmological moduli fields on cosmic microwave
  background}.
\newblock {\em Phys. Lett. B}, 522:215--221, 2001.
\newblock [Erratum: Phys.Lett.B 539, 303--303 (2002)].

\bibitem{Lyth:2001nq}
David~H. Lyth and David Wands.
\newblock {Generating the curvature perturbation without an inflaton}.
\newblock {\em Phys. Lett. B}, 524:5--14, 2002.

\bibitem{Lyth:2002my}
David~H. Lyth, Carlo Ungarelli, and David Wands.
\newblock {The Primordial density perturbation in the curvaton scenario}.
\newblock {\em Phys. Rev. D}, 67:023503, 2003.

\bibitem{Meerburg:2019qqi}
P.~Daniel Meerburg et~al.
\newblock {Primordial Non-Gaussianity}.
\newblock 3 2019.

\bibitem{CosmicVisions21cm:2018rfq}
R\'eza Ansari et~al.
\newblock {Inflation and Early Dark Energy with a Stage II Hydrogen Intensity
  Mapping experiment}.
\newblock 10 2018.

\bibitem{PUMA:2019jwd}
An\v{z}e Slosar et~al.
\newblock {Packed Ultra-wideband Mapping Array (PUMA): A Radio Telescope for
  Cosmology and Transients}.
\newblock {\em Bull. Am. Astron. Soc.}, 51:53, 2019.

\bibitem{Ivanov:2019pdj}
Mikhail~M. Ivanov, Marko Simonovi\'c, and Matias Zaldarriaga.
\newblock {Cosmological Parameters from the BOSS Galaxy Power Spectrum}.
\newblock {\em JCAP}, 05:042, 2020.

\bibitem{DAmico:2019fhj}
Guido D'Amico, J\'er\^ome Gleyzes, Nickolas Kokron, Katarina Markovic, Leonardo
  Senatore, Pierre Zhang, Florian Beutler, and H\'ector Gil-Mar\'\i{}n.
\newblock {The Cosmological Analysis of the SDSS/BOSS data from the Effective
  Field Theory of Large-Scale Structure}.
\newblock {\em JCAP}, 05:005, 2020.

\bibitem{Snowmass2021:EFT}
Giovanni Cabass, Mehrdad Mirbabayi, and Marko Simonovi\'{c}.
\newblock Eft of cosmology.
\newblock \emph{Snowmass 2021 White Paper}, 2022.

\bibitem{Castorina:2019wmr}
Emanuele Castorina et~al.
\newblock {Redshift-weighted constraints on primordial non-Gaussianity from the
  clustering of the eBOSS DR14 quasars in Fourier space}.
\newblock {\em JCAP}, 09:010, 2019.

\bibitem{Mueller:2021tqa}
Eva-Maria Mueller et~al.
\newblock {The clustering of galaxies in the completed SDSS-IV extended Baryon
  Oscillation Spectroscopic Survey: Primordial non-Gaussianity in Fourier
  Space}.
\newblock 6 2021.

\bibitem{Cabass:2022wjy}
Giovanni Cabass, Mikhail~M. Ivanov, Oliver H.~E. Philcox, Marko Simonovi\'c,
  and Matias Zaldarriaga.
\newblock {Constraints on Single-Field Inflation from the BOSS Galaxy Survey}.
\newblock 1 2022.

\bibitem{DAmico:2022gki}
Guido D'Amico, Matthew Lewandowski, Leonardo Senatore, and Pierre Zhang.
\newblock {Limits on primordial non-Gaussianities from BOSS galaxy-clustering
  data}.
\newblock 1 2022.

\bibitem{Philcox:2021ukg}
Oliver H.~E. Philcox.
\newblock {Cosmology without window functions. II. Cubic estimators for the
  galaxy bispectrum}.
\newblock {\em Phys. Rev. D}, 104(12):123529, 2021.

\bibitem{Philcox:2021kcw}
Oliver H.~E. Philcox and Mikhail~M. Ivanov.
\newblock {BOSS DR12 full-shape cosmology: \ensuremath{\Lambda}CDM constraints
  from the large-scale galaxy power spectrum and bispectrum monopole}.
\newblock {\em Phys. Rev. D}, 105(4):043517, 2022.

\bibitem{Wadekar:2020hax}
Digvijay Wadekar, Mikhail~M. Ivanov, and Roman Scoccimarro.
\newblock {Cosmological constraints from BOSS with analytic covariance
  matrices}.
\newblock {\em Phys. Rev. D}, 102:123521, 2020.

\bibitem{Philcox:2020zyp}
Oliver H.~E. Philcox, Mikhail~M. Ivanov, Matias Zaldarriaga, Marko Simonovic,
  and Marcel Schmittfull.
\newblock {Fewer Mocks and Less Noise: Reducing the Dimensionality of
  Cosmological Observables with Subspace Projections}.
\newblock {\em Phys. Rev. D}, 103(4):043508, 2021.

\bibitem{Chudaykin:2020aoj}
Anton Chudaykin, Mikhail~M. Ivanov, Oliver H.~E. Philcox, and Marko
  Simonovi\'c.
\newblock {Nonlinear perturbation theory extension of the Boltzmann code
  CLASS}.
\newblock {\em Phys. Rev. D}, 102(6):063533, 2020.

\bibitem{Chen:2020zjt}
Shi-Fan Chen, Zvonimir Vlah, Emanuele Castorina, and Martin White.
\newblock {Redshift-Space Distortions in Lagrangian Perturbation Theory}.
\newblock {\em JCAP}, 03:100, 2021.

\bibitem{DAmico:2020kxu}
Guido D'Amico, Leonardo Senatore, and Pierre Zhang.
\newblock {Limits on $w$CDM from the EFTofLSS with the PyBird code}.
\newblock {\em JCAP}, 01:006, 2021.

\bibitem{SDSS-III:2015hof}
Shadab Alam et~al.
\newblock {The Eleventh and Twelfth Data Releases of the Sloan Digital Sky
  Survey: Final Data from SDSS-III}.
\newblock {\em Astrophys. J. Suppl.}, 219(1):12, 2015.

\bibitem{Sailer:2021yzm}
Noah Sailer, Emanuele Castorina, Simone Ferraro, and Martin White.
\newblock {Cosmology at high redshift \textemdash{} a probe of fundamental
  physics}.
\newblock {\em JCAP}, 12(12):049, 2021.

\bibitem{Ferraro:2019uce}
Simone Ferraro et~al.
\newblock {Inflation and Dark Energy from Spectroscopy at $z > 2$}.
\newblock 3 2019.

\bibitem{Ferraro:2022ee}
Simone Ferraro, Noah Sailer, Anze Slosar, Martin White, et~al.
\newblock {Snowmass2021 Cosmic Frontier White Paper: cosmology and fundamental
  physics from the three-dimensional Large Scale Structure}.
\newblock {\em Contribution to Snowmass 2021}, March 2022.

\bibitem{Dore:2014cca}
Olivier Dor\'e et~al.
\newblock {Cosmology with the SPHEREX All-Sky Spectral Survey}.
\newblock 12 2014.

\bibitem{Rezaie:2021voi}
Mehdi Rezaie et~al.
\newblock {Primordial non-Gaussianity from the completed SDSS-IV extended
  Baryon Oscillation Spectroscopic Survey \textendash{} I: Catalogue
  preparation and systematic mitigation}.
\newblock {\em Mon. Not. Roy. Astron. Soc.}, 506(3):3439--3454, 2021.

\bibitem{Schmittfull:2017ffw}
Marcel Schmittfull and Uros Seljak.
\newblock {Parameter constraints from cross-correlation of CMB lensing with
  galaxy clustering}.
\newblock {\em Phys. Rev. D}, 97(12):123540, 2018.

\bibitem{Munchmeyer:2018eey}
Moritz M\"unchmeyer, Mathew~S. Madhavacheril, Simone Ferraro, Matthew~C.
  Johnson, and Kendrick~M. Smith.
\newblock {Constraining local non-Gaussianities with kinetic
  Sunyaev-Zel\textquoteright{}dovich tomography}.
\newblock {\em Phys. Rev. D}, 100(8):083508, 2019.

\bibitem{Smith:2018bpn}
Kendrick~M. Smith, Mathew~S. Madhavacheril, Moritz M\"unchmeyer, Simone
  Ferraro, Utkarsh Giri, and Matthew~C. Johnson.
\newblock {KSZ tomography and the bispectrum}.
\newblock 10 2018.

\bibitem{Chen:2006xjb}
Xingang Chen, Richard Easther, and Eugene~A. Lim.
\newblock {Large Non-Gaussianities in Single Field Inflation}.
\newblock {\em JCAP}, 06:023, 2007.

\bibitem{Planck:2013wtn}
P.~A.~R. Ade et~al.
\newblock {Planck 2013 Results. XXIV. Constraints on primordial
  non-Gaussianity}.
\newblock {\em Astron. Astrophys.}, 571:A24, 2014.

\bibitem{Fergusson:2014hya}
J.~R. Fergusson, H.~F. Gruetjen, E.~P.~S. Shellard, and M.~Liguori.
\newblock {Combining power spectrum and bispectrum measurements to detect
  oscillatory features}.
\newblock {\em Phys. Rev. D}, 91(2):023502, 2015.

\bibitem{Fergusson:2014tza}
J.~R. Fergusson, H.~F. Gruetjen, E.~P.~S. Shellard, and B.~Wallisch.
\newblock {Polyspectra searches for sharp oscillatory features in cosmic
  microwave sky data}.
\newblock {\em Phys. Rev. D}, 91(12):123506, 2015.

\bibitem{Meerburg:2015owa}
P.~Daniel Meerburg, Moritz M\"unchmeyer, and Benjamin Wandelt.
\newblock {Joint resonant CMB power spectrum and bispectrum estimation}.
\newblock {\em Phys. Rev. D}, 93(4):043536, 2016.

\bibitem{Planck:2015zfm}
P.~A.~R. Ade et~al.
\newblock {Planck 2015 results. XVII. Constraints on primordial
  non-Gaussianity}.
\newblock {\em Astron. Astrophys.}, 594:A17, 2016.

\bibitem{Planck:2019kim}
Y.~Akrami et~al.
\newblock {Planck 2018 results. IX. Constraints on primordial non-Gaussianity}.
\newblock {\em Astron. Astrophys.}, 641:A9, 2020.

\bibitem{Bond:2009xx}
J.~Richard Bond, Andrei~V. Frolov, Zhiqi Huang, and Lev Kofman.
\newblock {Non-Gaussian Spikes from Chaotic Billiards in Inflation Preheating}.
\newblock {\em Phys. Rev. Lett.}, 103:071301, 2009.

\bibitem{Flauger:2016idt}
Raphael Flauger, Mehrdad Mirbabayi, Leonardo Senatore, and Eva Silverstein.
\newblock {Productive Interactions: heavy particles and non-Gaussianity}.
\newblock {\em JCAP}, 10:058, 2017.

\bibitem{Chen:2018uul}
Xingang Chen, Gonzalo~A. Palma, Walter Riquelme, Bruno Scheihing~Hitschfeld,
  and Spyros Sypsas.
\newblock {Landscape tomography through primordial non-Gaussianity}.
\newblock {\em Phys. Rev. D}, 98(8):083528, 2018.

\bibitem{Panagopoulos:2019ail}
George Panagopoulos and Eva Silverstein.
\newblock {Primordial Black Holes from non-Gaussian tails}.
\newblock 6 2019.

\bibitem{Panagopoulos:2020sxp}
George Panagopoulos and Eva Silverstein.
\newblock {Multipoint correlators in multifield cosmology}.
\newblock 3 2020.

\bibitem{Celoria:2021vjw}
Marco Celoria, Paolo Creminelli, Giovanni Tambalo, and Vicharit Yingcharoenrat.
\newblock {Beyond perturbation theory in inflation}.
\newblock {\em JCAP}, 06:051, 2021.

\bibitem{Munchmeyer:2019wlh}
Moritz M\"unchmeyer and Kendrick~M. Smith.
\newblock {Higher N-point function data analysis techniques for heavy particle
  production and WMAP results}.
\newblock {\em Phys. Rev. D}, 100(12):123511, 2019.

\bibitem{Baumann:2021ykm}
Daniel Baumann and Daniel Green.
\newblock {The Power of Locality: Primordial Non-Gaussianity at the Map Level}.
\newblock 12 2021.

\bibitem{Zwicky:1933gu}
F.~Zwicky.
\newblock {Die Rotverschiebung von extragalaktischen Nebeln}.
\newblock {\em Helv. Phys. Acta}, 6:110--127, 1933.

\bibitem{Rubin:1970zza}
Vera~C. Rubin and W.~Kent Ford, Jr.
\newblock {Rotation of the Andromeda Nebula from a Spectroscopic Survey of
  Emission Regions}.
\newblock {\em Astrophys. J.}, 159:379--403, 1970.

\bibitem{Rubin:1980zd}
V.~C. Rubin, N.~Thonnard, and W.~K. Ford, Jr.
\newblock {Rotational properties of 21 SC galaxies with a large range of
  luminosities and radii, from NGC 4605 /R = 4kpc/ to UGC 2885 /R = 122 kpc/}.
\newblock {\em Astrophys. J.}, 238:471, 1980.

\bibitem{Glashow:1961tr}
S.~L. Glashow.
\newblock {Partial Symmetries of Weak Interactions}.
\newblock {\em Nucl. Phys.}, 22:579--588, 1961.

\bibitem{Weinberg:1967tq}
Steven Weinberg.
\newblock {A Model of Leptons}.
\newblock {\em Phys. Rev. Lett.}, 19:1264--1266, 1967.

\bibitem{Salam:1968rm}
Abdus Salam.
\newblock {Weak and Electromagnetic Interactions}.
\newblock {\em Conf. Proc. C}, 680519:367--377, 1968.

\bibitem{Huchra:1983wy}
J.~Huchra, M.~Davis, D.~Latham, and J.~Tonry.
\newblock {A survey of galaxy redshifts: 4. The data.}
\newblock {\em Astrophys. J. Suppl.}, 52:L89--L119, 1983.

\bibitem{Peebles:1982ib}
P.~J.~E. Peebles.
\newblock {PRIMEVAL ADIABATIC PERTURBATIONS: EFFECT OF MASSIVE NEUTRINOS}.
\newblock {\em Astrophys. J.}, 258:415--424, 1982.

\bibitem{White:1983fcs}
Simon D.~M. White, C.~S. Frenk, and M.~Davis.
\newblock {Clustering in a Neutrino Dominated Universe}.
\newblock {\em Astrophys. J. Lett.}, 274:L1--L5, 1983.

\bibitem{Geller:1989da}
Margaret~J. Geller and John~P. Huchra.
\newblock {Mapping the universe}.
\newblock {\em Science}, 246:897--903, 1989.

\bibitem{Blumenthal:1984bp}
George~R. Blumenthal, S.~M. Faber, Joel~R. Primack, and Martin~J. Rees.
\newblock {Formation of Galaxies and Large Scale Structure with Cold Dark
  Matter}.
\newblock {\em Nature}, 311:517--525, 1984.

\bibitem{Davis:1985rj}
Marc Davis, George Efstathiou, Carlos~S. Frenk, and Simon D.~M. White.
\newblock {The Evolution of Large Scale Structure in a Universe Dominated by
  Cold Dark Matter}.
\newblock {\em Astrophys. J.}, 292:371--394, 1985.

\bibitem{Pagels:1981ke}
Heinz Pagels and Joel~R. Primack.
\newblock {Supersymmetry, Cosmology and New TeV Physics}.
\newblock {\em Phys. Rev. Lett.}, 48:223, 1982.

\bibitem{Buckley:2017ijx}
Matthew~R. Buckley and Annika H.~G. Peter.
\newblock {Gravitational probes of dark matter physics}.
\newblock {\em Phys. Rept.}, 761:1--60, 2018.

\bibitem{bullock2017}
James~S Bullock and Michael Boylan-Kolchin.
\newblock {Small-Scale} challenges to the {$\Lambda$CDM} paradigm.
\newblock {\em Annu. Rev. Astron. Astrophys.}, 55(1):343--387, August 2017.

\bibitem{LSSTDarkMatterGroup:2019mwo}
Alex Drlica-Wagner et~al.
\newblock {Probing the Fundamental Nature of Dark Matter with the Large
  Synoptic Survey Telescope}.
\newblock 2 2019.

\bibitem{Green:2021jrr}
Anne~M. Green.
\newblock {Dark Matter in Astrophysics/Cosmology}.
\newblock In {\em {Les Houches summer school on Dark Matter}}, 9 2021.

\bibitem{2004MNRAS.350.1210Z}
M.~A. {Zwaan}, M.~J. {Meyer}, R.~L. {Webster}, L.~{Staveley-Smith}, M.~J.
  {Drinkwater}, D.~G. {Barnes}, R.~{Bhathal}, W.~J.~G. {de Blok}, M.~J.
  {Disney}, R.~D. {Ekers}, K.~C. {Freeman}, D.~A. {Garcia}, B.~K. {Gibson},
  J.~{Harnett}, P.~A. {Henning}, M.~{Howlett}, H.~{Jerjen}, M.~J. {Kesteven},
  V.~A. {Kilborn}, P.~M. {Knezek}, B.~S. {Koribalski}, S.~{Mader},
  M.~{Marquarding}, R.~F. {Minchin}, J.~{O'Brien}, T.~{Oosterloo}, M.~J.
  {Pierce}, R.~M. {Price}, M.~E. {Putman}, E.~{Ryan-Weber}, S.~D. {Ryder},
  E.~M. {Sadler}, J.~{Stevens}, I.~M. {Stewart}, F.~{Stootman}, M.~{Waugh}, and
  A.~E. {Wright}.
\newblock {The HIPASS catalogue - II. Completeness, reliability and parameter
  accuracy}.
\newblock {\em \mnras}, 350(4):1210--1219, June 2004.

\bibitem{2009ApJ...696.2179K}
Sergey~E. {Koposov}, Jaiyul {Yoo}, Hans-Walter {Rix}, David~H. {Weinberg},
  Andrea~V. {Macci{\`o}}, and Jordi~Miralda {Escud{\'e}}.
\newblock {A Quantitative Explanation of the Observed Population of Milky Way
  Satellite Galaxies}.
\newblock {\em \apj}, 696(2):2179--2194, May 2009.

\bibitem{2009AJ....137..450W}
S.~M. {Walsh}, B.~{Willman}, and H.~{Jerjen}.
\newblock {The Invisibles: A Detection Algorithm to Trace the Faintest Milky
  Way Satellites}.
\newblock {\em \aj}, 137(1):450--469, January 2009.

\bibitem{Fermi-LAT:2015att}
M.~Ackermann et~al.
\newblock {Searching for Dark Matter Annihilation from Milky Way Dwarf
  Spheroidal Galaxies with Six Years of Fermi Large Area Telescope Data}.
\newblock {\em Phys. Rev. Lett.}, 115(23):231301, 2015.

\bibitem{2020ApJ...893...47D}
A.~{Drlica-Wagner}, K.~{Bechtol}, S.~{Mau}, M.~{McNanna}, E.~O. {Nadler}, A.~B.
  {Pace}, T.~S. {Li}, A.~{Pieres}, E.~{Rozo}, J.~D. {Simon}, A.~R. {Walker},
  R.~H. {Wechsler}, T.~M.~C. {Abbott}, S.~{Allam}, J.~{Annis}, E.~{Bertin},
  D.~{Brooks}, D.~L. {Burke}, A.~Carnero {Rosell}, M.~{Carrasco Kind},
  J.~{Carretero}, M.~{Costanzi}, L.~N. {da Costa}, J.~{De Vicente}, S.~{Desai},
  H.~T. {Diehl}, P.~{Doel}, T.~F. {Eifler}, S.~{Everett}, B.~{Flaugher},
  J.~{Frieman}, J.~{Garc{\'\i}a-Bellido}, E.~{Gaztanaga}, D.~{Gruen}, R.~A.
  {Gruendl}, J.~{Gschwend}, G.~{Gutierrez}, K.~{Honscheid}, D.~J. {James},
  E.~{Krause}, K.~{Kuehn}, N.~{Kuropatkin}, O.~{Lahav}, M.~A.~G. {Maia}, J.~L.
  {Marshall}, P.~{Melchior}, F.~{Menanteau}, R.~{Miquel}, A.~{Palmese}, A.~A.
  {Plazas}, E.~{Sanchez}, V.~{Scarpine}, M.~{Schubnell}, S.~{Serrano},
  I.~{Sevilla-Noarbe}, M.~{Smith}, E.~{Suchyta}, G.~{Tarle}, and {DES
  Collaboration}.
\newblock {Milky Way Satellite Census. I. The Observational Selection Function
  for Milky Way Satellites in DES Y3 and Pan-STARRS DR1}.
\newblock {\em \apj}, 893(1):47, April 2020.

\bibitem{rocha2013}
Miguel Rocha, Annika H~G Peter, James~S Bullock, Manoj Kaplinghat, Shea
  Garrison-Kimmel, Jose O{\~n}orbe, and Leonidas~A Moustakas.
\newblock Cosmological simulations with self-interacting dark matter -- i.
  constant-density cores and substructure.
\newblock {\em Mon. Not. R. Astron. Soc.}, 430(1):81--104, March 2013.

\bibitem{Wang:2013rha}
Mei-Yu Wang, Rupert A.~C. Croft, Annika H.~G. Peter, Andrew~R. Zentner, and
  Chris~W. Purcell.
\newblock {Lyman-\ensuremath{\alpha} forest constraints on decaying dark
  matter}.
\newblock {\em Phys. Rev. D}, 88(12):123515, 2013.

\bibitem{Robertson:2016qef}
Andrew Robertson, Richard Massey, and Vincent Eke.
\newblock {Cosmic particle colliders: simulations of self-interacting dark
  matter with anisotropic scattering}.
\newblock {\em Mon. Not. Roy. Astron. Soc.}, 467(4):4719--4730, 2017.

\bibitem{Despali:2018zpw}
Giulia Despali, Martin Sparre, Simona Vegetti, Mark Vogelsberger, Jes\'us
  Zavala, and Federico Marinacci.
\newblock {The interplay of Self-Interacting Dark Matter and baryons in shaping
  the halo evolution}.
\newblock {\em Mon. Not. Roy. Astron. Soc.}, 484:4563, 2019.

\bibitem{2019MNRAS.490..962F}
Alex {Fitts}, Michael {Boylan-Kolchin}, Brandon {Bozek}, James~S. {Bullock},
  Andrew {Graus}, Victor {Robles}, Philip~F. {Hopkins}, Kareem {El-Badry}, Shea
  {Garrison-Kimmel}, Claude-Andr{\'e} {Faucher-Gigu{\`e}re}, Andrew {Wetzel},
  and Du{\v{s}}an {Kere{\v{s}}}.
\newblock {Dwarf galaxies in CDM, WDM, and SIDM: disentangling baryons and dark
  matter physics}.
\newblock {\em \mnras}, 490(1):962--977, November 2019.

\bibitem{2020JCAP...02..024B}
Arka {Banerjee}, Susmita {Adhikari}, Neal {Dalal}, Surhud {More}, and Andrey
  {Kravtsov}.
\newblock {Signatures of self-interacting dark matter on cluster density
  profile and subhalo distributions}.
\newblock {\em \jcap}, 2020(2):024, February 2020.

\bibitem{2020MNRAS.498..702L}
Mark~R. {Lovell}, Wojciech {Hellwing}, Aaron {Ludlow}, Jes{\'u}s {Zavala},
  Andrew {Robertson}, Azadeh {Fattahi}, Carlos~S. {Frenk}, and Jennifer
  {Hardwick}.
\newblock {Local group star formation in warm and self-interacting dark matter
  cosmologies}.
\newblock {\em \mnras}, 498(1):702--717, October 2020.

\bibitem{2020MNRAS.497.2393L}
Alexandres {Lazar}, James~S. {Bullock}, Michael {Boylan-Kolchin}, T.~K. {Chan},
  Philip~F. {Hopkins}, Andrew~S. {Graus}, Andrew {Wetzel}, Kareem {El-Badry},
  Coral {Wheeler}, Maria~C. {Straight}, Du{\v{s}}an {Kere{\v{s}}},
  Claude-Andr{\'e} {Faucher-Gigu{\`e}re}, Alex {Fitts}, and Shea
  {Garrison-Kimmel}.
\newblock {A dark matter profile to model diverse feedback-induced core sizes
  of {\ensuremath{\Lambda}}CDM haloes}.
\newblock {\em \mnras}, 497(2):2393--2417, September 2020.

\bibitem{2021ApJ...906...96A}
Elaad {Applebaum}, Alyson~M. {Brooks}, Charlotte~R. {Christensen}, Ferah
  {Munshi}, Thomas~R. {Quinn}, Sijing {Shen}, and Michael {Tremmel}.
\newblock {Ultrafaint Dwarfs in a Milky Way Context: Introducing the Mint
  Condition DC Justice League Simulations}.
\newblock {\em \apj}, 906(2):96, January 2021.

\bibitem{2021MNRAS.507.4826L}
Mark~R. {Lovell}, Marius {Cautun}, Carlos~S. {Frenk}, Wojciech~A. {Hellwing},
  and Oliver {Newton}.
\newblock {The spatial distribution of Milky Way satellites, gaps in streams,
  and the nature of dark matter}.
\newblock {\em \mnras}, 507(4):4826--4839, November 2021.

\bibitem{2021ApJ...923...35M}
Ferah {Munshi}, Alyson~M. {Brooks}, Elaad {Applebaum}, Charlotte~R.
  {Christensen}, T.~{Quinn}, and Serena {Sligh}.
\newblock {Quantifying Scatter in Galaxy Formation at the Lowest Masses}.
\newblock {\em \apj}, 923(1):35, December 2021.

\bibitem{2021arXiv211101158S}
Isabel M.~E. {Santos-Santos}, Laura~V. {Sales}, Azadeh {Fattahi}, and Julio~F.
  {Navarro}.
\newblock {Satellite mass functions and the faint end of the galaxy mass-halo
  mass relation in LCDM}.
\newblock {\em arXiv e-prints}, page arXiv:2111.01158, November 2021.

\bibitem{2021MNRAS.500.1531C}
Kun Ting~Eddie {Chua}, Karia {Dibert}, Mark {Vogelsberger}, and Jes{\'u}s
  {Zavala}.
\newblock {The impact of inelastic self-interacting dark matter on the dark
  matter structure of a Milky Way halo}.
\newblock {\em \mnras}, 500(1):1531--1546, January 2021.

\bibitem{2022MNRAS.tmp..348M}
Stuart {McAlpine}, John~C. {Helly}, Matthieu {Schaller}, Till {Sawala}, Guilhem
  {Lavaux}, Jens {Jasche}, Carlos~S. {Frenk}, Adrian {Jenkins}, John~R.
  {Lucey}, and Peter~H. {Johansson}.
\newblock {SIBELIUS-DARK: a galaxy catalogue of the Local Volume from a
  constrained realisation simulation}.
\newblock {\em \mnras}, February 2022.

\bibitem{Kennedy:2013uta}
Rachel Kennedy, Carlos Frenk, Shaun Cole, and Andrew Benson.
\newblock {Constraining the warm dark matter particle mass with Milky Way
  satellites}.
\newblock {\em Mon. Not. Roy. Astron. Soc.}, 442(3):2487--2495, 2014.

\bibitem{Kim:2017iwr}
Stacy~Y. Kim, Annika H.~G. Peter, and Jonathan~R. Hargis.
\newblock {Missing Satellites Problem: Completeness Corrections to the Number
  of Satellite Galaxies in the Milky Way are Consistent with Cold Dark Matter
  Predictions}.
\newblock {\em Phys. Rev. Lett.}, 121(21):211302, 2018.

\bibitem{Gilman:2019nap}
Daniel Gilman, Simon Birrer, Anna Nierenberg, Tommaso Treu, Xiaolong Du, and
  Andrew Benson.
\newblock {Warm dark matter chills out: constraints on the halo mass function
  and the free-streaming length of dark matter with eight quadruple-image
  strong gravitational lenses}.
\newblock {\em Mon. Not. Roy. Astron. Soc.}, 491(4):6077--6101, 2020.

\bibitem{Banik:2019smi}
Nilanjan Banik, Jo~Bovy, Gianfranco Bertone, Denis Erkal, and T.~J.~L. de~Boer.
\newblock {Novel constraints on the particle nature of dark matter from stellar
  streams}.
\newblock {\em JCAP}, 10:043, 2021.

\bibitem{Nadler:2021dft}
Ethan~O. Nadler, Simon Birrer, Daniel Gilman, Risa~H. Wechsler, Xiaolong Du,
  Andrew Benson, Anna~M. Nierenberg, and Tommaso Treu.
\newblock {Dark Matter Constraints from a Unified Analysis of Strong
  Gravitational Lenses and Milky Way Satellite Galaxies}.
\newblock {\em Astrophys. J.}, 917(1):7, 2021.

\bibitem{Dekker:2021scf}
Ariane Dekker, Shin'ichiro Ando, Camila~A. Correa, and Kenny C.~Y. Ng.
\newblock {Warm Dark Matter Constraints Using Milky-Way Satellite Observations
  and Subhalo Evolution Modeling}.
\newblock 11 2021.

\bibitem{StenDelos:2019xdk}
M.~Sten Delos, Tim Linden, and Adrienne~L. Erickcek.
\newblock {Breaking a dark degeneracy: The gamma-ray signature of early matter
  domination}.
\newblock {\em Phys. Rev. D}, 100(12):123546, 2019.

\bibitem{Newton:2020cog}
Oliver Newton, Matteo Leo, Marius Cautun, Adrian Jenkins, Carlos~S. Frenk,
  Mark~R. Lovell, John~C. Helly, Andrew~J. Benson, and Shaun Cole.
\newblock {Constraints on the properties of warm dark matter using the
  satellite galaxies of the Milky Way}.
\newblock {\em JCAP}, 08:062, 2021.

\bibitem{Nguyen:2021cnb}
David~V. Nguyen, Dimple Sarnaaik, Kimberly~K. Boddy, Ethan~O. Nadler, and Vera
  Gluscevic.
\newblock {Observational constraints on dark matter scattering with electrons}.
\newblock {\em Phys. Rev. D}, 104(10):103521, 2021.

\bibitem{Kim:2021zzw}
Stacy~Y. Kim and Annika H.~G. Peter.
\newblock {The Milky Way satellite velocity function is a sharp probe of
  small-scale structure problems}.
\newblock 6 2021.

\bibitem{Mau:2022sbf}
S.~Mau et~al.
\newblock {Milky Way Satellite Census. IV. Constraints on Decaying Dark Matter
  from Observations of Milky Way Satellite Galaxies}.
\newblock 1 2022.

\bibitem{DMfacilitiesWP}
Sukanya Chakrabarti et~al.
\newblock {Snowmass2021 Cosmic Frontier White Paper: Observational Facilities
  to Study Dark Matter}.
\newblock {\em arXiv e-prints}, page arXiv:2203.06200, March 2022.

\bibitem{DMRubinWP}
Yao-Yuan Mao, Annika H.~G. Peter, et~al.
\newblock {Snowmass2021: Vera C. Rubin Observatory as a Flagship Dark Matter
  Experiment}.
\newblock {\em arXiv e-prints}, page arXiv:2203.07252, March 2022.

\bibitem{DMDESIWP}
Monica Valluri et~al.
\newblock {Snowmass2021 Cosmic Frontier White Paper: Dark Matter Constraints
  from DESI}.

\bibitem{DMCMB64WP}
Cora Dvorkin et~al.
\newblock {Dark Matter Physics from the CMB-S4 Experiment}.

\bibitem{DMsimsWP}
Arka Banerjee et~al.
\newblock {Snowmass2021 Cosmic Frontier White Paper: Cosmological Simulations
  for Dark Matter Physics}.
\newblock {\em arXiv e-prints}, page arXiv:2203.07049, March 2022.

\bibitem{DMhalosWP}
Keith Bechtol et~al.
\newblock {Snowmass2021 Cosmic Frontier White Paper: Dark Matter Physics from
  Halo Measurements}.
\newblock {\em arXiv e-prints}, page arXiv:2203.07354, March 2022.

\bibitem{DMPBHWP}
Andrea Albert et~al.
\newblock {Snowmass2021: Cosmic Frontier Primordial Black Hole Dark Matter and
  the Early Universe}.

\bibitem{DMextremeWP}
Masha Baryakhtar et~al.
\newblock {Snowmass2021 Cosmic Frontier White Paper: Dark Matter In Extreme
  Astrophysical Environments}.

\bibitem{TFastroDMWP}
Kimberly~K. Boddy et~al.
\newblock {Astrophysical and Cosmological Probes of Dark Matter}.
\newblock {\em arXiv e-prints}, page arXiv:2203.06380, March 2022.

\bibitem{Cyr-Racine:2015ihg}
Francis-Yan Cyr-Racine, Kris Sigurdson, Jesus Zavala, Torsten Bringmann, Mark
  Vogelsberger, and Christoph Pfrommer.
\newblock {ETHOS\textemdash{}an effective theory of structure formation: From
  dark particle physics to the matter distribution of the Universe}.
\newblock {\em Phys. Rev. D}, 93(12):123527, 2016.

\bibitem{Vogelsberger:2015gpr}
Mark Vogelsberger, Jesus Zavala, Francis-Yan Cyr-Racine, Christoph Pfrommer,
  Torsten Bringmann, and Kris Sigurdson.
\newblock {ETHOS \textendash{} an effective theory of structure formation: dark
  matter physics as a possible explanation of the small-scale CDM problems}.
\newblock {\em Mon. Not. Roy. Astron. Soc.}, 460(2):1399--1416, 2016.

\bibitem{Tulin:2017ara}
Sean Tulin and Hai-Bo Yu.
\newblock {Dark Matter Self-interactions and Small Scale Structure}.
\newblock {\em Phys. Rept.}, 730:1--57, 2018.

\bibitem{Brooks2017}
Alyson~M. {Brooks}, Emmanouil {Papastergis}, Charlotte~R. {Christensen}, Fabio
  {Governato}, Adrienne {Stilp}, Thomas~R. {Quinn}, and James {Wadsley}.
\newblock {How to Reconcile the Observed Velocity Function of Galaxies with
  Theory}.
\newblock {\em \apj}, 850(1):97, November 2017.

\bibitem{Benson:2010kx}
Andrew~J. Benson.
\newblock {Galacticus: A Semi-Analytic Model of Galaxy Formation}.
\newblock {\em New Astron.}, 17:175--197, 2012.

\bibitem{Pullen:2014gna}
Anthony~R. Pullen, Andrew~J. Benson, and Leonidas~A. Moustakas.
\newblock {Nonlinear evolution of dark matter subhalos and applications to warm
  dark matter}.
\newblock {\em Astrophys. J.}, 792:24, 2014.

\bibitem{2016MNRAS.461...60L}
Mark~R. {Lovell}, Sownak {Bose}, Alexey {Boyarsky}, Shaun {Cole}, Carlos~S.
  {Frenk}, Violeta {Gonzalez-Perez}, Rachel {Kennedy}, Oleg {Ruchayskiy}, and
  Alex {Smith}.
\newblock {Satellite galaxies in semi-analytic models of galaxy formation with
  sterile neutrino dark matter}.
\newblock {\em \mnras}, 461(1):60--72, September 2016.

\bibitem{Sameie:2018juk}
Omid Sameie, Andrew~J. Benson, Laura~V. Sales, Hai-Bo Yu, Leonidas~A.
  Moustakas, and Peter Creasey.
\newblock {The Effect of Dark Matter\textendash{}Dark Radiation Interactions on
  Halo Abundance: A Press\textendash{}Schechter Approach}.
\newblock {\em Astrophys. J.}, 874(1):101, 2019.

\bibitem{Kaplinghat:2015aga}
Manoj Kaplinghat, Sean Tulin, and Hai-Bo Yu.
\newblock {Dark Matter Halos as Particle Colliders: Unified Solution to
  Small-Scale Structure Puzzles from Dwarfs to Clusters}.
\newblock {\em Phys. Rev. Lett.}, 116(4):041302, 2016.

\bibitem{Enzi:2020ieg}
Wolfgang Enzi et~al.
\newblock {Joint constraints on thermal relic dark matter from strong
  gravitational lensing, the Ly\,\ensuremath{\alpha} forest, and Milky Way
  satellites}.
\newblock {\em Mon. Not. Roy. Astron. Soc.}, 506(4):5848--5862, 2021.

\bibitem{Cruz:2020rit}
A.~Cruz, A.~Pontzen, M.~Volonteri, T.~R. Quinn, M.~Tremmel, A.~M. Brooks, N.~N.
  Sanchez, F.~Munshi, and A.~Di~Cintio.
\newblock {Self-interacting dark matter and the delay of supermassive black
  hole growth}.
\newblock {\em Mon. Not. Roy. Astron. Soc.}, 500(2):2177--2187, 2020.

\bibitem{2021Natur.592..534C}
Charlie {Conroy}, Rohan~P. {Naidu}, Nicol{\'a}s {Garavito-Camargo}, Gurtina
  {Besla}, Dennis {Zaritsky}, Ana {Bonaca}, and Benjamin~D. {Johnson}.
\newblock {All-sky dynamical response of the Galactic halo to the Large
  Magellanic Cloud}.
\newblock {\em \nat}, 592(7855):534--536, April 2021.

\bibitem{Read:2008fh}
J.~I. Read, G.~Lake, O.~Agertz, and Victor~P. Debattista.
\newblock {Thin, thick and dark discs in LCDM}.
\newblock {\em Mon. Not. Roy. Astron. Soc.}, 389:1041--1057, 2008.

\bibitem{Necib:2018iwb}
Lina Necib, Mariangela Lisanti, and Vasily Belokurov.
\newblock {Inferred Evidence For Dark Matter Kinematic Substructure with
  SDSS-Gaia}.
\newblock 7 2018.

\bibitem{Besla:2019xbx}
Gurtina Besla, Annika Peter, and Nicolas Garavito-Camargo.
\newblock {The highest-speed local dark matter particles come from the Large
  Magellanic Cloud}.
\newblock {\em JCAP}, 11:013, 2019.

\bibitem{LRWP}
Cora Dvorkin et~al.
\newblock {The Physics of Light Relics}.

\bibitem{NSWP}
Martina Gerbino et~al.
\newblock {Synergy between cosmological and laboratory searches in neutrino
  physics: a white paper}.

\bibitem{Gnedin:1997vn}
Nickolay~Y. Gnedin and Oleg~Y. Gnedin.
\newblock {Cosmological neutrino background revisited}.
\newblock {\em Astrophys. J.}, 509:11--15, 1998.

\bibitem{Mangano:2001iu}
G.~Mangano, G.~Miele, S.~Pastor, and M.~Peloso.
\newblock {A Precision calculation of the effective number of cosmological
  neutrinos}.
\newblock {\em Phys. Lett. B}, 534:8--16, 2002.

\bibitem{Mangano:2005cc}
Gianpiero Mangano, Gennaro Miele, Sergio Pastor, Teguayco Pinto, Ofelia
  Pisanti, and Pasquale~D. Serpico.
\newblock {Relic neutrino decoupling including flavor oscillations}.
\newblock {\em Nucl. Phys. B}, 729:221--234, 2005.

\bibitem{deSalas:2016ztq}
Pablo~F. de~Salas and Sergio Pastor.
\newblock {Relic neutrino decoupling with flavour oscillations revisited}.
\newblock {\em JCAP}, 07:051, 2016.

\bibitem{Froustey:2020mcq}
Julien Froustey, Cyril Pitrou, and Maria~Cristina Volpe.
\newblock {Neutrino decoupling including flavour oscillations and primordial
  nucleosynthesis}.
\newblock {\em JCAP}, 12:015, 2020.

\bibitem{Planck:2018vyg}
N.~Aghanim et~al.
\newblock {Planck 2018 results. VI. Cosmological parameters}.
\newblock {\em Astron. Astrophys.}, 641:A6, 2020.

\bibitem{Baumann:2017gkg}
Daniel Baumann, Daniel Green, and Benjamin Wallisch.
\newblock {Searching for light relics with large-scale structure}.
\newblock {\em JCAP}, 08:029, 2018.

\bibitem{Baumann:2017lmt}
Daniel Baumann, Daniel Green, and Matias Zaldarriaga.
\newblock {Phases of New Physics in the BAO Spectrum}.
\newblock {\em JCAP}, 11:007, 2017.

\bibitem{CFCMBWP}
Clarence Chang et~al.
\newblock {Snowmass2021 Cosmic Frontier: CMB Measurements White Paper}.

\bibitem{CFCCMBS4WP}
The CMB-S4 Collaboration.
\newblock {Snomass CMB-S4 White Paper}.

\bibitem{Saikawa:2018rcs}
Ken'ichi Saikawa and Satoshi Shirai.
\newblock {Primordial gravitational waves, precisely: The role of
  thermodynamics in the Standard Model}.
\newblock {\em JCAP}, 05:035, 2018.

\bibitem{Cyr-Racine:2013jua}
Francis-Yan Cyr-Racine and Kris Sigurdson.
\newblock {Limits on Neutrino-Neutrino Scattering in the Early Universe}.
\newblock {\em Phys. Rev. D}, 90(12):123533, 2014.

\bibitem{Lancaster:2017ksf}
Lachlan Lancaster, Francis-Yan Cyr-Racine, Lloyd Knox, and Zhen Pan.
\newblock {A tale of two modes: Neutrino free-streaming in the early universe}.
\newblock {\em JCAP}, 07:033, 2017.

\bibitem{Song:2018zyl}
Ningqiang Song, M.~C. Gonzalez-Garcia, and Jordi Salvado.
\newblock {Cosmological constraints with self-interacting sterile neutrinos}.
\newblock {\em JCAP}, 10:055, 2018.

\bibitem{Kreisch:2019yzn}
Christina~D. Kreisch, Francis-Yan Cyr-Racine, and Olivier Dor\'e.
\newblock {Neutrino puzzle: Anomalies, interactions, and cosmological
  tensions}.
\newblock {\em Phys. Rev. D}, 101(12):123505, 2020.

\bibitem{Das:2020xke}
Anirban Das and Subhajit Ghosh.
\newblock {Flavor-specific Interaction Favors Strong Neutrino Self-coupling in
  the Early Universe}.
\newblock 11 2020.

\bibitem{RoyChoudhury:2020dmd}
Shouvik Roy~Choudhury, Steen Hannestad, and Thomas Tram.
\newblock {Updated constraints on massive neutrino self-interactions from
  cosmology in light of the $H_0$ tension}.
\newblock {\em JCAP}, 03:084, 2021.

\bibitem{Brinckmann:2020bcn}
Thejs Brinckmann, Jae~Hyeok Chang, and Marilena LoVerde.
\newblock {Self-interacting neutrinos, the Hubble parameter tension, and the
  Cosmic Microwave Background}.
\newblock 12 2020.

\bibitem{Berryman:2022hds}
Jeffrey~M. Berryman et~al.
\newblock {Neutrino Self-Interactions: A White Paper}.
\newblock In {\em {2022 Snowmass Summer Study}}, 3 2022.

\bibitem{Baumann:2016wac}
Daniel Baumann, Daniel Green, and Benjamin Wallisch.
\newblock {New Target for Cosmic Axion Searches}.
\newblock {\em Phys. Rev. Lett.}, 117(17):171301, 2016.

\bibitem{Blinov:2020hmc}
Nikita Blinov and Gustavo Marques-Tavares.
\newblock {Interacting radiation after Planck and its implications for the
  Hubble Tension}.
\newblock {\em JCAP}, 09:029, 2020.

\bibitem{Choi:2018gho}
Gongjun Choi, Chi-Ting Chiang, and Marilena LoVerde.
\newblock {Probing Decoupling in Dark Sectors with the Cosmic Microwave
  Background}.
\newblock {\em JCAP}, 06:044, 2018.

\bibitem{Hou:2011ec}
Z.~{Hou}, R.~{Keisler}, L.~{Knox}, M.~{Millea}, and C.~{Reichardt}.
\newblock {How massless neutrinos affect the cosmic microwave background
  damping tail}.
\newblock {\em \prd}, 87(8):083008, April 2013.

\bibitem{Bashinsky:2003tk}
Sergei Bashinsky and Uros Seljak.
\newblock {Neutrino perturbations in CMB anisotropy and matter clustering}.
\newblock {\em Phys. Rev. D}, 69:083002, 2004.

\bibitem{2011JCAP...07..034B}
Diego {Blas}, Julien {Lesgourgues}, and Thomas {Tram}.
\newblock {The Cosmic Linear Anisotropy Solving System (CLASS). Part II:
  Approximation schemes}.
\newblock {\em \jcap}, 2011(7):034, July 2011.

\bibitem{Lewis:1999camb}
Antony Lewis, Anthony Challinor, and Anthony Lasenby.
\newblock {Efficient computation of CMB anisotropies in closed FRW models}.
\newblock {\em Astrophys. J.}, 538:473--476, 2000.

\bibitem{Green:2016cjr}
Daniel Green, Joel Meyers, and Alexander van Engelen.
\newblock {CMB Delensing Beyond the B Modes}.
\newblock {\em JCAP}, 12:005, 2017.

\bibitem{Gonzalez-Garcia:2021dve}
Maria~Concepcion Gonzalez-Garcia, Michele Maltoni, and Thomas Schwetz.
\newblock {NuFIT: Three-Flavour Global Analyses of Neutrino Oscillation
  Experiments}.
\newblock {\em Universe}, 7(12):459, 2021.

\bibitem{Lesgourgues:2012uu}
Julien Lesgourgues and Sergio Pastor.
\newblock {Neutrino mass from Cosmology}.
\newblock {\em Adv. High Energy Phys.}, 2012:608515, 2012.

\bibitem{2009arXiv0912.0201L}
{LSST Science Collaboration}, Paul~A. {Abell}, Julius {Allison}, Scott~F.
  {Anderson}, John~R. {Andrew}, J.~Roger~P. {Angel}, Lee {Armus}, David
  {Arnett}, S.~J. {Asztalos}, Tim~S. {Axelrod}, Stephen {Bailey}, D.~R.
  {Ballantyne}, Justin~R. {Bankert}, Wayne~A. {Barkhouse}, Jeffrey~D. {Barr},
  L.~Felipe {Barrientos}, Aaron~J. {Barth}, James~G. {Bartlett}, Andrew~C.
  {Becker}, Jacek {Becla}, Timothy~C. {Beers}, Joseph~P. {Bernstein}, Rahul
  {Biswas}, Michael~R. {Blanton}, Joshua~S. {Bloom}, John~J. {Bochanski}, Pat
  {Boeshaar}, Kirk~D. {Borne}, Marusa {Bradac}, W.~N. {Brandt}, Carrie~R.
  {Bridge}, Michael~E. {Brown}, Robert~J. {Brunner}, James~S. {Bullock},
  Adam~J. {Burgasser}, James~H. {Burge}, David~L. {Burke}, Phillip~A.
  {Cargile}, Srinivasan {Chandrasekharan}, George {Chartas}, Steven~R.
  {Chesley}, You-Hua {Chu}, David {Cinabro}, Mark~W. {Claire}, Charles~F.
  {Claver}, Douglas {Clowe}, A.~J. {Connolly}, Kem~H. {Cook}, Jeff {Cooke},
  Asantha {Cooray}, Kevin~R. {Covey}, Christopher~S. {Culliton}, Roelof {de
  Jong}, Willem~H. {de Vries}, Victor~P. {Debattista}, Francisco {Delgado},
  Ian~P. {Dell'Antonio}, Saurav {Dhital}, Rosanne {Di Stefano}, Mark
  {Dickinson}, Benjamin {Dilday}, S.~G. {Djorgovski}, Gregory {Dobler}, Ciro
  {Donalek}, Gregory {Dubois-Felsmann}, Josef {Durech}, Ardis {Eliasdottir},
  Michael {Eracleous}, Laurent {Eyer}, Emilio~E. {Falco}, Xiaohui {Fan},
  Christopher~D. {Fassnacht}, Harry~C. {Ferguson}, Yanga~R. {Fernandez},
  Brian~D. {Fields}, Douglas {Finkbeiner}, Eduardo~E. {Figueroa}, Derek~B.
  {Fox}, Harold {Francke}, James~S. {Frank}, Josh {Frieman}, Sebastien
  {Fromenteau}, Muhammad {Furqan}, Gaspar {Galaz}, A.~{Gal-Yam}, Peter
  {Garnavich}, Eric {Gawiser}, John {Geary}, Perry {Gee}, Robert~R. {Gibson},
  Kirk {Gilmore}, Emily~A. {Grace}, Richard~F. {Green}, William~J. {Gressler},
  Carl~J. {Grillmair}, Salman {Habib}, J.~S. {Haggerty}, Mario {Hamuy}, Alan~W.
  {Harris}, Suzanne~L. {Hawley}, Alan~F. {Heavens}, Leslie {Hebb}, Todd~J.
  {Henry}, Edward {Hileman}, Eric~J. {Hilton}, Keri {Hoadley}, J.~B. {Holberg},
  Matt~J. {Holman}, Steve~B. {Howell}, Leopoldo {Infante}, Zeljko {Ivezic},
  Suzanne~H. {Jacoby}, Bhuvnesh {Jain}, {R}, {Jedicke}, M.~James {Jee},
  J.~{Garrett Jernigan}, Saurabh~W. {Jha}, Kathryn~V. {Johnston}, R.~Lynne
  {Jones}, Mario {Juric}, Mikko {Kaasalainen}, {Styliani}, {Kafka}, Steven~M.
  {Kahn}, Nathan~A. {Kaib}, Jason {Kalirai}, Jeff {Kantor}, Mansi~M.
  {Kasliwal}, Charles~R. {Keeton}, Richard {Kessler}, Zoran {Knezevic}, Adam
  {Kowalski}, Victor~L. {Krabbendam}, K.~Simon {Krughoff}, Shrinivas
  {Kulkarni}, Stephen {Kuhlman}, Mark {Lacy}, Sebastien {Lepine}, Ming {Liang},
  Amy {Lien}, Paulina {Lira}, Knox~S. {Long}, Suzanne {Lorenz}, Jennifer~M.
  {Lotz}, R.~H. {Lupton}, Julie {Lutz}, Lucas~M. {Macri}, Ashish~A. {Mahabal},
  Rachel {Mandelbaum}, Phil {Marshall}, Morgan {May}, Peregrine~M. {McGehee},
  Brian~T. {Meadows}, Alan {Meert}, Andrea {Milani}, Christopher~J. {Miller},
  Michelle {Miller}, David {Mills}, Dante {Minniti}, David {Monet}, Anjum~S.
  {Mukadam}, Ehud {Nakar}, Douglas~R. {Neill}, Jeffrey~A. {Newman}, Sergei
  {Nikolaev}, Martin {Nordby}, Paul {O'Connor}, Masamune {Oguri}, John
  {Oliver}, Scot~S. {Olivier}, Julia~K. {Olsen}, Knut {Olsen}, Edward~W.
  {Olszewski}, Hakeem {Oluseyi}, Nelson~D. {Padilla}, Alex {Parker}, Joshua
  {Pepper}, John~R. {Peterson}, Catherine {Petry}, Philip~A. {Pinto}, James~L.
  {Pizagno}, Bogdan {Popescu}, Andrej {Prsa}, Veljko {Radcka}, M.~Jordan
  {Raddick}, Andrew {Rasmussen}, Arne {Rau}, Jeonghee {Rho}, James~E. {Rhoads},
  Gordon~T. {Richards}, Stephen~T. {Ridgway}, Brant~E. {Robertson}, Rok
  {Roskar}, Abhijit {Saha}, Ata {Sarajedini}, Evan {Scannapieco}, Terry
  {Schalk}, Rafe {Schindler}, Samuel {Schmidt}, Sarah {Schmidt}, Donald~P.
  {Schneider}, German {Schumacher}, Ryan {Scranton}, Jacques {Sebag}, Lynn~G.
  {Seppala}, Ohad {Shemmer}, Joshua~D. {Simon}, M.~{Sivertz}, Howard~A.
  {Smith}, J.~{Allyn Smith}, Nathan {Smith}, Anna~H. {Spitz}, Adam {Stanford},
  Keivan~G. {Stassun}, Jay {Strader}, Michael~A. {Strauss}, Christopher~W.
  {Stubbs}, Donald~W. {Sweeney}, Alex {Szalay}, Paula {Szkody}, Masahiro
  {Takada}, Paul {Thorman}, David~E. {Trilling}, Virginia {Trimble}, Anthony
  {Tyson}, Richard {Van Berg}, Daniel {Vanden Berk}, Jake {VanderPlas}, Licia
  {Verde}, Bojan {Vrsnak}, Lucianne~M. {Walkowicz}, Benjamin~D. {Wandelt},
  Sheng {Wang}, Yun {Wang}, Michael {Warner}, Risa~H. {Wechsler}, Andrew~A.
  {West}, Oliver {Wiecha}, Benjamin~F. {Williams}, Beth {Willman}, David
  {Wittman}, Sidney~C. {Wolff}, W.~Michael {Wood-Vasey}, Przemek {Wozniak},
  Patrick {Young}, Andrew {Zentner}, and Hu~{Zhan}.
\newblock {LSST Science Book, Version 2.0}.
\newblock {\em arXiv e-prints}, page arXiv:0912.0201, December 2009.

\bibitem{Sherwin:2016tyf}
Blake~D. Sherwin et~al.
\newblock {Two-season Atacama Cosmology Telescope polarimeter lensing power
  spectrum}.
\newblock {\em Phys. Rev. D}, 95(12):123529, 2017.

\bibitem{SPT:2019fqo}
F.~Bianchini et~al.
\newblock {Constraints on Cosmological Parameters from the 500 deg$^2$ SPTpol
  Lensing Power Spectrum}.
\newblock {\em Astrophys. J.}, 888:119, 2020.

\bibitem{Euclid:2019clj}
A.~Blanchard et~al.
\newblock {Euclid preparation: VII. Forecast validation for Euclid cosmological
  probes}.
\newblock {\em Astron. Astrophys.}, 642:A191, 2020.

\bibitem{Font-Ribera:2013rwa}
Andreu Font-Ribera, Patrick McDonald, Nick Mostek, Beth~A. Reid, Hee-Jong Seo,
  and An~Slosar.
\newblock {DESI and other dark energy experiments in the era of neutrino mass
  measurements}.
\newblock {\em JCAP}, 05:023, 2014.

\bibitem{Xu:2021rwg}
Weishuang~Linda Xu, Julian~B. Mu\~noz, and Cora Dvorkin.
\newblock {Cosmological Constraints on Light (but Massive) Relics}.
\newblock 7 2021.

\bibitem{Ivanov:2019hqk}
Mikhail~M. Ivanov, Marko Simonovi\'c, and Matias Zaldarriaga.
\newblock {Cosmological Parameters and Neutrino Masses from the Final Planck
  and Full-Shape BOSS Data}.
\newblock {\em Phys. Rev. D}, 101(8):083504, 2020.

\bibitem{bird2012massive}
Simeon Bird, Matteo Viel, and Martin~G Haehnelt.
\newblock Massive neutrinos and the non-linear matter power spectrum.
\newblock {\em Monthly Notices of the Royal Astronomical Society},
  420(3):2551--2561, 2012.

\bibitem{villaescusa2014cosmology}
Francisco Villaescusa-Navarro, Federico Marulli, Matteo Viel, Enzo Branchini,
  Emanuele Castorina, Emiliano Sefusatti, and Shun Saito.
\newblock Cosmology with massive neutrinos i: towards a realistic modeling of
  the relation between matter, haloes and galaxies.
\newblock {\em Journal of Cosmology and Astroparticle Physics}, 2014(03):011,
  2014.

\bibitem{villaescusa2018imprint}
Francisco Villaescusa-Navarro, Arka Banerjee, Neal Dalal, Emanuele Castorina,
  Roman Scoccimarro, Raul Angulo, and David~N Spergel.
\newblock The imprint of neutrinos on clustering in redshift space.
\newblock {\em The Astrophysical Journal}, 861(1):53, 2018.

\bibitem{rossi2020sejong}
Graziano Rossi.
\newblock The sejong suite: Cosmological hydrodynamical simulations with
  massive neutrinos, dark radiation, and warm dark matter.
\newblock {\em The Astrophysical Journal Supplement Series}, 249(2):19, 2020.

\bibitem{adamek2017relativistic}
Julian Adamek, Ruth Durrer, and Martin Kunz.
\newblock Relativistic n-body simulations with massive neutrinos.
\newblock {\em Journal of Cosmology and Astroparticle Physics}, 2017(11):004,
  2017.

\bibitem{brandbyge2008effect}
Jacob Brandbyge, Steen Hannestad, Troels Haugb{\o}lle, and Bjarne Thomsen.
\newblock The effect of thermal neutrino motion on the non-linear cosmological
  matter power spectrum.
\newblock {\em Journal of Cosmology and Astroparticle Physics}, 2008(08):020,
  2008.

\bibitem{villaescusa2013non}
Francisco Villaescusa-Navarro, Simeon Bird, Carlos Pena-Garay, and Matteo Viel.
\newblock Non-linear evolution of the cosmic neutrino background.
\newblock {\em Journal of Cosmology and Astroparticle Physics}, 2013(03):019,
  2013.

\bibitem{castorina2015demnuni}
Emanuele Castorina, Carmelita Carbone, Julien Bel, Emiliano Sefusatti, and
  Klaus Dolag.
\newblock Demnuni: The clustering of large-scale structures in the presence of
  massive neutrinos.
\newblock {\em Journal of Cosmology and Astroparticle Physics}, 2015(07):043,
  2015.

\bibitem{Emberson:2016ecv}
J.~D. Emberson et~al.
\newblock {Cosmological neutrino simulations at extreme scale}.
\newblock {\em Res. Astron. Astrophys.}, 17(8):085, 2017.

\bibitem{baldi2014cosmic}
Marco Baldi, Francisco Villaescusa-Navarro, Matteo Viel, Ewald Puchwein, Volker
  Springel, and Lauro Moscardini.
\newblock Cosmic degeneracies--i. joint n-body simulations of modified gravity
  and massive neutrinos.
\newblock {\em Monthly Notices of the Royal Astronomical Society},
  440(1):75--88, 2014.

\bibitem{viel2010effect}
Matteo Viel, Martin~G Haehnelt, and Volker Springel.
\newblock The effect of neutrinos on the matter distribution as probed by the
  intergalactic medium.
\newblock {\em Journal of Cosmology and Astroparticle Physics}, 2010(06):015,
  2010.

\bibitem{LoVerde:2013lta}
Marilena LoVerde and Matias Zaldarriaga.
\newblock {Neutrino clustering around spherical dark matter halos}.
\newblock {\em Phys. Rev. D}, 89(6):063502, 2014.

\bibitem{chen2021cosmic}
Joe~Zhiyu Chen, Amol Upadhye, and Yvonne~YY Wong.
\newblock The cosmic neutrino background as a collection of fluids in
  large-scale structure simulations.
\newblock {\em Journal of Cosmology and Astroparticle Physics}, 2021(03):065,
  2021.

\bibitem{de2021neutrinos}
Caio~Bastos de~Senna~Nascimento and Marilena Loverde.
\newblock Neutrinos in n-body simulations.
\newblock {\em Physical Review D}, 104(4):043512, 2021.

\bibitem{banerjee2018reducing}
Arka Banerjee, Devon Powell, Tom Abel, and Francisco Villaescusa-Navarro.
\newblock Reducing noise in cosmological n-body simulations with neutrinos.
\newblock {\em Journal of Cosmology and Astroparticle Physics}, 2018(09):028,
  2018.

\bibitem{elbers2021optimal}
Willem Elbers, Carlos~S Frenk, Adrian Jenkins, Baojiu Li, and Silvia Pascoli.
\newblock An optimal non-linear method for simulating relic neutrinos.
\newblock {\em Monthly Notices of the Royal Astronomical Society},
  507(2):2614--2631, 2021.

\bibitem{Inman:2020oda}
Derek Inman and Hao-ran Yu.
\newblock {Simulating the Cosmic Neutrino Background using Collisionless
  Hydrodynamics}.
\newblock {\em Astrophys. J. Suppl.}, 250(1):21, 2020.

\bibitem{Yoshikawa:2020ehd}
Kohji Yoshikawa, Satoshi Tanaka, Naoki Yoshida, and Shun Saito.
\newblock {Cosmological Vlasov\textendash{}Poisson Simulations of Structure
  Formation with Relic Neutrinos: Nonlinear Clustering and the Neutrino Mass}.
\newblock {\em Astrophys. J.}, 904(2):159, 2020.

\bibitem{Natarajan:2014xba}
Aravind Natarajan, Andrew~R. Zentner, Nicholas Battaglia, and Hy~Trac.
\newblock {Systematic errors in the measurement of neutrino masses due to
  baryonic feedback processes: Prospects for stage IV lensing surveys}.
\newblock {\em Phys. Rev. D}, 90(6):063516, 2014.

\bibitem{brando2021relativistic}
Guilherme Brando, Kazuya Koyama, and David Wands.
\newblock Relativistic corrections to the growth of structure in modified
  gravity.
\newblock {\em Journal of Cosmology and Astroparticle Physics}, 2021(01):013,
  2021.

\bibitem{tram2019fully}
Thomas Tram, Jacob Brandbyge, Jeppe Dakin, and Steen Hannestad.
\newblock Fully relativistic treatment of light neutrinos in n-body
  simulations.
\newblock {\em Journal of Cosmology and Astroparticle Physics}, 2019(03):022,
  2019.

\bibitem{chiang2019first}
Chi-Ting Chiang, Marilena LoVerde, and Francisco Villaescusa-Navarro.
\newblock First detection of scale-dependent linear halo bias in n-body
  simulations with massive neutrinos.
\newblock {\em Physical Review Letters}, 122(4):041302, 2019.

\bibitem{brandbyge2009grid}
Jacob Brandbyge and Steen Hannestad.
\newblock Grid based linear neutrino perturbations in cosmological n-body
  simulations.
\newblock {\em Journal of Cosmology and Astroparticle Physics}, 2009(05):002,
  2009.

\bibitem{archidiacono2015efficient}
Maria Archidiacono and Steen Hannestad.
\newblock Efficient calculation of cosmological neutrino clustering with both
  linear and non-linear gravity.
\newblock {\em arXiv preprint arXiv:1510.02907}, 2015.

\bibitem{ali2013efficient}
Yacine Ali-Haimoud and Simeon Bird.
\newblock An efficient implementation of massive neutrinos in non-linear
  structure formation simulations.
\newblock {\em Monthly Notices of the Royal Astronomical Society},
  428(4):3375--3389, 2013.

\bibitem{heuschling2022minimal}
Pol Heuschling, Christian Partmann, and Christian Fidler.
\newblock A minimal model for massive neutrinos in newtonian n-body
  simulations.
\newblock {\em arXiv preprint arXiv:2201.13186}, 2022.

\bibitem{fidler2019new}
Christian Fidler, Alexander Kleinjohann, Thomas Tram, Cornelius Rampf, and
  Kazuya Koyama.
\newblock A new approach to cosmological structure formation with massive
  neutrinos.
\newblock {\em Journal of Cosmology and Astroparticle Physics}, 2019(01):025,
  2019.

\bibitem{partmann2020fast}
Christian Partmann, Christian Fidler, Cornelius Rampf, and Oliver Hahn.
\newblock Fast simulations of cosmic large-scale structure with massive
  neutrinos.
\newblock {\em Journal of Cosmology and Astroparticle Physics}, 2020(09):018,
  2020.

\bibitem{Chiang:2017vuk}
Chi-Ting Chiang, Wayne Hu, Yin Li, and Marilena Loverde.
\newblock {Scale-dependent bias and bispectrum in neutrino separate universe
  simulations}.
\newblock {\em Phys. Rev. D}, 97(12):123526, 2018.

\bibitem{bird2018efficient}
Simeon Bird, Yacine Ali-Ha{\"\i}moud, Yu~Feng, and Jia Liu.
\newblock An efficient and accurate hybrid method for simulating non-linear
  neutrino structure.
\newblock {\em Monthly Notices of the Royal Astronomical Society},
  481(2):1486--1500, 2018.

\bibitem{brandbyge2010resolving}
Jacob Brandbyge and Steen Hannestad.
\newblock Resolving cosmic neutrino structure: a hybrid neutrino n-body scheme.
\newblock {\em Journal of Cosmology and Astroparticle Physics}, 2010(01):021,
  2010.

\bibitem{Inman:2016qmg}
Derek Inman and Ue-Li Pen.
\newblock {Cosmic neutrinos: A dispersive and nonlinear fluid}.
\newblock {\em Phys. Rev. D}, 95(6):063535, 2017.

\bibitem{Shoji:2010hm}
Masatoshi Shoji and Eiichiro Komatsu.
\newblock {Massive Neutrinos in Cosmology: Analytic Solutions and Fluid
  Approximation}.
\newblock {\em Phys. Rev. D}, 81:123516, 2010.
\newblock [Erratum: Phys.Rev.D 82, 089901 (2010)].

\bibitem{Lesgourgues:2009am}
Julien Lesgourgues, Sabino Matarrese, Massimo Pietroni, and Antonio Riotto.
\newblock {Non-linear Power Spectrum including Massive Neutrinos: the Time-RG
  Flow Approach}.
\newblock {\em JCAP}, 06:017, 2009.

\bibitem{Dupuy:2013jaa}
Helene Dupuy and Francis Bernardeau.
\newblock {Describing massive neutrinos in cosmology as a collection of
  independent flows}.
\newblock {\em JCAP}, 01:030, 2014.

\bibitem{Fuhrer:2014zka}
Florian F\"uhrer and Yvonne Y.~Y. Wong.
\newblock {Higher-order massive neutrino perturbations in large-scale
  structure}.
\newblock {\em JCAP}, 03:046, 2015.

\bibitem{Blas:2014hya}
Diego Blas, Mathias Garny, Thomas Konstandin, and Julien Lesgourgues.
\newblock {Structure formation with massive neutrinos: going beyond linear
  theory}.
\newblock {\em JCAP}, 11:039, 2014.

\bibitem{Peloso:2015jua}
Marco Peloso, Massimo Pietroni, Matteo Viel, and Francisco Villaescusa-Navarro.
\newblock {The effect of massive neutrinos on the BAO peak}.
\newblock {\em JCAP}, 07:001, 2015.

\bibitem{Levi:2016tlf}
Michele Levi and Zvonimir Vlah.
\newblock {Massive neutrinos in nonlinear large scale structure: A consistent
  perturbation theory}.
\newblock 5 2016.

\bibitem{Senatore:2017hyk}
Leonardo Senatore and Matias Zaldarriaga.
\newblock {The Effective Field Theory of Large-Scale Structure in the presence
  of Massive Neutrinos}.
\newblock 7 2017.

\bibitem{Ichiki:2011ue}
Kiyotomo Ichiki and Masahiro Takada.
\newblock {The impact of massive neutrinos on the abundance of massive
  clusters}.
\newblock {\em Phys. Rev. D}, 85:063521, 2012.

\bibitem{LoVerde:2014rxa}
Marilena LoVerde.
\newblock {Spherical collapse in $\nu \Lambda$CDM}.
\newblock {\em Phys. Rev. D}, 90(8):083518, 2014.

\bibitem{LoVerde:2014pxa}
Marilena LoVerde.
\newblock {Halo bias in mixed dark matter cosmologies}.
\newblock {\em Phys. Rev. D}, 90(8):083530, 2014.

\bibitem{Kuijken:2019gsa}
K.~Kuijken et~al.
\newblock {The fourth data release of the Kilo-Degree Survey: ugri imaging and
  nine-band optical-IR photometry over 1000 square degrees}.
\newblock {\em Astron. Astrophys.}, 625:A2, 2019.

\bibitem{Aihara:2021jwb}
Hiroaki Aihara et~al.
\newblock {Third Data Release of the Hyper Suprime-Cam Subaru Strategic
  Program}.
\newblock 8 2021.

\bibitem{DES:2021wwk}
T.~M.~C. Abbott et~al.
\newblock {Dark Energy Survey Year 3 results: Cosmological constraints from
  galaxy clustering and weak lensing}.
\newblock {\em Phys. Rev. D}, 105(2):023520, 2022.

\bibitem{eBOSS:2020yzd}
Shadab Alam et~al.
\newblock {Completed SDSS-IV extended Baryon Oscillation Spectroscopic Survey:
  Cosmological implications from two decades of spectroscopic surveys at the
  Apache Point Observatory}.
\newblock {\em Phys. Rev. D}, 103(8):083533, 2021.

\bibitem{ACT:2020frw}
Steve~K. Choi et~al.
\newblock {The Atacama Cosmology Telescope: a measurement of the Cosmic
  Microwave Background power spectra at 98 and 150 GHz}.
\newblock {\em JCAP}, 12:045, 2020.

\bibitem{ACT:2020gnv}
Simone Aiola et~al.
\newblock {The Atacama Cosmology Telescope: DR4 Maps and Cosmological
  Parameters}.
\newblock {\em JCAP}, 12:047, 2020.

\bibitem{tec14}
Masahiro {Takada}, Richard~S. {Ellis}, Masashi {Chiba}, Jenny~E. {Greene},
  Hiroaki {Aihara}, Nobuo {Arimoto}, Kevin {Bundy}, Judith {Cohen}, Olivier
  {Dor{\'e}}, Genevieve {Graves}, James~E. {Gunn}, Timothy {Heckman},
  Christopher~M. {Hirata}, Paul {Ho}, Jean-Paul {Kneib}, Olivier {Le
  F{\`e}vre}, Lihwai {Lin}, Surhud {More}, Hitoshi {Murayama}, Tohru {Nagao},
  Masami {Ouchi}, Michael {Seiffert}, John~D. {Silverman}, Laerte {Sodr{\'e}},
  David~N. {Spergel}, Michael~A. {Strauss}, Hajime {Sugai}, Yasushi {Suto},
  Hideki {Takami}, and Rosemary {Wyse}.
\newblock {Extragalactic science, cosmology, and Galactic archaeology with the
  Subaru Prime Focus Spectrograph}.
\newblock {\em \pasj}, 66(1):R1, Feb 2014.

\bibitem{LSST19}
{\v{Z}}eljko {Ivezi{\'c}}, Steven~M. {Kahn}, J.~Anthony {Tyson}, Bob {Abel},
  Emily {Acosta}, Robyn {Allsman}, David {Alonso}, Yusra {AlSayyad}, Scott~F.
  {Anderson}, John {Andrew}, James Roger~P. {Angel}, George~Z. {Angeli}, Reza
  {Ansari}, Pierre {Antilogus}, Constanza {Araujo}, Robert {Armstrong}, Kirk~T.
  {Arndt}, Pierre {Astier}, {\'E}ric {Aubourg}, Nicole {Auza}, Tim~S.
  {Axelrod}, Deborah~J. {Bard}, Jeff~D. {Barr}, Aurelian {Barrau}, James~G.
  {Bartlett}, Amanda~E. {Bauer}, Brian~J. {Bauman}, Sylvain {Baumont}, Ellen
  {Bechtol}, Keith {Bechtol}, Andrew~C. {Becker}, Jacek {Becla}, Cristina
  {Beldica}, Steve {Bellavia}, Federica~B. {Bianco}, Rahul {Biswas}, Guillaume
  {Blanc}, Jonathan {Blazek}, Roger~D. {Bland ford}, Josh~S. {Bloom}, Joanne
  {Bogart}, Tim~W. {Bond}, Michael~T. {Booth}, Anders~W. {Borgland}, Kirk
  {Borne}, James~F. {Bosch}, Dominique {Boutigny}, Craig~A. {Brackett}, Andrew
  {Bradshaw}, William~Nielsen {Brand t}, Michael~E. {Brown}, James~S.
  {Bullock}, Patricia {Burchat}, David~L. {Burke}, Gianpietro {Cagnoli}, Daniel
  {Calabrese}, Shawn {Callahan}, Alice~L. {Callen}, Jeffrey~L. {Carlin},
  Erin~L. {Carlson}, Srinivasan {Chand rasekharan}, Glenaver {Charles-Emerson},
  Steve {Chesley}, Elliott~C. {Cheu}, Hsin-Fang {Chiang}, James {Chiang}, Carol
  {Chirino}, Derek {Chow}, David~R. {Ciardi}, Charles~F. {Claver}, Johann
  {Cohen-Tanugi}, Joseph~J. {Cockrum}, Rebecca {Coles}, Andrew~J. {Connolly},
  Kem~H. {Cook}, Asantha {Cooray}, Kevin~R. {Covey}, Chris {Cribbs}, Wei {Cui},
  Roc {Cutri}, Philip~N. {Daly}, Scott~F. {Daniel}, Felipe {Daruich}, Guillaume
  {Daubard}, Greg {Daues}, William {Dawson}, Francisco {Delgado}, Alfred
  {Dellapenna}, Robert {de Peyster}, Miguel {de Val-Borro}, Seth~W. {Digel},
  Peter {Doherty}, Richard {Dubois}, Gregory~P. {Dubois-Felsmann}, Josef
  {Durech}, Frossie {Economou}, Tim {Eifler}, Michael {Eracleous}, Benjamin~L.
  {Emmons}, Angelo {Fausti Neto}, Henry {Ferguson}, Enrique {Figueroa}, Merlin
  {Fisher-Levine}, Warren {Focke}, Michael~D. {Foss}, James {Frank}, Michael~D.
  {Freemon}, Emmanuel {Gangler}, Eric {Gawiser}, John~C. {Geary}, Perry {Gee},
  Marla {Geha}, Charles J.~B. {Gessner}, Robert~R. {Gibson}, D.~Kirk {Gilmore},
  Thomas {Glanzman}, William {Glick}, Tatiana {Goldina}, Daniel~A. {Goldstein},
  Iain {Goodenow}, Melissa~L. {Graham}, William~J. {Gressler}, Philippe {Gris},
  Leanne~P. {Guy}, Augustin {Guyonnet}, Gunther {Haller}, Ron {Harris},
  Patrick~A. {Hascall}, Justine {Haupt}, Fabio {Hernand ez}, Sven {Herrmann},
  Edward {Hileman}, Joshua {Hoblitt}, John~A. {Hodgson}, Craig {Hogan},
  James~D. {Howard}, Dajun {Huang}, Michael~E. {Huffer}, Patrick {Ingraham},
  Walter~R. {Innes}, Suzanne~H. {Jacoby}, Bhuvnesh {Jain}, Fabrice {Jammes},
  M.~James {Jee}, Tim {Jenness}, Garrett {Jernigan}, Darko {Jevremovi{\'c}},
  Kenneth {Johns}, Anthony~S. {Johnson}, Margaret W.~G. {Johnson}, R.~Lynne
  {Jones}, Claire {Juramy-Gilles}, Mario {Juri{\'c}}, Jason~S. {Kalirai},
  Nitya~J. {Kallivayalil}, Bryce {Kalmbach}, Jeffrey~P. {Kantor}, Pierre
  {Karst}, Mansi~M. {Kasliwal}, Heather {Kelly}, Richard {Kessler}, Veronica
  {Kinnison}, David {Kirkby}, Lloyd {Knox}, Ivan~V. {Kotov}, Victor~L.
  {Krabbendam}, K.~Simon {Krughoff}, Petr {Kub{\'a}nek}, John {Kuczewski}, Shri
  {Kulkarni}, John {Ku}, Nadine~R. {Kurita}, Craig~S. {Lage}, Ron {Lambert},
  Travis {Lange}, J.~Brian {Langton}, Laurent {Le Guillou}, Deborah {Levine},
  Ming {Liang}, Kian-Tat {Lim}, Chris~J. {Lintott}, Kevin~E. {Long}, Margaux
  {Lopez}, Paul~J. {Lotz}, Robert~H. {Lupton}, Nate~B. {Lust}, Lauren~A.
  {MacArthur}, Ashish {Mahabal}, Rachel {Mand elbaum}, Thomas~W. {Markiewicz},
  Darren~S. {Marsh}, Philip~J. {Marshall}, Stuart {Marshall}, Morgan {May},
  Robert {McKercher}, Michelle {McQueen}, Joshua {Meyers}, Myriam {Migliore},
  Michelle {Miller}, David~J. {Mills}, Connor {Miraval}, Joachim {Moeyens},
  Fred~E. {Moolekamp}, David~G. {Monet}, Marc {Moniez}, Serge {Monkewitz},
  Christopher {Montgomery}, Christopher~B. {Morrison}, Fritz {Mueller}, Gary~P.
  {Muller}, Freddy {Mu{\~n}oz Arancibia}, Douglas~R. {Neill}, Scott~P.
  {Newbry}, Jean-Yves {Nief}, Andrei {Nomerotski}, Martin {Nordby}, Paul
  {O'Connor}, John {Oliver}, Scot~S. {Olivier}, Knut {Olsen}, William
  {O'Mullane}, Sandra {Ortiz}, Shawn {Osier}, Russell~E. {Owen}, Reynald
  {Pain}, Paul~E. {Palecek}, John~K. {Parejko}, James~B. {Parsons}, Nathan~M.
  {Pease}, J.~Matt {Peterson}, John~R. {Peterson}, Donald~L. {Petravick}, M.~E.
  {Libby Petrick}, Cathy~E. {Petry}, Francesco {Pierfederici}, Stephen
  {Pietrowicz}, Rob {Pike}, Philip~A. {Pinto}, Raymond {Plante}, Stephen
  {Plate}, Joel~P. {Plutchak}, Paul~A. {Price}, Michael {Prouza}, Veljko
  {Radeka}, Jayadev {Rajagopal}, Andrew~P. {Rasmussen}, Nicolas {Regnault},
  Kevin~A. {Reil}, David~J. {Reiss}, Michael~A. {Reuter}, Stephen~T. {Ridgway},
  Vincent~J. {Riot}, Steve {Ritz}, Sean {Robinson}, William {Roby}, Aaron
  {Roodman}, Wayne {Rosing}, Cecille {Roucelle}, Matthew~R. {Rumore}, Stefano
  {Russo}, Abhijit {Saha}, Benoit {Sassolas}, Terry~L. {Schalk}, Pim
  {Schellart}, Rafe~H. {Schindler}, Samuel {Schmidt}, Donald~P. {Schneider},
  Michael~D. {Schneider}, William {Schoening}, German {Schumacher}, Megan~E.
  {Schwamb}, Jacques {Sebag}, Brian {Selvy}, Glenn~H. {Sembroski}, Lynn~G.
  {Seppala}, Andrew {Serio}, Eduardo {Serrano}, Richard~A. {Shaw}, Ian
  {Shipsey}, Jonathan {Sick}, Nicole {Silvestri}, Colin~T. {Slater}, J.~Allyn
  {Smith}, R.~Chris {Smith}, Shahram {Sobhani}, Christine {Soldahl}, Lisa
  {Storrie-Lombardi}, Edward {Stover}, Michael~A. {Strauss}, Rachel~A.
  {Street}, Christopher~W. {Stubbs}, Ian~S. {Sullivan}, Donald {Sweeney},
  John~D. {Swinbank}, Alexander {Szalay}, Peter {Takacs}, Stephen~A. {Tether},
  Jon~J. {Thaler}, John~Gregg {Thayer}, Sandrine {Thomas}, Adam~J. {Thornton},
  Vaikunth {Thukral}, Jeffrey {Tice}, David~E. {Trilling}, Max {Turri}, Richard
  {Van Berg}, Daniel {Vanden Berk}, Kurt {Vetter}, Francoise {Virieux},
  Tomislav {Vucina}, William {Wahl}, Lucianne {Walkowicz}, Brian {Walsh},
  Christopher~W. {Walter}, Daniel~L. {Wang}, Shin-Yawn {Wang}, Michael
  {Warner}, Oliver {Wiecha}, Beth {Willman}, Scott~E. {Winters}, David
  {Wittman}, Sidney~C. {Wolff}, W.~Michael {Wood-Vasey}, Xiuqin {Wu}, Bo~{Xin},
  Peter {Yoachim}, and Hu~{Zhan}.
\newblock {LSST: From Science Drivers to Reference Design and Anticipated Data
  Products}.
\newblock {\em \apj}, 873(2):111, Mar 2019.

\bibitem{sgb15}
D.~{Spergel}, N.~{Gehrels}, C.~{Baltay}, D.~{Bennett}, J.~{Breckinridge},
  M.~{Donahue}, A.~{Dressler}, B.~S. {Gaudi}, T.~{Greene}, O.~{Guyon},
  C.~{Hirata}, J.~{Kalirai}, N.~J. {Kasdin}, B.~{Macintosh}, W.~{Moos},
  S.~{Perlmutter}, M.~{Postman}, B.~{Rauscher}, J.~{Rhodes}, Y.~{Wang},
  D.~{Weinberg}, D.~{Benford}, M.~{Hudson}, W.~S. {Jeong}, Y.~{Mellier},
  W.~{Traub}, T.~{Yamada}, P.~{Capak}, J.~{Colbert}, D.~{Masters}, M.~{Penny},
  D.~{Savransky}, D.~{Stern}, N.~{Zimmerman}, R.~{Barry}, L.~{Bartusek},
  K.~{Carpenter}, E.~{Cheng}, D.~{Content}, F.~{Dekens}, R.~{Demers},
  K.~{Grady}, C.~{Jackson}, G.~{Kuan}, J.~{Kruk}, M.~{Melton}, B.~{Nemati},
  B.~{Parvin}, I.~{Poberezhskiy}, C.~{Peddie}, J.~{Ruffa}, J.~K. {Wallace},
  A.~{Whipple}, E.~{Wollack}, and F.~{Zhao}.
\newblock {Wide-Field InfrarRed Survey Telescope-Astrophysics Focused Telescope
  Assets WFIRST-AFTA 2015 Report}.
\newblock {\em arXiv e-prints}, page arXiv:1503.03757, Mar 2015.

\bibitem{Alonso:2021nzr}
David Alonso et~al.
\newblock {Combining information from multiple cosmological surveys: inference
  and modeling challenges}.
\newblock 3 2021.

\bibitem{Battaglia:2020cch}
Nick Battaglia et~al.
\newblock {Report from the Tri-Agency Cosmological Simulation Task Force}.
\newblock 5 2020.

\bibitem{Banerjee:2022ljx}
Arka Banerjee, Simon Birrer, Salman Habib, Katrin Heitmann, Zarija Lukic,
  Julian~B. Munoz, Yuuki Omori, Hyunbae Park, Annika H.~G. Peter, and Yi-Ming
  Zhong.
\newblock {Snowmass2021 Computational Frontier White Paper: Cosmological
  Simulations and Modeling}.
\newblock 3 2022.

\bibitem{Riess:2021jrx}
Adam~G. Riess et~al.
\newblock {A Comprehensive Measurement of the Local Value of the Hubble
  Constant with 1 km/s/Mpc Uncertainty from the Hubble Space Telescope and the
  SH0ES Team}.
\newblock 12 2021.

\bibitem{SPT-3G:2021wgf}
L.~Balkenhol et~al.
\newblock {Constraints on \ensuremath{\Lambda}CDM extensions from the SPT-3G
  2018 EE and TE power spectra}.
\newblock {\em Phys. Rev. D}, 104(8):083509, 2021.

\bibitem{aiola:2020}
Simone Aiola, Erminia Calabrese, Lo{\"\i}c Maurin, Sigurd Naess, Benjamin~L
  Schmitt, Maximilian~H Abitbol, Graeme~E Addison, Peter~AR Ade, David Alonso,
  Mandana Amiri, et~al.
\newblock The atacama cosmology telescope: Dr4 maps and cosmological
  parameters.
\newblock {\em Journal of Cosmology and Astroparticle Physics}, 2020(12):047,
  2020.

\bibitem{Aylor:2018drw}
Kevin {Aylor}, Mackenzie {Joy}, Lloyd {Knox}, Marius {Millea}, Srinivasan
  {Raghunathan}, and W.~L. {Kimmy Wu}.
\newblock {Sounds Discordant: Classical Distance Ladder and
  {\ensuremath{\Lambda}}CDM-based Determinations of the Cosmological Sound
  Horizon}.
\newblock {\em \apj}, 874:4, Mar 2019.

\bibitem{Bernal:2016gxb}
Jose~Luis Bernal, Licia Verde, and Adam~G. Riess.
\newblock {The trouble with $H_0$}.
\newblock {\em JCAP}, 1610(10):019, 2016.

\bibitem{Lemos:2019}
Pablo {Lemos}, Elizabeth {Lee}, George {Efstathiou}, and Steven {Gratton}.
\newblock {Model independent H(z) reconstruction using the cosmic inverse
  distance ladder}.
\newblock {\em \mnras}, 483(4):4803--4810, March 2019.

\bibitem{Beaton:2016nsw}
Rachael~L. Beaton et~al.
\newblock {The Carnegie-Chicago Hubble Program. I. An Independent Approach to
  the Extragalactic Distance Scale Using only Population II Distance
  Indicators}.
\newblock {\em Astrophys. J.}, 832(2):210, 2016.

\bibitem{Hatt:2017rxl}
Dylan Hatt et~al.
\newblock {The Carnegie-Chicago Hubble Program. II. The Distance to IC 1613:
  The Tip of the Red Giant Branch and RR Lyrae Period\textendash{}luminosity
  Relations}.
\newblock {\em Astrophys. J.}, 845(2):146, 2017.

\bibitem{Hatt:2018opj}
Dylan Hatt et~al.
\newblock {The Carnegie-Chicago Hubble Program. IV. The Distance to NGC 4424,
  NGC 4526, and NGC 4356 via the Tip of the Red Giant Branch}.
\newblock {\em Astrophys. J.}, 861(2):104, 2018.

\bibitem{Hatt:2018zfv}
Dylan Hatt et~al.
\newblock {The Carnegie\textendash{}Chicago Hubble Program. V. The Distances to
  NGC 1448 and NGC 1316 via the Tip of the Red Giant Branch}.
\newblock {\em Astrophys. J.}, 866(2):145, 2018.

\bibitem{CSP:2018rag}
Christopher~R. Burns et~al.
\newblock {The Carnegie Supernova Project: Absolute Calibration and the Hubble
  Constant}.
\newblock {\em Astrophys. J.}, 869(1):56, 2018.

\bibitem{Hoyt_2019}
Taylor~J. Hoyt, Wendy~L. Freedman, Barry~F. Madore, Dylan Hatt, Rachael~L.
  Beaton, In~Sung Jang, Myung~Gyoon Lee, Andrew~J. Monson, Jillian~R. Neeley,
  Jeffrey~A. Rich, and et~al.
\newblock The carnegie chicago hubble program. vi. tip of the red giant branch
  distances to m66 and m96 of the leo i group.
\newblock {\em The Astrophysical Journal}, 882(2):150, Sep 2019.

\bibitem{Beaton_2019}
Rachael~L. Beaton, Mark Seibert, Dylan Hatt, Wendy~L. Freedman, Taylor~J. Hoyt,
  In~Sung Jang, Myung~Gyoon Lee, Barry~F. Madore, Andrew~J. Monson, Jillian~R.
  Neeley, and et~al.
\newblock The carnegie-chicago hubble program. vii. the distance to m101 via
  the optical tip of the red giant branch method.
\newblock {\em The Astrophysical Journal}, 885(2):141, Nov 2019.

\bibitem{freedman2019}
Wendy~L. {Freedman}, Barry~F. {Madore}, Dylan {Hatt}, Taylor~J. {Hoyt}, In~Sung
  {Jang}, Rachael~L. {Beaton}, Christopher~R. {Burns}, Myung~Gyoon {Lee},
  Andrew~J. {Monson}, Jillian~R. {Neeley}, M.~M. {Phillips}, Jeffrey~A. {Rich},
  and Mark {Seibert}.
\newblock {The Carnegie-Chicago Hubble Program. VIII. An Independent
  Determination of the Hubble Constant Based on the Tip of the Red Giant
  Branch}.
\newblock {\em \apj}, 882(1):34, September 2019.

\bibitem{Jang_2021}
In~Sung Jang, Taylor~J. Hoyt, Rachael~L. Beaton, Wendy~L. Freedman, Barry~F.
  Madore, Myung~Gyoon Lee, Jillian~R. Neeley, Andrew~J. Monson, Jeffrey~A.
  Rich, and Mark Seibert.
\newblock The carnegie–chicago hubble program. ix. calibration of the tip of
  the red giant branch method in the megamaser host galaxy, ngc 4258 (m106).
\newblock {\em The Astrophysical Journal}, 906(2):125, Jan 2021.

\bibitem{Hoyt_2021}
Taylor~J. Hoyt, Rachael~L. Beaton, Wendy~L. Freedman, In~Sung Jang, Myung~Gyoon
  Lee, Barry~F. Madore, Andrew~J. Monson, Jillian~R. Neeley, Jeffrey~A. Rich,
  and Mark Seibert.
\newblock The carnegie chicago hubble program x: Tip of the red giant branch
  distances to ngc 5643 and ngc 1404.
\newblock {\em The Astrophysical Journal}, 915(1):34, Jul 2021.

\bibitem{Riess:2011}
A.~G. {Riess}, L.~{Macri}, S.~{Casertano}, H.~{Lampeitl}, H.~C. {Ferguson},
  A.~V. {Filippenko}, S.~W. {Jha}, W.~{Li}, and R.~{Chornock}.
\newblock {A 3\% Solution: Determination of the Hubble Constant with the Hubble
  Space Telescope and Wide Field Camera 3}.
\newblock {\em \apj}, 730:119, April 2011.

\bibitem{Riess:2016jrr}
Adam~G. Riess et~al.
\newblock {A 2.4\% Determination of the Local Value of the Hubble Constant}.
\newblock {\em Astrophys. J.}, 826(1):56, 2016.

\bibitem{Riess2019}
Adam~G. Riess, Stefano Casertano, Wenlong Yuan, Lucas~M. Macri, and Dan
  Scolnic.
\newblock {Large Magellanic Cloud Cepheid Standards Provide a 1\% Foundation
  for the Determination of the Hubble Constant and Stronger Evidence for
  Physics beyond $\Lambda$CDM}.
\newblock {\em Astrophys. J.}, 876(1):85, 2019.

\bibitem{riess2021}
Adam~G. {Riess}, Stefano {Casertano}, Wenlong {Yuan}, J.~Bradley {Bowers},
  Lucas {Macri}, Joel~C. {Zinn}, and Dan {Scolnic}.
\newblock {Cosmic Distances Calibrated to 1\% Precision with Gaia EDR3
  Parallaxes and Hubble Space Telescope Photometry of 75 Milky Way Cepheids
  Confirm Tension with {\ensuremath{\Lambda}}CDM}.
\newblock {\em \apjl}, 908(1):L6, February 2021.

\bibitem{freedman2012}
Wendy~L. {Freedman}, Barry~F. {Madore}, Victoria {Scowcroft}, Chris {Burns},
  Andy {Monson}, S.~Eric {Persson}, Mark {Seibert}, and Jane {Rigby}.
\newblock {Carnegie Hubble Program: A Mid-infrared Calibration of the Hubble
  Constant}.
\newblock {\em \apj}, 758(1):24, October 2012.

\bibitem{Suyu:2016qxx}
S.~H. Suyu et~al.
\newblock {H0LiCOW \textendash{} I. H0 Lenses in COSMOGRAIL's Wellspring:
  program overview}.
\newblock {\em Mon. Not. Roy. Astron. Soc.}, 468(3):2590--2604, 2017.

\bibitem{Birrer:2018vtm}
S.~{Birrer}, T.~{Treu}, C.~E. {Rusu}, V.~{Bonvin}, C.~D. {Fassnacht}, J.~H.~H.
  {Chan}, A.~{Agnello}, A.~J. {Shajib}, G.~C.~F. {Chen}, M.~{Auger},
  F.~{Courbin}, S.~{Hilbert}, D.~{Sluse}, S.~H. {Suyu}, K.~C. {Wong},
  P.~{Marshall}, B.~C. {Lemaux}, and G.~{Meylan}.
\newblock {H0LiCOW - IX. Cosmographic analysis of the doubly imaged quasar SDSS
  1206+4332 and a new measurement of the Hubble constant}.
\newblock {\em \mnras}, 484:4726--4753, Apr 2019.

\bibitem{Wong:2019kwg}
Kenneth~C. Wong et~al.
\newblock {H0LiCOW \textendash{} XIII. A 2.4 per cent measurement of H0 from
  lensed quasars: 5.3\ensuremath{\sigma} tension between early- and
  late-Universe probes}.
\newblock {\em Mon. Not. Roy. Astron. Soc.}, 498(1):1420--1439, 2020.

\bibitem{Huang:2019yhh}
Caroline~D. Huang, Adam~G. Riess, Wenlong Yuan, Lucas~M. Macri, Nadia~L.
  Zakamska, Stefano Casertano, Patricia~A. Whitelock, Samantha~L. Hoffmann,
  Alexei~V. Filippenko, and Daniel Scolnic.
\newblock {Hubble Space Telescope Observations of Mira Variables in the Type Ia
  Supernova Host NGC 1559: An Alternative Candle to Measure the Hubble
  Constant}.
\newblock 8 2019.

\bibitem{Kourkchi:2020iyz}
Ehsan Kourkchi, R.~Brent Tully, Gagandeep~S. Anand, Helene~M. Courtois,
  Alexandra Dupuy, James~D. Neill, Luca Rizzi, and Mark Seibert.
\newblock {Cosmicflows-4: The Calibration of Optical and Infrared
  Tully\textendash{}Fisher Relations}.
\newblock {\em Astrophys. J.}, 896(1):3, 2020.

\bibitem{Reid:2019tiq}
M.~J. Reid, D.~W. Pesce, and A.~G. Riess.
\newblock {An Improved Distance to NGC 4258 and its Implications for the Hubble
  Constant}.
\newblock {\em Astrophys. J. Lett.}, 886(2):L27, 2019.

\bibitem{Freedman:2020dne}
Wendy~L. Freedman, Barry~F. Madore, Taylor Hoyt, In~Sung Jang, Rachael Beaton,
  Myung~Gyoon Lee, Andrew Monson, Jill Neeley, and Jeffrey Rich.
\newblock {Calibration of the Tip of the Red Giant Branch (TRGB)}.
\newblock 2 2020.

\bibitem{Freedman:2021ahq}
Wendy~L. Freedman.
\newblock {Measurements of the Hubble Constant: Tensions in Perspective}.
\newblock 6 2021.

\bibitem{Pesce:2020xfe}
D.~W. Pesce et~al.
\newblock {The Megamaser Cosmology Project. XIII. Combined Hubble constant
  constraints}.
\newblock {\em Astrophys. J. Lett.}, 891(1):L1, 2020.

\bibitem{Khetan:2020hmh}
Nandita Khetan et~al.
\newblock {A new measurement of the Hubble constant using Type Ia supernovae
  calibrated with surface brightness fluctuations}.
\newblock {\em Astron. Astrophys.}, 647:A72, 2021.

\bibitem{Blakeslee:2021rqi}
John~P. Blakeslee, Joseph~B. Jensen, Chung-Pei Ma, Peter~A. Milne, and Jenny~E.
  Greene.
\newblock {The Hubble Constant from Infrared Surface Brightness Fluctuation
  Distances}.
\newblock {\em Astrophys. J.}, 911(1):65, 2021.

\bibitem{DiValentino:2020vnx}
Eleonora Di~Valentino.
\newblock {A combined analysis of the $H_0$ late time direct measurements and
  the impact on the Dark Energy sector}.
\newblock {\em Mon. Not. Roy. Astron. Soc.}, 502(2):2065--2073, 2021.

\bibitem{Moresco:2022phi}
Michele Moresco et~al.
\newblock {Unveiling the Universe with Emerging Cosmological Probes}.
\newblock 1 2022.

\bibitem{Efstathiou:2020wxn}
G.~Efstathiou.
\newblock {A Lockdown Perspective on the Hubble Tension (with comments from the
  SH0ES team)}.
\newblock 7 2020.

\bibitem{Camarena:2021jlr}
David Camarena and Valerio Marra.
\newblock {On the use of the local prior on the absolute magnitude of Type Ia
  supernovae in cosmological inference}.
\newblock {\em Mon. Not. Roy. Astron. Soc.}, 504:5164--5171, 2021.

\bibitem{Efstathiou:2021ocp}
George Efstathiou.
\newblock {To H0 or not to H0?}
\newblock 3 2021.

\bibitem{Benevento:2020fev}
Giampaolo Benevento, Wayne Hu, and Marco Raveri.
\newblock {Can Late Dark Energy Transitions Raise the Hubble constant?}
\newblock {\em Phys. Rev. D}, 101(10):103517, 2020.

\bibitem{Greene:2021shv}
Kylar~L. Greene and Francis-Yan Cyr-Racine.
\newblock {Hubble distancing: Focusing on distance measurements in cosmology}.
\newblock 12 2021.

\bibitem{Scolnic:2017caz}
D.~M. Scolnic et~al.
\newblock {The Complete Light-curve Sample of Spectroscopically Confirmed SNe
  Ia from Pan-STARRS1 and Cosmological Constraints from the Combined Pantheon
  Sample}.
\newblock {\em Astrophys. J.}, 859(2):101, 2018.

\bibitem{Brout:2022vxf}
Dillon Brout et~al.
\newblock {The Pantheon+ Analysis: Cosmological Constraints}.
\newblock 2 2022.

\bibitem{Poulin:2018cxd}
Vivian Poulin, Tristan~L. Smith, Tanvi Karwal, and Marc Kamionkowski.
\newblock {Early Dark Energy Can Resolve The Hubble Tension}.
\newblock {\em Phys. Rev. Lett.}, 122(22):221301, 2019.

\bibitem{Agrawal:2019lmo}
Prateek Agrawal, Francis-Yan Cyr-Racine, David Pinner, and Lisa Randall.
\newblock {Rock 'n' Roll Solutions to the Hubble Tension}.
\newblock 4 2019.

\bibitem{Niedermann:2020dwg}
Florian Niedermann and Martin~S. Sloth.
\newblock {Resolving the Hubble tension with new early dark energy}.
\newblock {\em Phys. Rev. D}, 102(6):063527, 2020.

\bibitem{Hill:2021yec}
J.~Colin Hill et~al.
\newblock {The Atacama Cosmology Telescope: Constraints on Pre-Recombination
  Early Dark Energy}.
\newblock 9 2021.

\bibitem{Smith:2022hwi}
Tristan~L. Smith, Matteo Lucca, Vivian Poulin, Guillermo~F. Abellan, Lennart
  Balkenhol, Karim Benabed, Silvia Galli, and Riccardo Murgia.
\newblock {Hints of Early Dark Energy in Planck, SPT, and ACT data: new physics
  or systematics?}
\newblock 2 2022.

\bibitem{Jedamzik:2020abc}
Karsten {Jedamzik} and Levon {Pogosian}.
\newblock {Relieving the Hubble Tension with Primordial Magnetic Fields}.
\newblock {\em \prl}, 125(18):181302, October 2020.

\bibitem{Pogosian:2020ded}
Levon Pogosian, Gong-Bo Zhao, and Karsten Jedamzik.
\newblock {Recombination-independent determination of the sound horizon and the
  Hubble constant from BAO}.
\newblock {\em Astrophys. J. Lett.}, 904(2):L17, 2020.

\bibitem{Rashkovetskyi:2021rwg}
Michael Rashkovetskyi, Julian~B. Mu\~noz, Daniel~J. Eisenstein, and Cora
  Dvorkin.
\newblock {Small-scale clumping at recombination and the Hubble tension}.
\newblock {\em Phys. Rev. D}, 104(10):103517, 2021.

\bibitem{Cyr-Racine:2021alc}
Francis-Yan Cyr-Racine, Fei Ge, and Lloyd Knox.
\newblock {A Symmetry of Cosmological Observables, and a High Hubble Constant
  as an Indicator of a Mirror World Dark Sector}.
\newblock 7 2021.

\bibitem{Hart:2020}
Luke {Hart} and Jens {Chluba}.
\newblock {Updated fundamental constant constraints from Planck 2018 data and
  possible relations to the Hubble tension}.
\newblock {\em \mnras}, 493(3):3255--3263, April 2020.

\bibitem{Hart:2021kad}
Luke Hart and Jens Chluba.
\newblock {Varying fundamental constants principal component analysis:
  additional hints about the Hubble tension}.
\newblock {\em Mon. Not. Roy. Astron. Soc.}, 510(2):2206--2227, 2022.

\bibitem{Sekiguichi:2021}
Toyokazu {Sekiguchi} and Tomo {Takahashi}.
\newblock {Early recombination as a solution to the H$_{0}$ tension}.
\newblock {\em \prd}, 103(8):083507, April 2021.

\bibitem{Schoneberg:2021qvd}
Nils Sch\"oneberg, Guillermo~Franco Abell\'an, Andrea~P\'erez S\'anchez,
  Samuel~J. Witte, c.~Vivian Poulin, and Julien Lesgourgues.
\newblock {The $H_0$ Olympics: A fair ranking of proposed models}.
\newblock 7 2021.

\bibitem{Burgess:2021obw}
C.~P. Burgess, Danielle Dineen, and F.~Quevedo.
\newblock {Yoga Dark Energy: Natural Relaxation and Other Dark Implications of
  a Supersymmetric Gravity Sector}.
\newblock 11 2021.

\bibitem{Burgess:2021qti}
C.~P. Burgess and F.~Quevedo.
\newblock {Axion Homeopathy: Screening Dilaton Interactions}.
\newblock 10 2021.

\bibitem{Fields:2019pfx}
Brian~D. Fields, Keith~A. Olive, Tsung-Han Yeh, and Charles Young.
\newblock {Big-Bang Nucleosynthesis after Planck}.
\newblock {\em JCAP}, 03:010, 2020.
\newblock [Erratum: JCAP 11, E02 (2020)].

\bibitem{1999ApJ...522...82K}
Anatoly {Klypin}, Andrey~V. {Kravtsov}, Octavio {Valenzuela}, and Francisco
  {Prada}.
\newblock {Where Are the Missing Galactic Satellites?}
\newblock {\em \apj}, 522(1):82--92, September 1999.

\bibitem{1999ApJ...524L..19M}
Ben {Moore}, Sebastiano {Ghigna}, Fabio {Governato}, George {Lake}, Thomas
  {Quinn}, Joachim {Stadel}, and Paolo {Tozzi}.
\newblock {Dark Matter Substructure within Galactic Halos}.
\newblock {\em \apjl}, 524(1):L19--L22, October 1999.

\bibitem{2018PhRvL.121u1302K}
Stacy~Y. {Kim}, Annika H.~G. {Peter}, and Jonathan~R. {Hargis}.
\newblock {Missing Satellites Problem: Completeness Corrections to the Number
  of Satellite Galaxies in the Milky Way are Consistent with Cold Dark Matter
  Predictions}.
\newblock {\em \prl}, 121(21):211302, November 2018.

\bibitem{Nadler191203303}
E.~O. {Nadler}, R.~H. {Wechsler}, K.~{Bechtol}, Y.~Y. {Mao}, G.~{Green},
  A.~{Drlica-Wagner}, M.~{McNanna}, S.~{Mau}, A.~B. {Pace}, J.~D. {Simon},
  A.~{Kravtsov}, S.~{Dodelson}, T.~S. {Li}, A.~H. {Riley}, M.~Y. {Wang},
  T.~M.~C. {Abbott}, M.~{Aguena}, S.~{Allam}, J.~{Annis}, S.~{Avila}, G.~M.
  {Bernstein}, E.~{Bertin}, D.~{Brooks}, D.~L. {Burke}, A.~Carnero {Rosell},
  M.~Carrasco {Kind}, J.~{Carretero}, M.~{Costanzi}, L.~N. {da Costa}, J.~{De
  Vicente}, S.~{Desai}, A.~E. {Evrard}, B.~{Flaugher}, P.~{Fosalba},
  J.~{Frieman}, J.~{Garc{\'\i}a-Bellido}, E.~{Gaztanaga}, D.~W. {Gerdes},
  D.~{Gruen}, J.~{Gschwend}, G.~{Gutierrez}, W.~G. {Hartley}, S.~R. {Hinton},
  K.~{Honscheid}, E.~{Krause}, K.~{Kuehn}, N.~{Kuropatkin}, O.~{Lahav},
  M.~A.~G. {Maia}, J.~L. {Marshall}, F.~{Menanteau}, R.~{Miquel}, A.~{Palmese},
  F.~{Paz-Chinch{\'o}n}, A.~A. {Plazas}, A.~K. {Romer}, E.~{Sanchez},
  B.~{Santiago}, V.~{Scarpine}, S.~{Serrano}, M.~{Smith}, M.~{Soares-Santos},
  E.~{Suchyta}, G.~{Tarle}, D.~{Thomas}, T.~N. {Varga}, A.~R. {Walker}, and
  {DES Collaboration}.
\newblock {Milky Way Satellite Census. II. Galaxy-Halo Connection Constraints
  Including the Impact of the Large Magellanic Cloud}.
\newblock {\em \apj}, 893(1):48, April 2020.

\bibitem{spergel2000}
D~N Spergel and P~J Steinhardt.
\newblock Observational evidence for self-interacting cold dark matter.
\newblock {\em Phys. Rev. Lett.}, 84(17):3760--3763, April 2000.

\bibitem{Cooley:2021rws}
Jodi Cooley.
\newblock {Dark Matter Direct Detection of Classical WIMPs}.
\newblock In {\em {Les Houches summer school on Dark Matter}}, 10 2021.

\bibitem{Aalbers:2022dzr}
J.~Aalbers et~al.
\newblock {A Next-Generation Liquid Xenon Observatory for Dark Matter and
  Neutrino Physics}.
\newblock 3 2022.

\bibitem{Chang:2008aa}
J.~Chang et~al.
\newblock {An excess of cosmic ray electrons at energies of 300-800 GeV}.
\newblock {\em Nature}, 456:362--365, 2008.

\bibitem{PAMELA:2008gwm}
Oscar Adriani et~al.
\newblock {An anomalous positron abundance in cosmic rays with energies 1.5-100
  GeV}.
\newblock {\em Nature}, 458:607--609, 2009.

\bibitem{Pospelov:2008jd}
Maxim Pospelov and Adam Ritz.
\newblock {Astrophysical Signatures of Secluded Dark Matter}.
\newblock {\em Phys. Lett. B}, 671:391--397, 2009.

\bibitem{Cholis:2008qq}
Ilias Cholis, Douglas~P. Finkbeiner, Lisa Goodenough, and Neal Weiner.
\newblock {The PAMELA Positron Excess from Annihilations into a Light Boson}.
\newblock {\em JCAP}, 12:007, 2009.

\bibitem{2009PhRvD..79a5014A}
Nima {Arkani-Hamed}, Douglas~P. {Finkbeiner}, Tracy~R. {Slatyer}, and Neal
  {Weiner}.
\newblock {A theory of dark matter}.
\newblock {\em \prd}, 79(1):015014, January 2009.

\bibitem{Buckley:2009in}
Matthew~R. Buckley and Patrick~J. Fox.
\newblock {Dark Matter Self-Interactions and Light Force Carriers}.
\newblock {\em Phys. Rev. D}, 81:083522, 2010.

\bibitem{Feng:2009mn}
Jonathan~L. Feng, Manoj Kaplinghat, Huitzu Tu, and Hai-Bo Yu.
\newblock {Hidden Charged Dark Matter}.
\newblock {\em JCAP}, 07:004, 2009.

\bibitem{2011PhRvL.106q1302L}
Abraham {Loeb} and Neal {Weiner}.
\newblock {Cores in Dwarf Galaxies from Dark Matter with a Yukawa Potential}.
\newblock {\em \prl}, 106(17):171302, April 2011.

\bibitem{Tulin:2012wi}
Sean Tulin, Hai-Bo Yu, and Kathryn~M. Zurek.
\newblock {Resonant Dark Forces and Small Scale Structure}.
\newblock {\em Phys. Rev. Lett.}, 110(11):111301, 2013.

\bibitem{Tulin:2013teo}
Sean Tulin, Hai-Bo Yu, and Kathryn~M. Zurek.
\newblock {Beyond Collisionless Dark Matter: Particle Physics Dynamics for Dark
  Matter Halo Structure}.
\newblock {\em Phys. Rev. D}, 87(11):115007, 2013.

\bibitem{Vogelsberger:2012ku}
Mark Vogelsberger, Jesus Zavala, and Abraham Loeb.
\newblock {Subhaloes in Self-Interacting Galactic Dark Matter Haloes}.
\newblock {\em Mon. Not. Roy. Astron. Soc.}, 423:3740, 2012.

\bibitem{Dooley:2016ajo}
Gregory~A. Dooley, Annika H.~G. Peter, Mark Vogelsberger, Jes\'us Zavala, and
  Anna Frebel.
\newblock {Enhanced Tidal Stripping of Satellites in the Galactic Halo from
  Dark Matter Self-Interactions}.
\newblock {\em Mon. Not. Roy. Astron. Soc.}, 461(1):710--727, 2016.

\bibitem{Correa:2020qam}
Camila~A. Correa.
\newblock {Constraining velocity-dependent self-interacting dark matter with
  the Milky Way\textquoteright{}s dwarf spheroidal galaxies}.
\newblock {\em Mon. Not. Roy. Astron. Soc.}, 503(1):920--937, 2021.

\bibitem{essig2009}
Rouven Essig, Neelima Sehgal, and Louis~E. Strigari.
\newblock Bounds on cross sections and lifetimes for dark matter annihilation
  and decay into charged leptons from gamma-ray observations of dwarf galaxies.
\newblock {\em Phys. Rev. D}, 80:023506, Jul 2009.

\bibitem{DelNobile:2015uua}
Eugenio Del~Nobile, Manoj Kaplinghat, and Hai-Bo Yu.
\newblock {Direct Detection Signatures of Self-Interacting Dark Matter with a
  Light Mediator}.
\newblock {\em JCAP}, 10:055, 2015.

\bibitem{Bjorken:2009mm}
James~D. Bjorken, Rouven Essig, Philip Schuster, and Natalia Toro.
\newblock {New Fixed-Target Experiments to Search for Dark Gauge Forces}.
\newblock {\em Phys. Rev. D}, 80:075018, 2009.

\bibitem{Essig:2009nc}
Rouven Essig, Philip Schuster, and Natalia Toro.
\newblock {Probing Dark Forces and Light Hidden Sectors at Low-Energy e+e-
  Colliders}.
\newblock {\em Phys. Rev. D}, 80:015003, 2009.

\bibitem{Battaglieri:2017aum}
Marco Battaglieri et~al.
\newblock {US Cosmic Visions: New Ideas in Dark Matter 2017: Community Report}.
\newblock In {\em {U.S. Cosmic Visions: New Ideas in Dark Matter}}, 7 2017.

\bibitem{Cyr-Racine:2013fsa}
Francis-Yan Cyr-Racine, Roland de~Putter, Alvise Raccanelli, and Kris
  Sigurdson.
\newblock {Constraints on Large-Scale Dark Acoustic Oscillations from
  Cosmology}.
\newblock {\em Phys. Rev. D}, 89(6):063517, 2014.

\bibitem{Buckley:2014hja}
Matthew~R. Buckley, Jes\'us Zavala, Francis-Yan Cyr-Racine, Kris Sigurdson, and
  Mark Vogelsberger.
\newblock {Scattering, Damping, and Acoustic Oscillations: Simulating the
  Structure of Dark Matter Halos with Relativistic Force Carriers}.
\newblock {\em Phys. Rev. D}, 90(4):043524, 2014.

\bibitem{Ren:2018jpt}
Tao Ren, Anna Kwa, Manoj Kaplinghat, and Hai-Bo Yu.
\newblock {Reconciling the Diversity and Uniformity of Galactic Rotation Curves
  with Self-Interacting Dark Matter}.
\newblock {\em Phys. Rev. X}, 9(3):031020, 2019.

\bibitem{2019PhRvD.100l3006P}
K.~{Pardo}, H.~{Desmond}, and P.~G. {Ferreira}.
\newblock {Testing self-interacting dark matter with galaxy warps}.
\newblock {\em \prd}, 100(12):123006, December 2019.

\bibitem{Turner:2020vlf}
Hannah~C. Turner, Mark~R. Lovell, Jes\'us Zavala, and Mark Vogelsberger.
\newblock {The onset of gravothermal core collapse in velocity-dependent
  self-interacting dark matter subhaloes}.
\newblock {\em Mon. Not. Roy. Astron. Soc.}, 505(4):5327--5339, 2021.

\bibitem{Gilman:2021sdr}
Daniel Gilman, Jo~Bovy, Tommaso Treu, Anna Nierenberg, Simon Birrer, Andrew
  Benson, and Omid Sameie.
\newblock {Strong lensing signatures of self-interacting dark matter in
  low-mass haloes}.
\newblock {\em Mon. Not. Roy. Astron. Soc.}, 507(2):2432--2447, 2021.

\bibitem{2020ApJ...896..112N}
Ethan~O. {Nadler}, Arka {Banerjee}, Susmita {Adhikari}, Yao-Yuan {Mao}, and
  Risa~H. {Wechsler}.
\newblock {Signatures of Velocity-dependent Dark Matter Self-interactions in
  Milky Way-mass Halos}.
\newblock {\em \apj}, 896(2):112, June 2020.

\bibitem{2021MNRAS.507..720S}
Omid {Sameie}, Michael {Boylan-Kolchin}, Robyn {Sanderson}, Drona {Vargya},
  Philip~F. {Hopkins}, Andrew {Wetzel}, James {Bullock}, Andrew {Graus}, and
  Victor~H. {Robles}.
\newblock {The central densities of Milky Way-mass galaxies in cold and
  self-interacting dark matter models}.
\newblock {\em \mnras}, 507(1):720--729, October 2021.

\bibitem{Zeng:2021ldo}
Zhichao~Carton Zeng, Annika H.~G. Peter, Xiaolong Du, Andrew Benson, Stacy Kim,
  Fangzhou Jiang, Francis-Yan Cyr-Racine, and Mark Vogelsberger.
\newblock {Core-collapse, evaporation and tidal effects: the life story of a
  self-interacting dark matter subhalo}.
\newblock 10 2021.

\end{thebibliography}

\end{document}